\documentclass[12pt]{article}
\usepackage{sectsty}
\allsectionsfont{\sffamily}
\newcommand{\x}{arXiv:}

\usepackage[nosort]{cite}
\usepackage{graphicx}
\usepackage[font={scriptsize,singlespacing}]{caption}
\usepackage{epstopdf}
\usepackage{amsmath}
\usepackage{amssymb}
\usepackage{subfigure}
\usepackage{hyperref}
\usepackage{url}
\usepackage{xcolor}
\usepackage{color}
\definecolor{amaranth}{rgb}{0.9, 0.17, 0.31}
\definecolor{purple(munsell)}{rgb}{0.62, 0.0, 0.77}
\definecolor{americanrose}{rgb}{1.0, 0.01, 0.24}
\definecolor{palatinateblue}{rgb}{0.15, 0.23, 0.89}
\definecolor{royalblue(web)}{rgb}{0.25, 0.41, 0.88}
\definecolor{hanpurple}{rgb}{0.32, 0.09, 0.98}
\definecolor{beaublue}{rgb}{0.74, 0.83, 0.9}
\definecolor{carminered}{rgb}{1.0, 0.0, 0.22}
\definecolor{brightpink}{rgb}{1.0, 0.0, 0.5}
\hypersetup{ linktoc=all,
    colorlinks, linkcolor={palatinateblue},
    citecolor={brightpink}, urlcolor={purple(munsell)}
}

\def\sideremark#1{\ifvmode\leavevmode\fi\vadjust{\vbox to0pt{\vss
 \hbox to 0pt{\hskip\hsize\hskip1em
 \vbox{\hsize2cm\tiny\raggedright\pretolerance10000
 \noindent #1\hfill}\hss}\vbox to8pt{\vfil}\vss}}}%
\begin{document}
\thispagestyle{empty}
\begin{center}

\null \vskip-1truecm \vskip2truecm

{\Large{\bf \textsf{Black Hole Remnants and the Information Loss Paradox}}}

\vskip1truecm
\textbf{\textsf{Pisin Chen}}
{\small \textsf{\\(1) Leung Center for Cosmology and Particle Astrophysics \& \\ Graduate Institute of Astrophysics \& Department of Physics,\\  National Taiwan University,
Taipei 10617, Taiwan\\ (2) Kavli Institute for Particle Astrophysics and Cosmology, \\SLAC National Accelerator Laboratory, Stanford University, CA 94305, U.S.A}\\
\tt{Email: pisinchen@phys.ntu.edu.tw}}\\

\vskip0.4truecm
\textbf{\textsf{Yen Chin Ong}}\\
{\small \textsf{Nordita, KTH Royal Institute of Technology and Stockholm University, \\ Roslagstullsbacken 23,
SE-106 91 Stockholm, Sweden}\\
\tt{Email: yenchin.ong@nordita.org}}\\

\vskip0.4truecm
\textbf{\textsf{Dong-han Yeom}}\\
{\small \textsf{Leung Center for Cosmology and Particle Astrophysics,\\ National Taiwan University,
Taipei 10617, Taiwan}\\
\tt{Email: innocent.yeom@gmail.com}}\\

\end{center}
\vskip1truecm \centerline{\textsf{ABSTRACT}} \baselineskip=15pt

\medskip
Forty years after the discovery of Hawking radiation, its exact nature remains elusive. If Hawking radiation does not carry any information out from the ever shrinking black hole, it seems that unitarity is violated once the black hole completely evaporates. On the other hand, attempts to recover information via quantum entanglement lead to the firewall controversy. Amid the confusions, the possibility that black hole evaporation stops with a ``remnant'' has remained unpopular and is often dismissed due to some ``undesired properties'' of such an object. Nevertheless, as in any scientific debate, the pros and cons of any proposal must be carefully scrutinized. We fill in the void of the literature by providing a timely review of various types of black hole remnants, and provide some new thoughts regarding the challenges that black hole remnants face in the context of the information loss paradox and its latest incarnation, namely the firewall controversy. The importance of understanding the role of curvature singularity is also emphasized, after all there remains a possibility that the singularity cannot be cured even by quantum gravity. In this context a black hole remnant conveniently serves as a cosmic censor.  
We conclude that a remnant remains a possible end state of Hawking evaporation, and \emph{if it contains large interior geometry}, may help to ameliorate the information loss paradox and the firewall controversy.  
We hope that this will raise some interests in the community to investigate remnants more critically but also more thoroughly.

\newpage
\tableofcontents

\section{Information Loss and Firewall vs. Black Hole Remnant}

It has been forty years since the discovery of Hawking radiation \cite{Hawking1, Hawking2}, but we are still unsure of its exact nature --- does the radiation come from a purely thermal state or does the spectrum only look thermal with the highly scrambled information encoded in the radiation via subtle quantum entanglement? [See e.g., the recent discussion in \cite{visser01} for an attempt to clarify the precise meaning of ``thermality''.] The first scenario leads to the loss of unitarity, in the sense that if one starts from a pure state collapsing to form a black hole, the end result of the evolution [after the black hole completely evaporates away] would be a mixed state [see, however, \cite{myers01}].

\begin{figure}
\begin{center}
\includegraphics[scale=0.72]{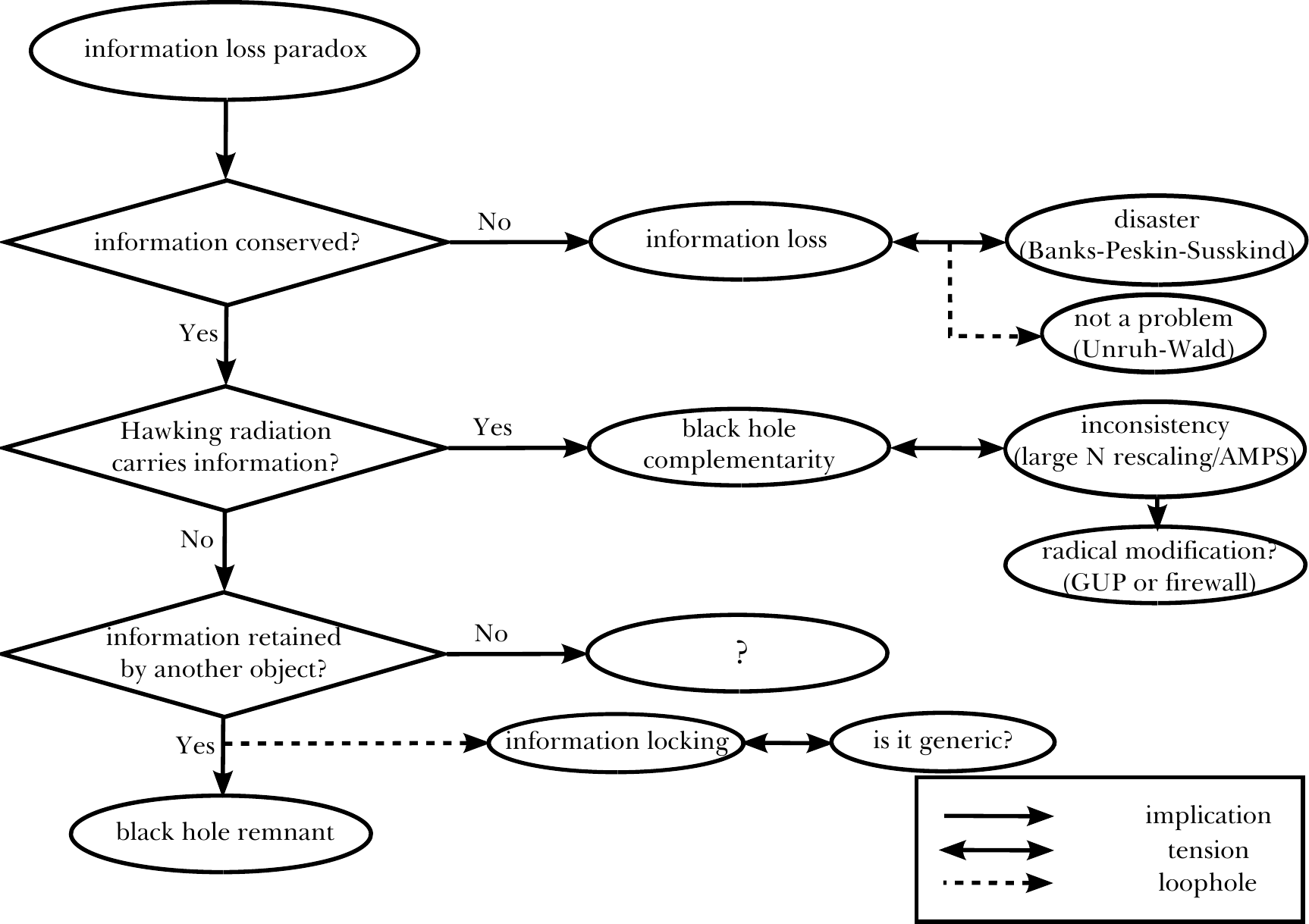}
\caption{\label{flowchart1} The information loss paradox at a glance. In this paper we are concerned with the feasibility of a black hole remnant as a possible solution to the information loss paradox. The remnant proposal is of course not without problems, see the second flowchart in Fig.(\ref{flowchart2}).}
\end{center}
\end{figure}

This apparent loss of unitary at the  end of black hole evaporation has been dubbed the \emph{information loss paradox}\cite{hawking1976}. A great deal of effort has been spent in devising possible scenarios to preserve unitarity in the context of black hole evaporation [see Fig.(\ref{flowchart1}) for a flowchart]. However, central to this debate is a crucial question that needs to be answered: \emph{what does it mean to resolve the information loss paradox?} First of all, the paradox is only a paradox if information loss is indeed a problem. Indeed there are people -- especially the relativists -- who remain unconvinced that this is a problem \cite{Wald1, HP}. Recently, it has also been found that even string theory admits non-unitary holography \cite{vafa}. Banks, Peskin and Susskind had argued in their classic work \cite{BPS0}  that generically a non-unitary evolution would lead to violations of  either causality or energy conservation. 
Of course in general relativity [GR] one usually does \emph{not} worry about energy conservation since energy is in fact \emph{not} conserved unless there is a timelike Killing vector field. However here the concern is that even \emph{locally} at the laboratory scale there would be drastic effect arising from the non-unitary evolution, leading to the temperature of the vacuum being $10^{32}$ K. This concern was, however, countered by Unruh and Wald \cite{WU}, asserting that such a violation would be negligibly small if the evolution is not assumed to be a Markov process. [See also the more recent work by Nikolic \cite{nikolic}.]

It is perhaps fair to say that most workers in the field are not comfortable with information being truly lost in black holes, and are therefore more receptive to the idea that Hawking radiation would somehow encode information. The leading proposal of Susskind, Thorlacius and Uglum \cite{STU}, known as the ``black hole complementarity principle'', asserts that as long as the infalling observer Alice [who carries some information into the black hole] cannot compare notes with the exterior observer Bob [who saw Alice burned up at the horizon due to the high energy particles emitted by the Hawking process], there is no true paradox. So while Alice carries her information into the black hole interior, this information can still be recovered by Bob, by reading the information encoded in the Hawking radiation.

However, recently AMPS [short for Almheiri-Marolf-Polchinski-Sully] \cite{amps, apologia} argued that the effort to get information out via purifying the early Hawking radiation by maximally entangling it with the late Hawking radiation would result in a ``firewall''  [see also \cite{sam}] that would burn up anyone who attempts to cross the horizon of the black hole. Essentially, the argument goes as follows: the requirement that the late time Hawking radiation must purify the earlier radiation means that they must be maximally entangled; and this breaks the entanglement between the late time Hawking radiation and the interior of the black hole. Recall that the Hamiltonian for a quantum field $\varphi$ contains terms like $(\partial_x \varphi)^2$, where $x$ denotes a spatial direction. Therefore, if the field configuration is not smooth over the horizon [since fields outside and inside are not correlated, there is no guarantee of continuity], this implies the presence of highly energetic particles in the neighborhood of the horizon --- the firewall. [For a detailed pedagogical introduction to the information paradox and quantum information, see \cite{harlow}.]

The firewall arises when the black hole already lost about half of its entropy \cite{page1, page2}, which means that the black hole can still be quite large when the firewall descends onto its horizon. This is in stark contradiction with the expectation from general relativity that one should not expect anything out of the ordinary at the horizon when the curvature is negligibly small [for a sufficiently large black hole]. Thus it would seem that if firewall exists, our understanding of quantum gravity has led us to a theory that does \emph{not} reduce back to general relativity [and QFT on curved background] at the energy scale that it should be valid. This is somewhat unpalatable and thus the debate ensues. 

One of the reasons why black hole evaporation is a difficult problem especially when information loss is concerned, is that the temperature of an asymptotically flat Schwarzschild black hole -- which has negative specific heat -- becomes unbounded as its mass monotonically decreases toward zero. This is sometimes taken to mean that new physics may come in at some sufficiently high temperature and our semi-classical effective field theory ceases to be valid. However, it has been argued that the use of the macrocanonical treatment of black hole evaporation
breaks down when the typical energy of the particles emitted
[i.e. the temperature] becomes comparable to the energy of the
object left behind. In this case the microcanonical ensemble
should be used, which removes the unphysical divergence in the
temperature \cite{ham1,ham2,ham3, 0412265}. Nevertheless, problems remain since the curvature [the Kretschmann scalar $R^{\alpha\beta\gamma\delta}R_{\alpha\beta\gamma\delta}$] at the horizon becomes larger and larger as the black hole shrinks. This is related to the curvature blow-up of the singularity, which should, as is often argued, be resolved by a quantum theory of gravity.
This opens up other possible ways to realize an eventual finite Hawking temperature. For example, the fate of the black hole may not be complete evaporation but a ``remnant'' \cite{pisin}, the idea which at least dates back to the work of Aharonov, Casher, and Nussinov in 1987 \cite{ACN}. 

The common assumptions regarding black hole physics, which are embodied in the complementarity principle \cite{kn:stu}, are:
\begin{itemize}
\item[(1)] Unitarity is preserved: Information is recovered by the exterior observers after a black hole evaporates away.
\item[(2)] ``No Drama'': An infalling observer does not feel anything out of the ordinary when crossing the horizon, as demanded by general relativity.
\item[(3)] Standard Quantum Field Theory: Local effective field theory is valid near the horizon, as well as at asymptotic infinity. [That is, in particular, Hawking's derivation of Hawking temperature is correct.]
\item[(4)] The statistical entropy of a black hole is the same as its Bekenstein-Hawking entropy. [That is, the horizon area is a measure of the underlying quantum gravitational degrees of freedom of the black hole. We will briefly review the different notions of entropy in \ref{A}.] 
\item[(5)] The existence of an observer who can collect and read the information contained in the Hawking radiation.
 \end{itemize}
What AMPS argued was that these assumptions cannot all be true. Prior to the AMPS argument, it has also been pointed out that if one introduced a large number of scalar fields, then under the so-called large $N$ rescaling, complementarity principle can be violated \cite{Yeom:2008qw1,Yeom:2008qw0}, although this can be avoided if we take into account the generalized uncertainty principle \cite{COY}. AMPS suggested that the ``most conservative'' solution would be to abandon Assumption (2), and introduced the notion of a firewall. However, there are other possibilities. As we shall see, the argument that black hole remnants may solve the information paradox would depend on dropping Assumption (4). Indeed, Aharonov et al. \cite{ACN} wrote:
\begin{quote}
\emph{``If the residual mini black hole disappears completely by decaying into ordinary particles we will be deprived of the large reservoir of states in the black hole which is required for the construction of the overall wave function.''}
\end{quote}
By ``overall wave function'' they meant the wave function that maintains unitarity by correlating the states in the black hole interior to the states of the Hawking radiation outside:
\begin{equation}
\left|\psi^{\text{total}} \right\rangle = \sum_n C_n \left|\psi_n^{\text{radiation}} \right\rangle  \otimes \left|\phi_n^{\text{interior}}\right\rangle.
\end{equation}
A remnant thus seems to contain a large amount of quantum states, far beyond their storage capability suggested by the Bekenstein-Hawking entropy, which is proportional to its small surface area.

The idea that a remnant can store information, or that information has simply gone elsewhere, e.g. into another universe, is not a very popular one. The usual notion of ``resolving'' the information loss paradox is that an exterior observer should be able to recover all the information that falls into a black hole. This perhaps is partially due to the common acceptance of the AdS/CFT correspondence. However, we feel that not enough attention has been paid to examine the prospects and challenges faced by the remnant scenario as a solution to the information loss paradox. This would therefore be the central theme of this work.

Let us start with discussing what a remnant actually is.
The term ``remnant'' has been used rather loosely without a clear-cut definition, which is understandable and -- to some extent -- justified, since we know next to nothing about what physics looks like at that energy scale. In this paper we would take a rather broad definition for a remnant: 
\begin{quote}
\textbf{Definition:} A \emph{remnant} is a localized late stage of a black hole under Hawking evaporation, which is either (i) absolutely stable, or (ii) long-lived.
\end{quote}
By long-lived, or meta-stable remnant, we mean the lifetime of the remnant is longer than the lifetime of a black hole in the remnant-less case. For example, consider an asymptotically flat Schwarzschild black hole. 
The time scale for a complete evaporation is of the order $M^3$, where $M$ is the initial mass of the hole. If the black hole becomes a long-lived remnant, its lifetime is of the order $M^n$, where $n \geqslant 4$.

\begin{figure}
\begin{center}
\includegraphics[scale=1.00]{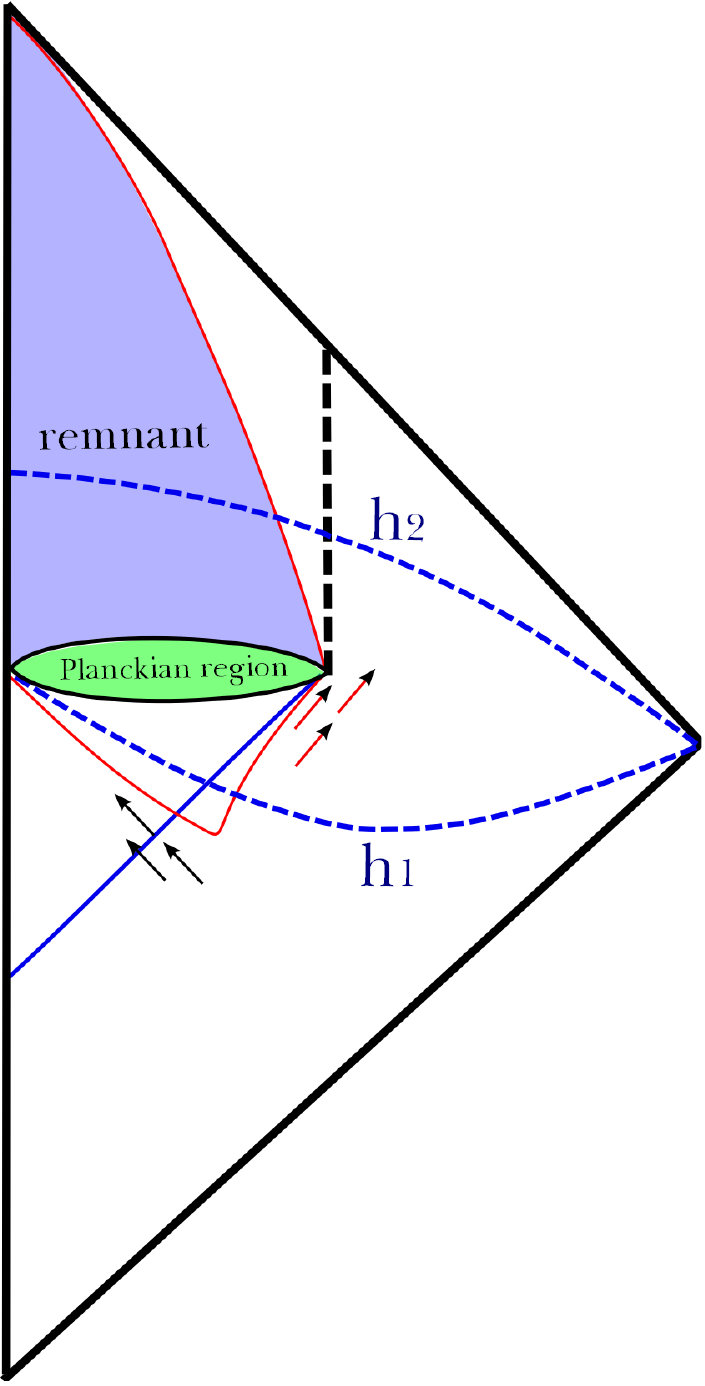}
\caption{\label{fig:remnant} (Color online) Penrose diagram for the black hole remnant, assuming that the lifetime of the remnant is infinite. The solid curve outside the boundary of the ``Planckian region'' [where quantum gravity becomes important] denotes the apparent horizon. Outgoing arrows represent the Hawking radiation, while ingoing arrows denote the in-falling matter. Here $h_1$ and $h_2$ are two Cauchy hypersurfaces. Note that in classical GR, if the matter field satisfies the null energy condition [NEC], then an apparent horizon is always inside the event horizon. This is no longer true if NEC is violated, e.g., in the presence of Hawking radiation, or in modified theories of gravity. See \cite{1309.4915} for further discussions.}
\end{center}
\end{figure}

Here, ``localized'' means that the remnant is confined to a small region [at least as seen from the outside], this excludes Hawking radiation itself from being included in the definition.
Long-lived remnants are configurations that slowly emit particles through other non-Hawking emission channels and slowly lose their mass, although their life times are so long that for all practical purposes they can be treated as ``stable''.
Note that it is possible that even an ``absolutely stable'' state can only be reached asymptotically. 
That is to say, starting with a black hole of mass $M$, it can eventually radiate down to some mass $m < M$ [in many cases the remnant mass would be \emph{much less} than the original black hole mass; and by mass here we mean the ADM mass in asymptotically flat geometry, and its analog in the case of other geometries, suitably defined] and then either ceases to radiate completely, or it can radiate at increasingly slower rate, never really reaching its ``end state'', though can come arbitrarily close to it. The second possibility is similar to a charged black hole that obeys the third law of black hole mechanics, never reaching its extremal state. Indeed, as we shall see in this review, many [though not all] remnant scenarios turn out to be some kind of extremal black holes. Fig.(\ref{fig:remnant}) shows the Penrose diagram of a remnant, with the assumption that the remnant is completely stable and has an infinite lifetime [see more discussion in subsec.(\ref{overproduce})].

Note that a remnant may or may not have a well-defined horizon. In fact, it is not obvious how much geometrical notions such as area and volume still make sense at the Planck scale. Even its size is a matter of debate -- some remnants are Planck size or the size of an elementary particle [at least as seen from the exterior], e.g., ``maximon'', ``friedmon'', and ``planckon'' \cite{ACN, markov1, markov2, markov3, MM0, MM}, while some ``remnants'' can be astrophysical \cite{ellis}, which we will not discuss much in this review [the proposal by Ellis in \cite{ellis} is that Hawking radiation and the cosmic microwave background photons would interact in such a way that most Hawking particles end up \emph{inside} the horizon and never escape from the black hole].  
Indeed there are many other types of black hole remnants arising from either modifications to the uncertainty principle or modified theories of gravity, and almost all of them had been proposed as a possible solution to the information loss paradox, with the information either locked inside the remnant indefinitely [the ``stable remnant'' scenario], or the information slowly leaking out after the remnant stage is reached [the ``long-lived/meta-stable remnant'' scenario]. Nevertheless, there are various problems with these proposals, and a black hole remnant is often dismissed as a viable alternative solution to the information loss paradox. However, remnant scenarios deserve a closer scrutiny, especially amid the current confusion regarding firewalls. 

In this review, we will start with a broad and brief overview of the various remnant scenarios arisen from both modified quantum mechanics and modified gravity, as well as their roles
in resolving the information loss paradox.
To be more specific, we will present a review on the reasons against remnants, as well as the counter-arguments against these objections. In addition, a remnant has the virtue that it shrouds the curvature singularity of the black hole, in agreement with the cosmic censorship conjecture \cite{penrose}. One usually expects that the singularity is resolved in a complete theory of quantum gravity. However, it is uncertain whether this notion will indeed happen.
The role of the singularity is important since information can presumably hit the singularity and be lost \cite{Wald1, HP, ellis}; or if the singularity is replaced by something else in the full theory of quantum gravity, be transferred to there. It thus seems somewhat perplexing that much effort was spent looking at the event horizon while information loss requires also the full understanding of the interior of the black hole\footnote{String theoretic approaches, \emph{\`a la} Papadodimas-Raju \cite{raju1, raju2, raju3, monica}, of probing the interior of black holes may eventually offer some clues. Another recent work that concerns black hole interior in quantum gravity could also be of interest \cite{1412.7539}.}, including but not limited to its singularity \cite{sabine}. The true meaning of unitarity in quantum mechanics, after all, should include all spacetime regions, including the interior of a black hole.
We would therefore re-emphasize this crucial point in our review, and examine the possibility that the interior of a black hole consists of non-trivial geometry, such as Wheeler's ``bag-of-gold'' spacetime \cite{wheeler}, which in turn may lead to the possibility that remnants are only point-like from the outside, but should nevertheless not be treated as a point particle.   

The purpose of this review paper is to give an overview of the issue from the viewpoint of the information loss paradox and the firewall controversy, and its connection with the notion of black hole remnants. This is roughly summarized in Fig.(\ref{flowchart2}) below.
It is nevertheless more than just a review, as we also give a preliminary but new analysis of the issue, with the conclusion that remnants might well be a possible end state for black holes undergoing Hawking evaporation. This
can indeed ameliorate the information loss paradox and the firewall problem, \emph{provided} it contains non-trivial interior geometries, in which case the Bekenstein-Hawking entropy cannot be the true measure of the information contained in a black hole. 
We hope that this will revive some interests in the community to investigate remnants not only more critically but also more thoroughly. 

\begin{figure}
\begin{center}
\includegraphics[scale=0.72]{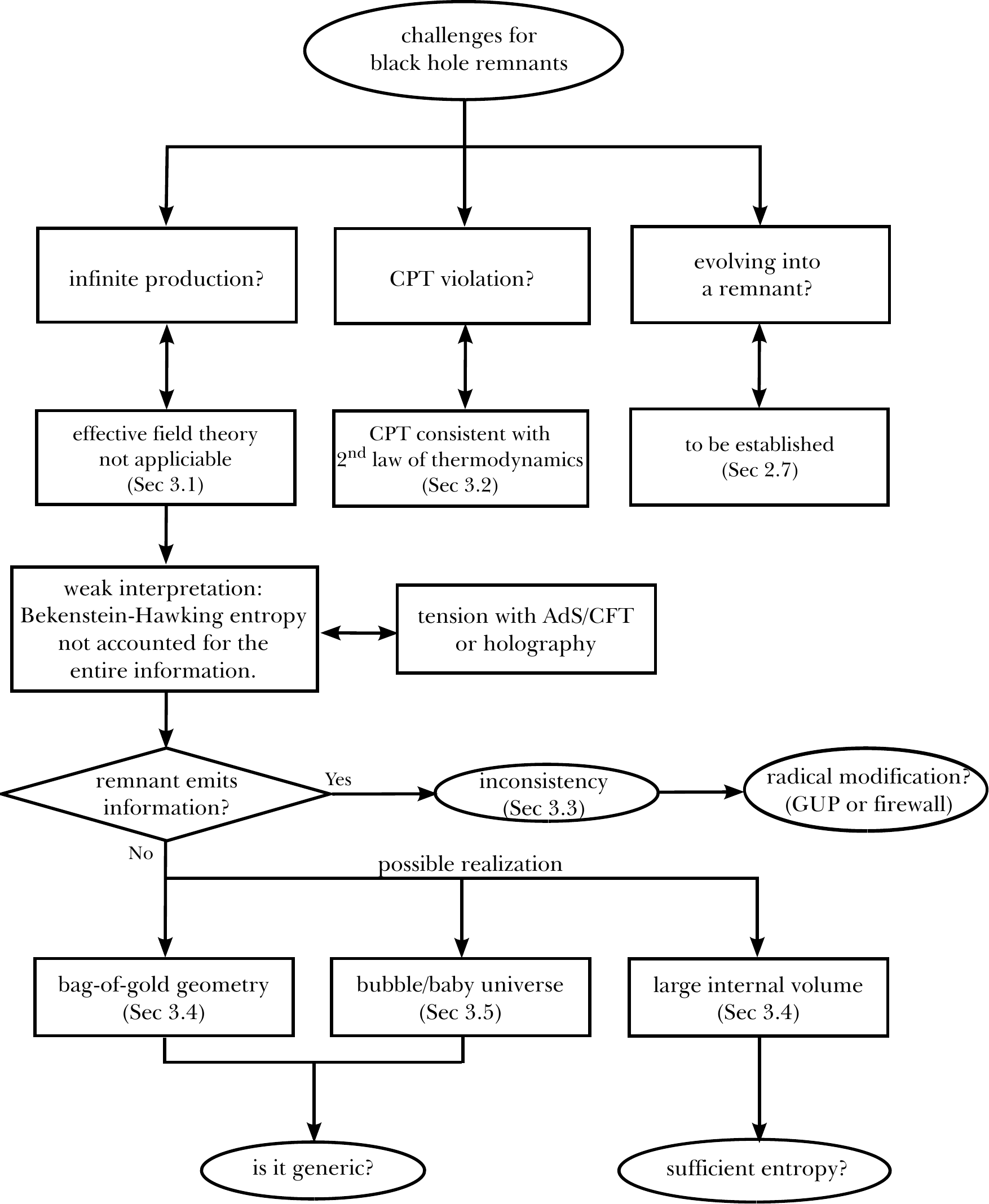}
\caption{\label{flowchart2} This flowchart shows the challenges faced by the black hole remnant proposal, and the respective sections in this paper in which the issues are discussed.}
\end{center}
\end{figure}

\section{The Various Roads to the Remnant Scenario}

Let us start by briefly reviewing the different models that lead to black hole remnants. With a large amount of literature now available, we do not claim to be exhaustive in our coverage, however we shall attempt to present the big picture that many different approaches to quantum gravity phenomenology have hinted at the existence of remnants, though not necessarily of the same type. These approaches include the following: the generalized uncertainty principle [of various types], regular black holes, quantum gravitational effects [e.g. loop quantum gravity and non-commutative geometry], modified theories of gravity, as well as modifications of semi-classical gravity [either by modifying the geometry or the matter sector]. 

\subsection{Generalized Uncertainty Principle: Quantum Physics at the Planck Scale}

The main idea of a black hole remnant is that as a black hole gets smaller and smaller when it gradually Hawking evaporates away, eventually at some minimal length -- possibly near the Planck length $L_\text{p} := \sqrt{G\hbar/c^3}$ -- new physics beyond the effective field theory used in the low energy regime kicks in, and stops the black hole from evaporating any further. One such proposal is to consider a modification to the Heisenberg uncertainty principle. Such modification can be obtained by quite general considerations of quantum mechanics and gravity \cite{1, 2, 3, 4}, but string theory also supports this scenario \cite{5, 6, 7, 8, 9}. 

The generalized uncertainty principle [GUP] is given by
\begin{equation}\label{GUP}
\Delta x\Delta p \geqslant \frac{1}{2}\left(\hbar + \alpha L_{\text{P}}^2\frac{(\Delta p)^2}{\hbar}\right),
\end{equation}
where $\alpha$ is a dimensionless parameter of order unity\footnote{This parameter can be constrained via observations and experiments. See for example, \cite{SE1, FabioRoberto}.}. Note that the $\alpha \to 0$ limit simply recovers the ordinary uncertainty principle.
GUP implies the following inequality:
\begin{eqnarray}
\frac{M_{\mathrm{P}}^{2}}{\alpha} \Delta x \left( 1 - \sqrt{1 - \frac{\alpha}{M_{\mathrm{P}}^{2} \Delta x^{2}}} \right) \leqslant \Delta p \leqslant \frac{M_{\mathrm{P}}^{2}}{\alpha} \Delta x \left( 1 + \sqrt{1 - \frac{\alpha}{M_{\mathrm{P}}^{2} \Delta x^{2}}} \right),
\end{eqnarray}
from which we note that the square-root sign imposes $\Delta x \geqslant \Delta x_{\mathrm{min}}$, where
\begin{eqnarray}
\Delta x_{\mathrm{min}} := \frac{\sqrt{\alpha}}{M_{\mathrm{P}}}.
\end{eqnarray}
That is to say, GUP incorporates the idea that there is a ``minimal length''.

By considering GUP instead of the conventional Heisenberg's uncertainty principle, it was suggested in \cite{pisin} that the modified Hawking temperature should take the following form:
\begin{equation}\label{ACS}
T_{\text{GUP}} = \frac{M}{4\pi}\left(1-\sqrt{1-\frac{M_p^2}{M^2}}\right).
\end{equation}
From Eq.(\ref{ACS}) onward, we will use the units in which $c=\hbar=k_B=\alpha=G=1$ [although sometimes it would be useful to restore them]. 

The entropy of the black hole can be derived by integrating the first law of black hole mechanics $dS= T dM$ [which does \emph{not} require any energy condition; see the discussion in \cite{primer}]. The result, upon normalizing so that the Bekenstein-Hawking entropy is zero when the black hole reaches the Planck mass, is \cite{pisin}:
\begin{equation}
S_{\text{GUP}} = 2\pi \left[\frac{M^2}{M_\text{P}^2}\left(1-\frac{M_\text{P}^2}{M^2}+\sqrt{1-\frac{M_\text{P}}{M^2}}\right)-\log\left(\frac{M+\sqrt{M^2-M_\text{P}^2}}{M_\text{P}}\right)\right].
\end{equation}

\begin{figure}[!h]
\centering
\mbox{\subfigure{\includegraphics[width=2.1in]{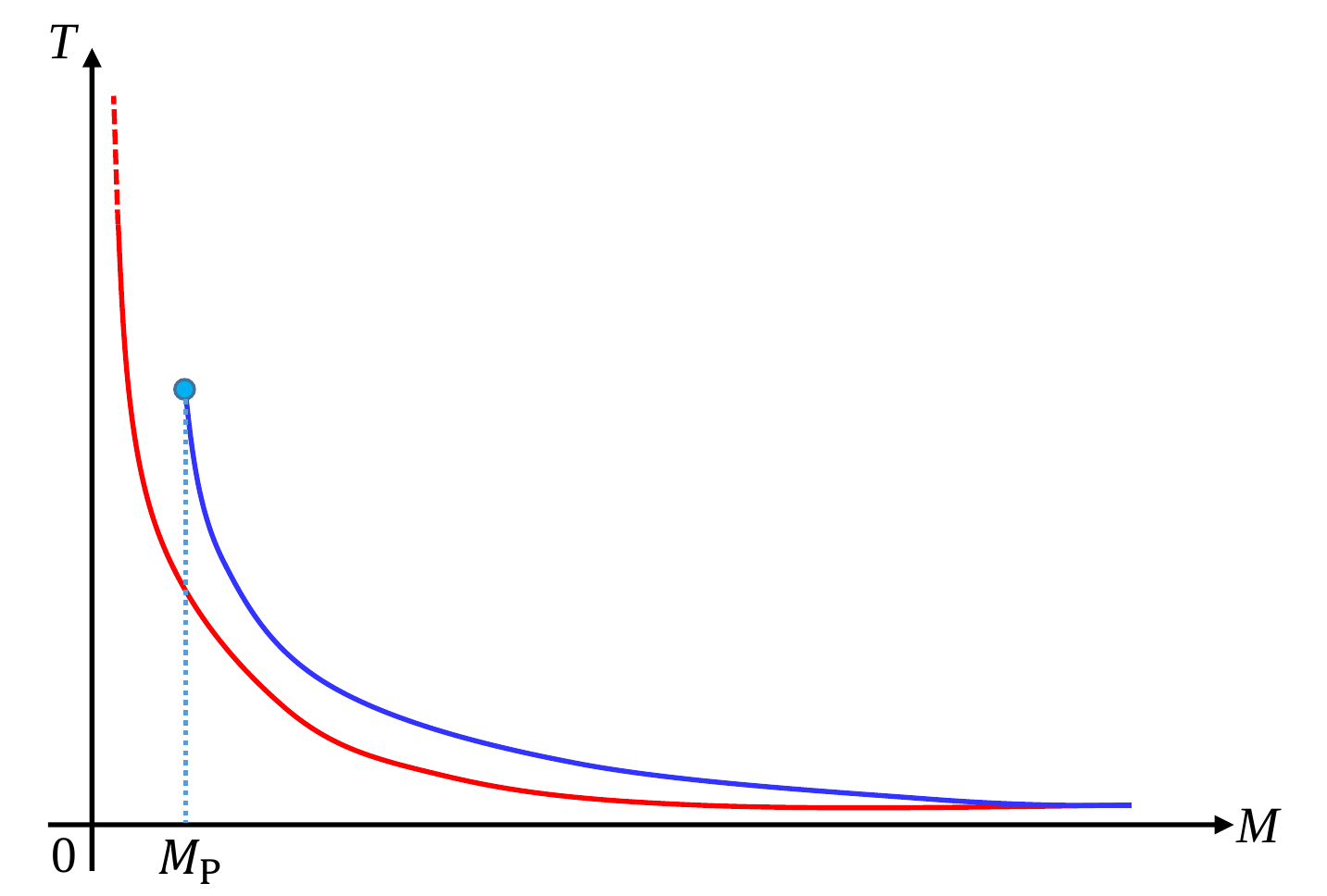}}\quad
\subfigure{\includegraphics[width=2.1in]{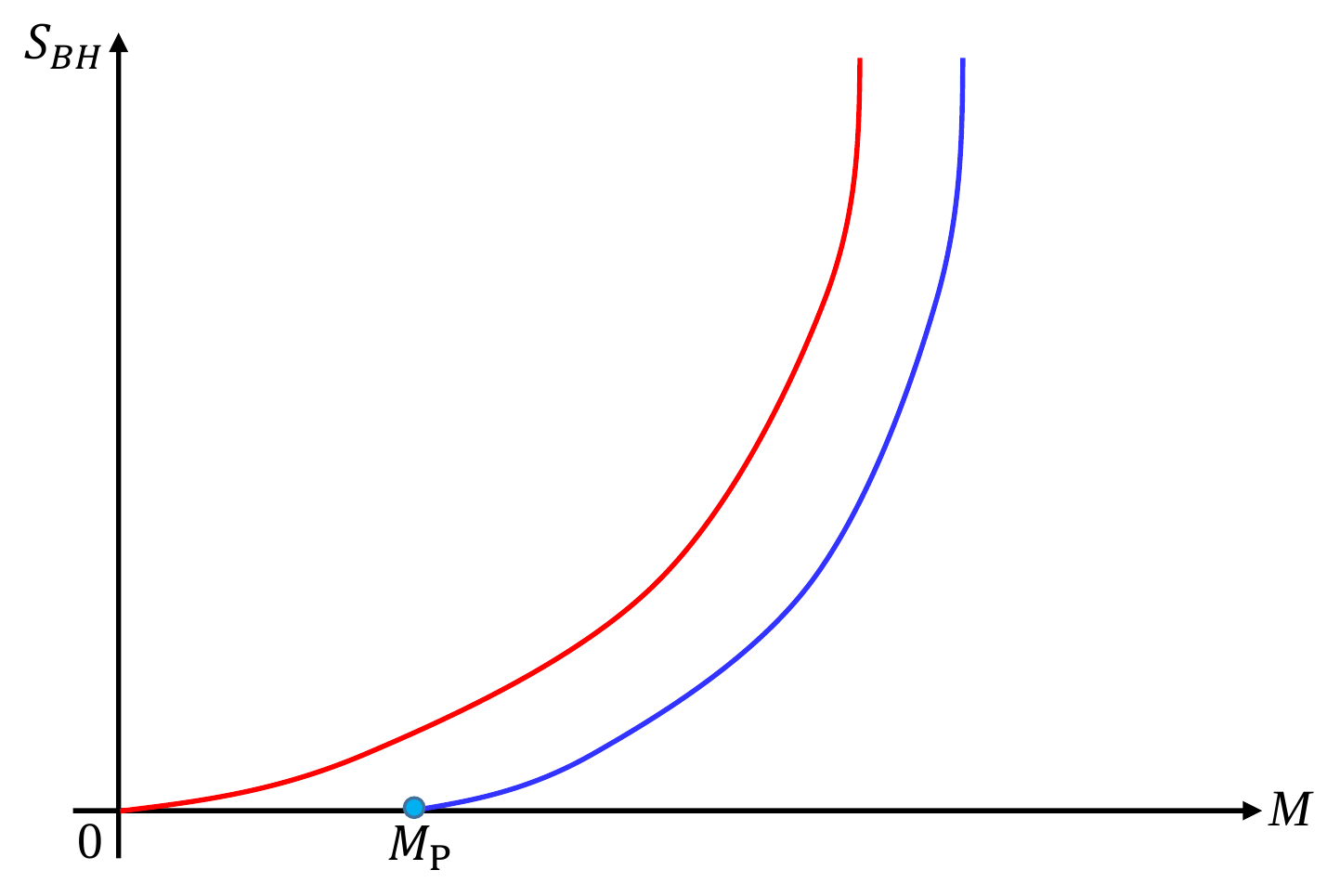} }}
\caption{\textbf{Left:} (Color online) A schematic sketch of the temperature of a Schwarzschild black hole as a function of its mass. In the conventional picture [bottom curve], the temperature is unbounded above and becomes arbitrarily hot as the black hole gets arbitrarily small [Keeping in mind, however, our discussion on page 5 about the difference between the microcanonical and canonical treatments of the thermodynamics.]. Generalized uncertainty principle suggests that the evaporation stops when the black hole reaches the Planck mass, $M_\text{P}$ [top curve]. \textbf{Right:} A schematic sketch of the Bekenstein-Hawking entropy of a Schwarzschild black hole as a function of its mass. In the conventional picture [top curve], the entropy steadily decreases to zero as the black hole completely evaporates away. In the GUP picture [bottom curve], the entropy can be normalized to zero as the black hole reached the Planck mass. \label{GUPeffect}}
\end{figure}

The evolution of the Hawking temperature and the Bekenstein-Hawking entropy is illustrated schematically in Fig.(\ref{GUPeffect}). We note that the derivative $dM/dT$ is of course the definition of the specific heat. Therefore we observe that a Schwarzschild black hole still has negative specific heat even under GUP modification. However $dT/dM \to -\infty$ as $M \to M_p$, so that when the mass of the hole reduces to Planck mass, the specific heat becomes zero\footnote{The notion of a fundamental length or discreteness does not necessarily lead to remnants. For example, in the approach using crystal-like lattice \cite{JKS}, in the final state of the evaporation, the black hole reduces to zero rest mass -- there is no remnant.}.  Note that this GUP remnant has nonzero finite temperature, this is different from many remnants that are extremal black holes [and thus are zero temperature objects]. Of course, not all GUP models yield remnants of the same thermodynamical behavior.

Note that with the logarithmic correction term [see also \cite{0611028}], the Bekenstein-Hawking entropy is no longer strictly equal to a quarter of the horizon area. This is one of the first hints that entropy content in a black hole could be different from its area measure. We will discuss more about this later. At the Planck scale, it is no longer clear that the concept of a horizon, or even the metric itself still makes sense\footnote{Such a metric is constructed recently in \cite{IMN}, by mimicking the effects of the GUP via a short scale modified version of general relativity. }.  Over the years, progresses have been made in both understanding GUP as well as applying GUP to various black holes [which consistently yield remnant pictures] \cite{0511054, 0704.1261, 1002.2302, DCXZ, 1306.5298, 1307.7045, 1307.1894, 1307.0172, 1312.3781, 1402.2133, 1404.6375, HAK, AHM}. Note that, the form of GUP could vary depending on the models considered, and as such the black holes of various GUP models could have different thermodynamical properties. For example, a correction term of the form $\sqrt{A}$ to the Bekenstein-Hawking entropy $S_{\text{bh}}=A/4$ can be obtained \cite{1402.2133} by assuming that GUP has a linear term in the momentum, i.e., with some parameter $\mathcal{A}$,
\begin{equation}
\Delta x\Delta p \geqslant\frac{1}{2}\left(\hbar + \mathcal{A}  L_{\text{P}} \Delta p+ \alpha L_{\text{P}}^2\frac{(\Delta p)^2}{\hbar}\right).
\end{equation}
In yet some other models, it is possible to obtain the change of sign in the specific heat of the black hole in addition to a minimum mass, and so a phase transition could occur that  may render black hole ``remnants'' unstable \cite{1205.3680}. 

In addition to the generalized uncertainty principle, which suggests that Hawking evaporation stops at some minimal length scale $\Delta x_{\mathrm{min}} = \sqrt{\alpha}/M_{\mathrm{P}}$, 
the notion of black hole remnants are also supported by various approaches to quantum gravity with a similar notion of minimal length. [For a review on minimal length in quantum gravity, see \cite{sabine2}; and for the implications of a minimal length to black hole physics in particular, see \cite{CMN}. In addition, a review on GUP is also available \cite{tawfik}.]  It has however been pointed out in \cite{DMZ} that the existence of a minimal length is not necessary for the occurrence of a black hole remnant. 

There have been some proposals that by taking GUP into account, the Hawking radiation is modified to be non-thermal and therefore one may resolve the information loss paradox. For such an approach, see e.g., \cite{0804.4221, 0905.0948}.

\subsection{Singularity, Regular Black Hole, and the Remnant Connection}\label{regular}

\begin{figure}
\begin{center}
\includegraphics[scale=0.75]{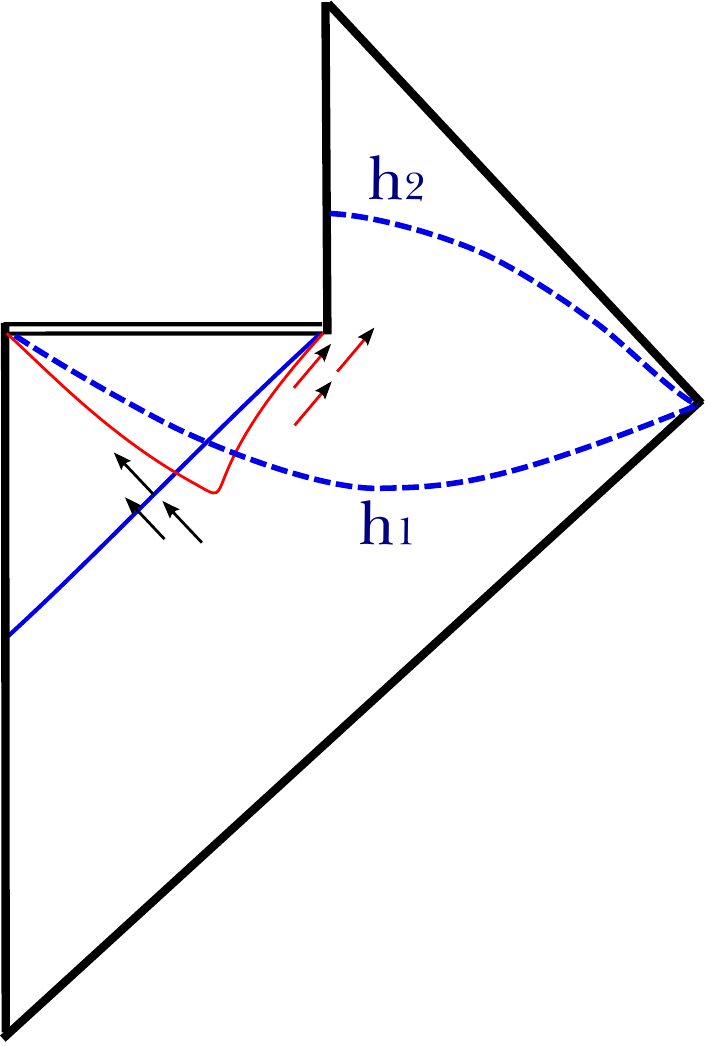}
\caption{\label{fig:Schwarzschild}(Color online) The usual causal structure of an evaporating Schwarzschild black hole. The solid curve denotes the apparent horizon. Outgoing arrows denote the Hawking radiation, while ingoing arrows denote the in-falling matter. Here, $h_1$ and $h_2$ are two Cauchy hypersurfaces. Double line denotes the singularity.}
\end{center}
\end{figure}

Traditionally, the information loss paradox is related to the existence of the singularity and the event horizon. If a piece of information falls into a Schwarszchild black hole, it will eventually crash into the singularity, which is spacelike\footnote{\color{black}A realistic black hole will not be completely neutral and static. One might then suspect that the singularity will become timelike. In that case information will not necessarily hit the singularity. However, such black holes also come with an inner Cauchy horizon, beyond which we lose predictability. Therefore the scenario is analogous to that of a singularity. Of course the inner horizon is not expected to be stable, and thus realistic singularities may still be generically spacelike, but this raises the question: what is the metric of a realistic black hole?}. This is a problem since the singularity is not ``predictable'' in classical general relativity, i.e. the theory breaks down at the singularity and there is no ``natural'' boundary conditions that one could impose for the field equations there. In this sense, information loss is trivial since the spacetime manifold is \emph{incomplete} and after the evaporation we lose the Cauchy data that were initially in the black hole. In Fig.(\ref{fig:Schwarzschild}), $h_1$ denotes an initial Cauchy surface, a part of which is inside the black hole horizon. The information on this portion eventually crashes into the singularity, and is no longer present on $h_2$, a Cauchy surface at a later time after the black hole has completely evaporated away.

On the other hand, if we can resolve the singularity, then even after the evaporation, we can retain a complete Cauchy surface so that the causal past and future of the spacetime remains fully determined. If this is indeed the case, then there is no room for information loss in this complete spacetime.

Naively, one could obtain a complete spacetime by ``resolving the singularity'' [whatever that means]. After resolving the singularity, if one can still write down a classical metric [this is a very non-trivial assumption], then this metric is called a \emph{regular black hole}. The first known model is the so-called Bardeen black hole \cite{Bardeen}:
\begin{eqnarray}
ds^{2} = - F(r) dt^{2} + \frac{1}{F(r)} dr^{2} + r^{2} d\Omega^{2},
\end{eqnarray}
where
\begin{eqnarray}
F(r) = 1 - \frac{2Mr^{2}}{(r^{2} + a^{2})^{3/2}},
\end{eqnarray}
in which $a$ is a constant. 
Note that $F(r) \approx 1 - 2M/r$ for the large $r$ limit and hence asymptotically the geometry appears the same as the Schwarzschild solution. The equation $F(r) = 0$ has two real solutions and this implies that this black hole has two horizons -- one outer and one inner horizon. Near the $r = 0$ region, the nonzeroness of $a$ prevents the geometry  from forming a singularity. Hence, $r=0$ becomes a regular and time-like center. These three properties, that the geometry (1) asymptotically approaches a known solution, (2) possesses two apparent horizons, and (3) has a regular and non-space-like center, are quite common properties among the various regular black hole models. 

By the virtue of  property (2), as the black hole evaporates, it is often argued that the two horizons will approach each other and eventually the black hole tends to a zero temperature\footnote{There are subtleties involved in this argument, which we will further discuss in subsec.\ref{concluding remarks}.}. Therefore, in many regular black hole models, the lifetime is much longer than that for semi-classical black holes, perhaps even infinite. [Similar -- but not entirely the same -- behavior also appears in the evolution of charged black holes in general relativity, see subsec.\ref{concluding remarks}.]

Note that as long as we rely on general relativity, we are subject to the singularity theorems. According to the Penrose singularity theorem \cite{P} [see also \cite{HE}, and the generalized theorem of Galloway-Senovilla \cite{GS}], if we assume the existence of a [future] trapped surface, the global hyperbolicity condition, and the null energy condition, then the singularity is unavoidable\footnote{It is important, however, to take note that the singularity here refers to geodesic incompleteness, it does not necessarily mean that scalar quantities constructed from the curvature tensor diverge.}. More precisely, 

\begin{quote}
\textbf{Singularity Theorem [Penrose (1965)]:} Suppose $(M, g_{ab})$ is a connected globally hyperbolic spacetime such that
\begin{itemize}
\item[(i)] there is a future trapped surface $\Sigma \subset M$,
\item[(ii)] the Ricci curvature satisfies $\text{Ric}(X,X)>0$ for all null vectors $X$, and
\item[(iii)] there exist a non-compact Cauchy surface $\mathcal{S} \subset M$, 
\end{itemize}
then there exists an inextendible null geodesic that has finite affine length. 
\end{quote}

\begin{figure}
\begin{center}
\includegraphics[scale=0.9]{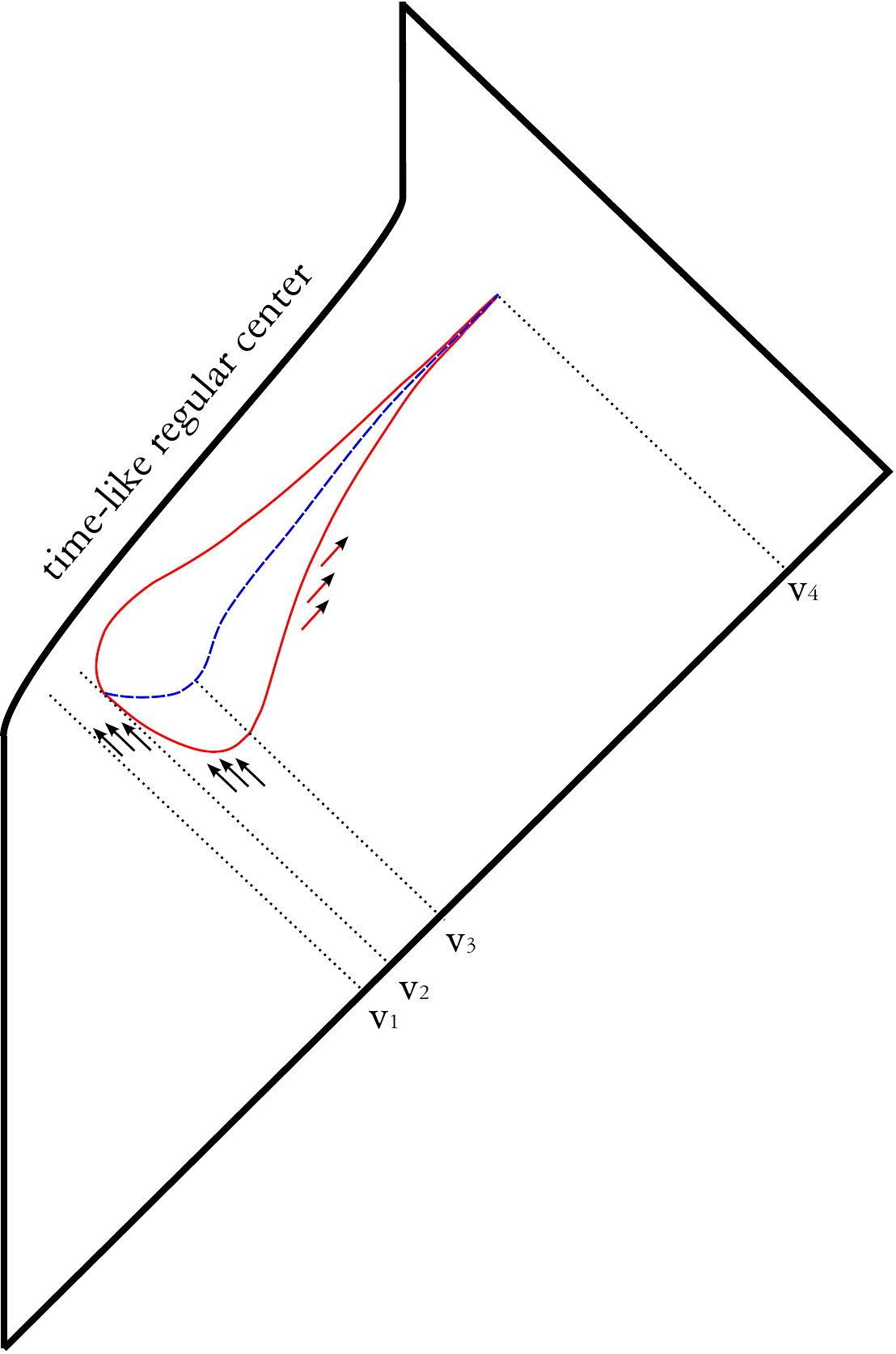}
\caption{\label{fig:regular_2}(Color online) An example of a regular black hole: the Hayward model \cite{Hayward:2005gi}. The solid curve denotes the apparent horizon. The critical mass is collapsed between $v_{1} \leqslant v \leqslant v_{2}$; after $v_{2}$, both the outer and inner apparent horizons are formed and the ``false vacuum lump'' is trapped by the horizons. The  dashed curve denotes the boundary of the false vacuum lump between the  horizons. Note that before $v_{2}$, the boundary should be time-like, although this is not depicted. Here $v_{3}$ is the time when the black hole stops to collapse and begins to evaporate; and $v_{4}$ is the time when the evaporation ends.}
\end{center}
\end{figure}

Note that condition (ii), i.e., the so-called \emph{null Ricci condition}, is precisely -- via the Einstein field equations -- equivalent to the null energy condition. However, it is really a \emph{geometric} condition, and is thus actually applicable to more general geometric theories than GR. 

\begin{figure}
\begin{center}
\includegraphics[scale=0.48]{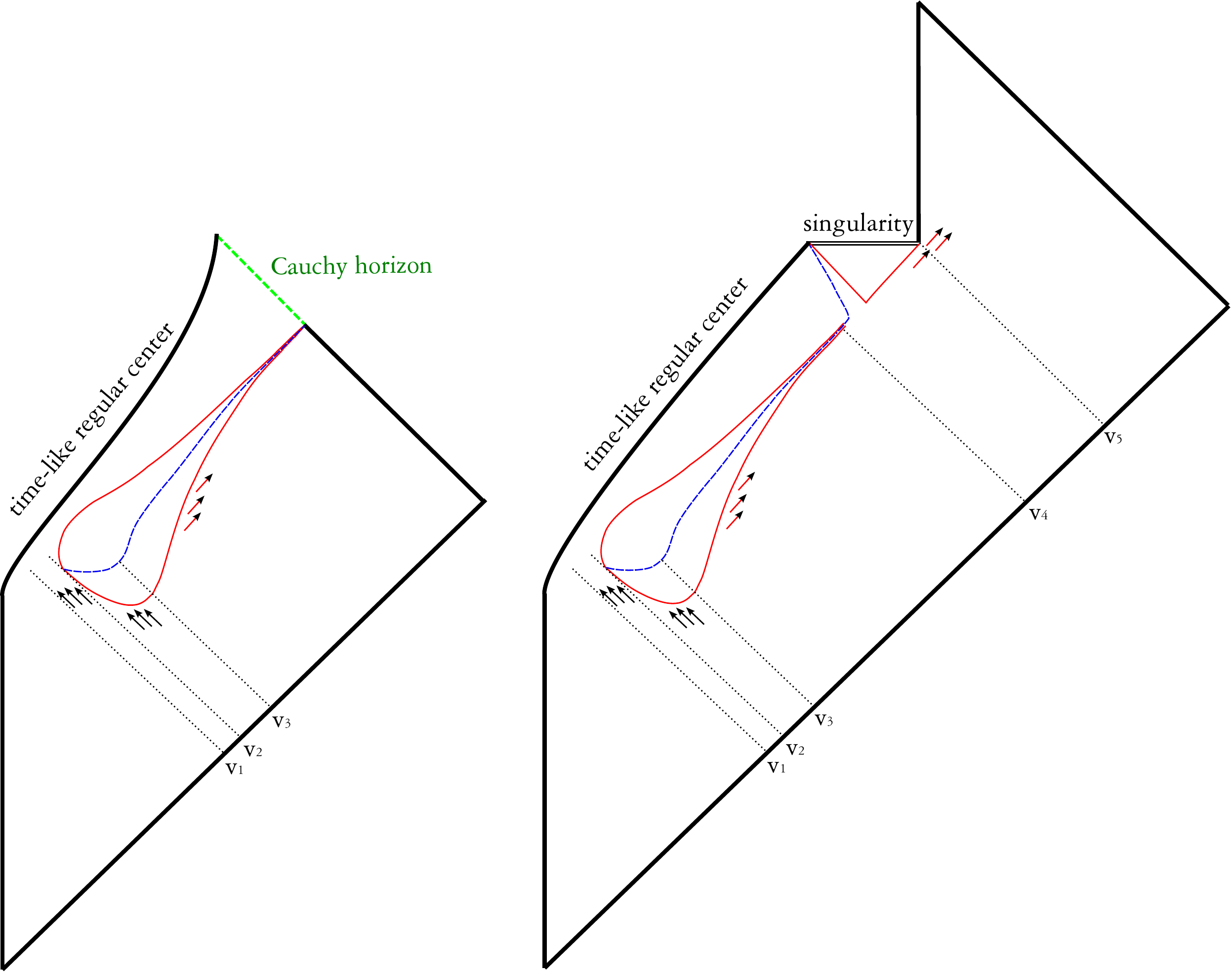}
\caption{ \label{fig:regular} (Color online) Another example of a regular black hole: the Frolov-Markov-Mukhanov model. The critical mass is collapsed between $v_{1} \leqslant v \leqslant v_{2}$; after $v_{2}$, both the outer and inner apparent horizons are formed [represented by solid curves] and the ``false vacuum lump'' is trapped by the horizons. The dashed curve denotes the boundary of the false vacuum lump between the  horizons. Note that before $v_{2}$, the boundary should be time-like, although this is not depicted.  Here $v_{3}$ is the time when the black hole stops to collapse and begins to evaporate. \textbf{Left:} If the shape of the black hole is maintained, then it will tend toward an extremal black hole, which would serve as a remnant. Nevertheless we cut off the diagram with a Cauchy horizon instead. \textbf{Right:} If the black hole is unstable, then the shell eventually [after $v_{4}$] forms a spacelike singularity [see \cite{Yeom:2008qw1}] and there is no remnant. The black hole completely evaporates at some later time $v_5$. This is numerically investigated in \cite{Hwang:2012nn}.}
\end{center}
\end{figure}

Of course, a [quasi-local] black hole by definition should have a trapped surface or an apparent horizon. Therefore, the existence of a regular black hole means one of the following:
\begin{itemize}
\item[(A)] \underline{Violation of the global hyperbolicity assumption:} static limits of Bardeen black hole \cite{Bardeen}, Frolov-Markov-Mukhanov model [see Fig.(\ref{fig:regular})]\cite{Frolov:1988vj}, Hayward model \cite{Hayward:2005gi} [see Fig.(\ref{fig:regular_2})] and its extension to the rotating case \cite{1410.4043}, charged black holes with non-linear electrodynamics \cite{AyonBeato:1999rg}, etc.
\item[(B)] \underline{Violation of the null energy condition:} phantom black holes \cite{Bronnikov:2005gm}, or some of the dynamical regular black holes [e.g., in the Vaidya approximation or numerical realization] \cite{Hayward:2005gi}.
\item[(C)] \underline{Modifications of Einstein field equations:} loop quantum black hole models \cite{Modesto:2005zm}, non-commutative regular black hole models \cite{Nicolini:2005vd}, etc.  
\end{itemize}

Take the case of the Bardeen black hole as an example, it corresponds to scenario (A) since it has an inner Cauchy horizon. However, for dynamical cases that include the formation and evaporation processes, the assumption of global hyperbolicity should be imposed at the beginning [so that we know how to evolve the system]; then the available option within general relativity would be (B), which is achieved via Hawking radiation. In such a case, one has to solve all the dynamical processes [such as the generation and disappearance of the outer and inner apparent horizons due to Hawking radiation, which is of course a monumental task. Such a check was carried out numerically in \cite{Hwang:2012nn}]. {\color{black}Nevertheless, it should be pointed out that one needs the Hawking effect to be strong enough to avoid the formation of singularities. It is not clear how generic -- or how physical -- such a singularity avoidance mechanism can be. See, for example, the discussion in \cite{roman}. We also remark that, in a large class of spacetimes that satisfy the weak energy condition, the existence of a regular black hole requires the change of the topology \cite{9612057}.}

We remark that in the Frolov-Markov-Mukhanov model, the Schwarzschild geometry inside the black hole is attached to a de Sitter one at some spacelike junction surface, which, the authors argued, may represent a short transition layer. In order to describe this layer, they employed Israel's thin shell method \cite{Israel}. However, once a Cauchy horizon is formed, the location of the thin shell becomes unclear, among other problems. We therefore introduce a Cauchy horizon as a cut-off in the left diagram of Fig.(\ref{fig:regular}). 

If the Hawking radiation does not contain information until the very late stage of the evaporation, then even though the spacetime is complete, the final object should contain all the information that has previously fallen into the black hole.
This means that after a black hole reduces to a very small size [perhaps of the order of the Planck scale], the particles emitted from the Hawking radiation should have sufficient entropy despite the fact that the energy content should be very small [at the order of the Planck energy]. In other words, since energy is required to carry the information, in order to emit a unit of information one now needs a long time. This would imply that the lifetime of the final object should be extended longer and longer as the hole becomes smaller and smaller. This seems to suggest that the final object does not really completely evaporate away. This property is naturally connected to the remnant picture, and we will discuss the challenges faced by such long-lived meta-stable remnant later in sec.\ref{challenges}. Here we would like to focus our discussion on another problem -- \emph{the singularity}.

It is actually not an easy task to simply ``replace the singularity'' of the black hole by ``something non-singular''. 
We first recall that in GR, the singularity of a Schwarzschild black hole lies \emph{in the future}, that is, $r=0$ is in no sense of the word a ``central singularity''. This means that the singularity should not be interpreted as a gravitational source, like how one could interpret the ``singularity'' of the $1/r$ potential in Newtonian gravity [see \cite{mcinnesessay} for more details about these under-appreciated subtleties in GR].
Therefore in order to resolve the singularity by an \emph{ad hoc} replacement of it with some kind of matter source, we have to assume a special kind of matter that modifies the ``would-be'' space-like singularity region. To remove the singularity, this ``new'' matter should be located within the space-like region. However, if the matter distribution is space-like, then it is very difficult to imagine that this matter distribution is stationary [though there is a discussion on the stability of space-like shells \cite{Balbinot:1990zz}]. [Note that the fact that regular black holes have two horizons does not quite help -- the regions between the horizons still have timelike $r$, and this presents the same problem for interpreting the matter configuration as the ``source of gravitation''.] Of course it is not out of possibility that this can be realized very non-trivially by some field combinations. For example, non-linear electrodynamics models \cite{AyonBeato:1999rg} may give an explicit field combination. [The situation is actually more natural if one starts from gravitational collapse and somehow \emph{prevents the formation} of singularity, rather than resolves an existing one; see the sub-subsec.\ref{smoothing}.] 

The question regarding whether a curvature singularity can be smeared and cured by quantum gravity is a valid one, although it is almost always taken as granted that it will be. Needless to say, lacking a fully working theory of quantum gravity we do not actually know this for certain\footnote{``\emph{Whether quantization of gravity will actually save spacetime from such singularities one cannot know until the `fiery marriage of general relativity with quantum physics has been consummated'.}'' \cite{MTW}}. 
On page 5 of \cite{HP}, Hawking's point of view is that if the singularity is indeed smeared out by quantum correction, things would be rather ``boring'' since gravity would then be ``just like any other field'', whereas gravity should be distinctively different since it is not just a player on a spacetime background, it is both a player \emph{and} the evolving stage\footnote{\emph{``Field? -- We have no field in the sense in which one has a Maxwell field. Whenever I have used the term `field' [in the context of the metric] I have done so as a matter of mere verbal convenience. [...] The classical fields -- the electrostatic field, for example -- in the first instance had, so to speak, a subjunctive existence. Let P be a particle satisfying all the criteria of being free except in as far as it carries an electric charge. Then if P were placed at some point it would be subject to a force depending on the value of the field intensity there. In particular, when the latter is zero there is no force. [...] the `true' fields subjunctively quantify the extent to which a given particle P is not free, granted that P would be free if any charges it may carry were neutralized [...] this field [the metric], unlike the `true' fields, cannot be absent, cannot be zero."} -- H. A. Buchdahl \cite{Buchdahl}.}. 

Indeed, given that GR is a geometric theory of gravity, one should pay more attention to the importance of geometry.  As a side remark and on a similar note, it has often been claimed that geometric quantities like torsion can be treated just like any other field, and thus there is no need to go beyond the Levi-Civita connection in a theory of gravity. Such a proposition, while not entirely wrong, can be misleading. In fact, from the point of view of well-posedness of the evolution equations, it is far simpler to work with torsion as what it truly is -- a geometric quantity \cite{nester}. 

In \cite{Hawking1966}, Hawking has argued that:
\begin{quote}
``\emph{The view has been expressed that singularities are so objectionable that if the Einstein equations were to predict their occurrence, this would be a compulsive reason for modifying them. However, the real test of a physical theory is not whether its predicted results are aesthetically attractive but whether they agree with observation. So far there are no observations which would show that singularities do not occur.}''
\end{quote}

We shall also make a side remark regarding ``thunderbolt'', which is not as well known as naked singularity, but remains an intriguing possibility: a thunderbolt [coined by Hawking \cite{Hawking1993}, but first described by Penrose \cite{Penrose1978}] is a ``wave of singularity'', much like an asymmetric PP-wave that propagates out from the collapse region. A thunderbolt would destroy the universe as it goes. As described by Penrose \cite{Penrose1999}, 
\begin{quote}
\emph{On this picture, the entire space-time could remain globally hyperbolic since everything beyond the domain of dependence of some initial hypersurface is cut off (‘destroyed’) by the singular wave. An observer, whether at infinity or in some finite location in the space-time, is destroyed just at the moment that the singularity would have become visible so that observer cannot actually ‘see’ the singularity.}
\end{quote}

Singularity may actually be ``useful''. Unlike the event horizon where according to GR, there should not be anything out of the ordinary, the singularity is where GR breaks down and one could imagine that new physics could appear. Indeed one proposal by Maldacena and Horowitz is to impose future boundary condition at the singularity, which allows quantum information to ``teleport'' to the black hole exterior \cite{MalHor} [post-selected quantum teleportation]. In other words, the singularity acts as a measurement device such that one always gets a particular outcome, a ``final state projection''. Such proposal of course has been criticized, and those criticisms in turn, countered. For related literature see e.g. \cite{MH1,MH2,MH3,MH4,MH5}. However, in our present work we only wanted to emphasize that one should pay more attention to the role of the singularities, and whether they are truly removed in a quantum theory of gravity. Indeed from the view point of string theory, not all singularities are bad, and there can be physically allowable naked singularities
in AdS/CFT correspondence \cite{singularity}. See also \cite{0306170} for an attempt to understand [quantum and stringy] physics in the vicinity of the black hole singularity, via AdS/CFT correspondence.

Even if one does successfully resolve the singularity, the solution in general still possesses an inner Cauchy horizon, and as mentioned before, it is well-known that the inner horizon has strong instability, the so-called ``mass inflation'' due to an extreme blue shift of the incoming radiation \cite{Poisson:1990eh}. To resolve the mass inflation singularity, we have to introduce more assumptions [e.g., including higher curvature terms \cite{Hwang:2011kg}], and this makes the model less natural. 

Finally, we remark that there might be a connection between regular black holes and the generalized uncertainty principle \cite{1305.3851}. For a more detailed review on regular black holes, the readers are referred to \cite{ansoldi}.

\subsection{Quantum Gravitational Effects}

To resolve the singularity problem we -- at the very least -- need quantum gravitational treatments [although, as previously discussed, there is the possibility that even quantum gravity cannot get rid of singularities]. Quantum gravity theories do provide some models that induce regular black holes as well as infinite life time objects. In the previous sections, we have only discussed models in the effective field theory of some underlying quantum gravity theories. In this section, we specifically focus on some quantum gravity inspired models: loop quantum black holes and non-commutative black holes. The string theory approach will be reviewed separately in subsect.\ref{grav}.

\subsubsection{Loop Quantum Black Holes}

To cure the singularity, both canonical quantization and loop quantum gravity [LQG] have to solve the Hamiltonian constraint or the Wheeler-DeWitt equation by using a certain choice of variables [e.g., Ashtekar variables in loop quantum gravity]. In a certain semi-classical approximation of the full loop quantum gravity theory \cite{Modesto:2008im}, the Hamiltonian constraint can allow a certain effective classical metric as a quantum modification of the Schwarzschild solution. {\color{black}In addition to the well-known Ashtekar-Bojowald singularity-free black holes \cite{BA}, which we will discuss later on, there are also other works concerning LQG black holes.}

The following is one expected form of the corrected Schwarzschild geometry\cite{Hossenfelder:2009fc}:
\begin{eqnarray}
ds^{2} = - g(r) dt^{2} + \frac{dr^{2}}{f(r)} + h(r) d\Omega^{2},
\end{eqnarray}
where
\begin{eqnarray}
g(r) &=& \frac{(r-r_{+})(r-r_{-})(r+r_{*})^{2}}{r^{4}+a_{0}^{2}},\\
f(r) &=& \frac{(r-r_{+})(r-r_{-})r^{4}}{(r+r_{*})^{2} (r^{4}+a_{0}^{2})}, ~~\text{and}\\
h(r) &=& r^{2} + \frac{a_{0}^{2}}{r^{2}}.
\end{eqnarray}
The outer horizon and inner horizon are given by $r_{+}=2m$ and $r_{-}=2m P^{2}$, respectively, in which $P \ll 1$ is a theory-dependent constant, $M = m (1+P)^{2}$ is the asymptotic mass, and $r_{*}=\sqrt{r_{+}r_{-}}$. Finally, $a_{0} = A_{\mathrm{min}}/8\pi$, where $A_{\mathrm{min}}$ is the minimum area gap in loop quantum gravity.

This model has some interesting properties:
\begin{itemize}
\item[--] In the asymptotic limit, this model tends to the standard Schwarzschild solution. [Note that $r$ is only asymptotically the usual areal radius.]
\item[--] The Schwarzschild solution is modified around $r \sim L_{\mathrm{P}}$. Specifically, we see from the function $h(r)$ that, as $r$ decreases, the areal radius is bounced around a certain minimum area $a_{0}$. Therefore, for an infalling observer, the metric looks like a wormhole and eventually the observer sees a second future infinity that is separated from the exterior future infinity.
\item[--] This model has two horizons [inner and outer apparent horizons]. Initially, the evaporating process is similar to the usual Schwarzschild black hole, and hence the temperature increases as time goes on. However, around the Planck radius, the temperature is modified and eventually approaches zero. In this sense, this model provides a zero-temperature remnant. Therefore, similar to the regular black hole picture, it has been argued that the end point of evaporation is a zero Hawking temperature object [although as we mentioned above, this is not obvious, and will be further discussed in subsec.\ref{concluding remarks}]. 
\end{itemize}
We can construct the whole causal structure by using the Vaidya approximation [i.e., we take the mass function $m$ as a function of the advanced time $v$ or the retarded time $u$] \cite{Hossenfelder:2009fc}.

Regarding this model, we wish to emphasize on some non-trivial properties:
\begin{itemize}
\item[--] In the $m \rightarrow 0$ limit, the model does \emph{not} reduce to Minkowski spacetime due to the non-vanishing $a_{0}$. Therefore, we need to assume that \emph{only} Planck scale effects turn on the parameter $a_{0}$.
\item[--] There exists a minimum areal radius $a_{0}$. In terms of the in-falling observer, the scale $a_{0}$ determines the area of the wormhole throat. In terms of the asymptotic observer, when the area of the black hole reduces to around $a_{0}$, the Hawking temperature changes drastically and the evaporation slows down. 
\item[--] This model has an inner apparent horizon, which, like its GR counterpart, is unstable in general due to mass inflation \cite{Brown:2011tv}. This problem should be resolved by some other independent arguments.
\end{itemize}

\begin{figure}
\begin{center}
\includegraphics[scale=0.95]{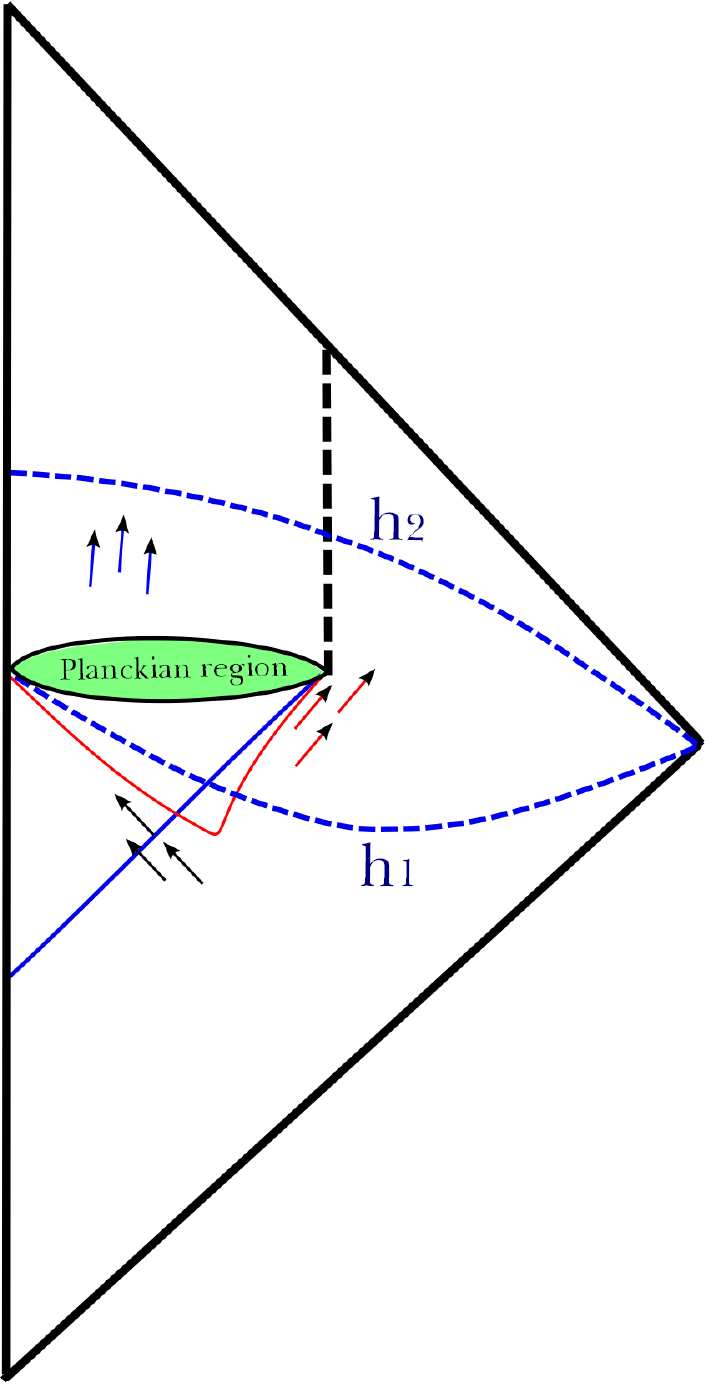}
\caption{\label{fig:AB_2}(Color online) Ashtekar-Bojowald picture. The solid curve below the boundary of the Planckian region denotes the apparent horizon, while $h_1$ and $h_2$ are two Cauchy hypersurfaces.}
\end{center}
\end{figure}

The most well-known scenario concerning black hole evaporation in LQG is probably the Ashtekar-Bojowald singularity-free black holes \cite{BA}, as shown in Fig.(\ref{fig:AB_2}).  Here the singularity is replaced by some ``Planckian region'', in which quantum gravitational effect becomes important. In fact, if one says nothing about which model of quantum gravity one works with, this picture is quite generic. For a recent discussion about Ashtekar-Bojowald model in the context of firewall, see \cite{1410.7062}.

In loop quantum cosmology [in which one first assumes that the universe is homogeneous and isotropic, and \emph{only after that} one proceeds to quantize the remaining degrees of freedom], one could also resolve the black hole singularity by replacing the Planckian region with a ``Euclidean core'' [This could be generic if one imposes the anomaly-free condition \cite{Bojowald}; see also the discussion about Euclidean core in the context of GR and higher derivative gravity theories \cite{hirayama}]. That is to say, the metric changes its signature from Lorentzian $(-,+,+,+)$ to Riemannian [Euclidean] $(+,+,+,+)$ once curvature becomes sufficiently strong near the ``would-be singularity''. However, in \cite{Bojowald} it is argued that this is problematic since the equations of motion cease to be hyperbolic and become elliptic instead in the signature change region, thus one would not be able to have full predictability of the system by just imposing initial conditions. 

We note that such a change of signature is of course familiar in the context of Hartle-Hawking no-boundary proposal, in which the universe started as a Euclidean manifold, say a 4-dimensional sphere \cite{HartleHawking}. In that context one only has to prescribe boundary condition once at the Big Bang, and the rest of the evolution will be determined from the dynamics\footnote{Maldacena-Horowitz's ``black hole final state'' \cite{MalHor} is also argued to be consistent with the Hartle-Hawking no-boundary proposal.}. Here, in the context of black hole singularity, one will have to prescribe boundary conditions at the Euclidean core of  every black hole throughout the lifetime of the universe. This is a less desirable -- although not necessarily disastrous -- requirement.

\subsubsection{Smoothing Matter Profile: Non-Commutativity and Non-Locality}\label{smoothing}

Perhaps, some quantum gravitational principles can be applied to smooth the singularity, in such a way that its back-reaction to the metric induces a regular space-time. Regarding this possibility, some radical principles that modify general relativity had been proposed, notably the ideas involving non-commutativity and non-locality.
\begin{description}
\item[-- Non-commutative quantum gravity:] A quantum gravity inspired model is related to non-commutative geometry \cite{Nicolini:2005vd,PN}. By quantizing spacetime via the non-commutative relation $[x^{\mu},x^{\nu}] = i \theta^{\mu\nu}$ with an anti-symmetric matrix $\theta^{\mu\nu}$, the black hole singularity can be avoided. At the intuitive level, this can be done because non-commutative geometry is a ``geometry without point''.

In a simple model, this is achieved by substituting the GR singularity by using the matter source with a Gaussian distribution of the form 
\begin{eqnarray}
\rho_{\theta}(r) = \frac{M}{(4\pi \theta)^{3/2}} e^{-r^{2}/4\theta}. 
\end{eqnarray}
In other words, the matter source is not concentrated at a point [like a delta function], but smeared over a region of width $\sqrt{\theta}$. 

Due to such a matter source, the induced metric becomes
\begin{eqnarray}
ds^{2} = - g(r) dt^{2} + \frac{dr^{2}}{g(r)} + r^{2} d\Omega^{2},
\end{eqnarray}
where
\begin{eqnarray}
g(r) = 1- \frac{4M}{r\sqrt{\pi}} \gamma\left( \frac{3}{2}, \frac{r^{2}}{4\theta}\right),
\end{eqnarray}
in which $\gamma(s,x):=\int_0^x t^{s-1}e^{-t} dt$ is the lower incomplete gamma function.

This model has some interesting properties:
\begin{itemize}
\item[--] In the asymptotic limit, the geometry approaches that of the standard Schwarzschild manifold. This result can be generalized to charged black holes \cite{Ansoldi:2006vg} and higher dimensional black holes \cite{Spallucci:2009zz}.
\item[--] The matter source $\rho$ works as a de Sitter-like core that replaces the black hole singularity. Therefore, this is comparable with the Frolov-Markov-Mukhanov model \cite{Frolov:1988vj}, where in this case they introduced a real de Sitter core inside the event horizon. 
\item[--] Such black hole also has two horizons \cite{0611130,1204.0143}. Therefore it is argued that as time goes on, the two horizons will approach each other [which, as we repeatedly mentioned, is not obvious, and will be discussed in sec.(\ref{concluding remarks})] and the Hawking temperature approaches zero; this means that the black hole will have infinite lifetime and hence becomes a kind of remnant. This remnant would have the length scale of the order of $\sqrt{\theta}$.
\end{itemize}


At the end of sec.(\ref{regular}) we discussed some problems regarding the interpretation of singularity as a gravitating source, and how replacing it with some kind of matter source is confusing at best. In the collapsing scenario the situation is slightly better --- in GR, comoving observers with the collapsing matter would crash into the singularity in a finite proper time [for an explicit calculation of some simple models, see \cite{ABCL}], although for an asymptotic observer, an apparent horizon would form and shroud the interior fate of the infalling observers. One could certainly discuss this collapse scenario with an infalling matter that does not get concentrated at a point, and thus possibly avoid the formation of the singularity\footnote{One should also keep in mind that a black hole singularity is guaranteed to form in GR due to Penrose's Singularity Theorem, which assumes the NEC. Once quantum effects are taken into consideration [such as in the presence of Hawking radiation], it is no longer obvious whether the singularity still forms during gravitational collapse. See the discussions in \cite{9506052,1012.6038}. For a recent review on the singularity theorem see \cite{Senovilla-Garfinkle}.}.  

Similar to the GUP approach, it has also been argued that non-commutativity can somehow allow information to escape the black hole, encoded in the Hawking radiation \cite{1312.3032}. 

{\color{black}Lastly, we would like to comment that the conventional non-commutative deformations of
Einstein gravity is based on the star product. However, it is possible to give an effective description of a non-commutative manifold based on coherent quantum states, see \cite{0406174}. The Gaussian profile originated from the mean values of the non-commutative position operators on the
coordinate coherent states, which provide the maximal possible resolution on the manifold. An alternative method to obtain a Gaussian profile is via the so-called Voros Product \cite{0911.2123}. It is also possible to arrive at the same result via a non-local deformation of gravity \cite{1010.0680}, similar to what we will review next in this subsection.}

\item[-- Non-locality:] The ideology is quite similar to the previous one, namely, if for some reasons the matter distribution is \emph{not} singularly localized as a delta-peak, then there is a hope to obtain a regular black hole. The idea of non-localization of the gravity sector \cite{1202.2102} is another such proposal. If one modifies the gravity action as
\begin{eqnarray}
S = \frac{1}{16\pi} \int d^{4}x \sqrt{-g} \left[ \int d^{4}y \sqrt{-g}\;\; \mathcal{A}^{2}\left( \ell^{2} \nabla_{x}^{2} \right) \delta^{(4)}\left( x - y \right) R(y)\right],
\end{eqnarray}
where $R(y)$ is the Ricci scalar at spacetime point $y$, $\ell$ is an arbitrary length scale, and $\mathcal{A}$ is some function of $\ell^{2} \nabla_{x}^{2}$, with $\nabla^2 := g_{\mu\nu}\nabla^\mu \nabla^\nu$, then one would obtain the modified Einstein equation as
\begin{eqnarray}
G_{\mu\nu} = 8\pi G \mathcal{A}^{-2}\left({\ell^{2} \nabla_{x}^{2}}\right) T_{\mu\nu},
\end{eqnarray}
and hence one obtains an effective non-localization of the matter distribution. If one starts initially with a localized matter distribution [hence the geometry has a singularity]
\begin{eqnarray}
T^{0}_{~0} = - \frac{M}{4\pi r^{2}} \delta(r),
\end{eqnarray}
the effective matter upon the non-locality correction becomes
\begin{eqnarray}
\mathcal{A}^{-2}\left({\ell^{2} \nabla_{x}^{2}}\right) T^{0}_{~0} = - M \frac{\ell}{\pi^{2} (\vec{x}^{2} + \ell^{2})^{2}}.
\end{eqnarray}
After solving the Einstein equation, one would obtain the metric
\begin{eqnarray}
ds^{2} = - \left(1 - \frac{2 \mathcal{G}(r) M}{r} \right) dt^{2} + \left(1 - \frac{2 \mathcal{G}(r) M}{r} \right)^{-1} dr^{2} + r^{2} d\Omega^{2},
\end{eqnarray}
where
\begin{eqnarray}
\mathcal{G}(r) = \frac{2G}{\pi} \left[ \tan^{-1} \left(\frac{r}{\ell}\right) - \frac{r/\ell}{1+(r/\ell)^{2}} \right].
\end{eqnarray}
For the regime in which $r \ll \ell$, this model is similar to the de Sitter metric and hence the properties are similar as the previous non-commutative quantum gravity models: no singularity, the existence of two horizons, and arguably tends toward a stable final state as it evaporates.
\end{description}

Finally, we summarize these three models [loop quantum black hole, non-commutative black hole, non-local black hole] by their shared properties:
\begin{itemize}
\item[(1)] There exists a critical radius where quantum gravity effects dominate, and this critical radius becomes the order of the size of the final objects.
\item[(2)] Since these models have two horizons, their corresponding black hole geometries may tend toward zero temperature in the limit, and finally become black hole remnants.
\item[(3)] Due to the presence of the inner apparent horizon, one still needs to take care of the mass inflation instability problem.
\end{itemize}

For more discussions, see also \cite{1410.1706}.

\subsubsection{Energy-Scale Dependence}

Another possibility is that [quantum] modifications of gravity may be related to the modification of the Newton's constant $G$ or the other coupling constants of nature. Perhaps, as the gravitational interaction energy scale increases, the gravitational coupling or other laws of nature also change. Along this line of thought, some examples are known in the literature. In this review,  we emphasize on three important directions.

\begin{description}
\item[-- RG modified gravity:]  
The idea that the effective description of quantum gravity may be non-perturbatively renormalizable via the notion of ``asymptotic safety'' dates back to Weinberg in the 1970's \cite{weinberg1, weinberg2}. [See also, \cite{weinberg3}.]
Simply put, the Newton's constant $G$ [as well as the cosmological constant $\Lambda$] could be treated as running coupling constants dictated by the RG flow, which approaches  some fixed point in the ultraviolet [UV] limit. Consideration of such running coupling constant may improve the high energy complications of Planck scale physics. 
Indeed, the Einstein-Hilbert truncation contains such a fixed point \cite{reuter}.
From the calculation of Donoghue \cite{Donoghue:1994dn}, we expect that the gravitation constant behaves like
\begin{eqnarray}
{G}(r) = \frac{G_0r^{3}}{r^{3} + \omega G_0 \left(r+ \gamma G_0M \right)},
\end{eqnarray}
where the numerical value of $\omega$ is $118/15\pi$, $\gamma = 9/2$ \cite{Bonanno:2000ep}, and
$G_0$ is the experimentally observed value of the Newton's constant.
 The Schwarzschild-analog solution in this model also has two horizons, a regular center, and approaches zero Hawking temperature configuration as it evaporates away. Nevertheless, doubt has been raised recently regarding this solution \cite{maziashvili}. 
For reviews on asymptotically safe quantum gravity, see \cite{0610018} and \cite{0709.3851}.

\item[-- Gravity's rainbow:] The idea known as gravity's rainbow was suggested by Magueijo and Smolin \cite{Magueijo:2002xx}. According to this proposal, the dispersion relation $E^{2} - p^{2} = m^{2}$ should be modified [as in doubly special relativity\footnote{In doubly special relativity, in addition to an observer-independent maximum velocity [namely, the speed of light], there is also a minimum length scale, namely the Planck length.}] even at the gravity level [hence,  ``doubly general relativity'']. This is achieved by introducing two functions of the energy $f$ and $g$:
\begin{eqnarray}
E^{2} f^{2}\left( \frac{E}{E_{\mathrm{Pl}}} \right) - p^{2} g^{2}\left( \frac{E}{E_{\mathrm{Pl}}} \right) = m^{2}.
\end{eqnarray}
If this is the case, then physical quantities like energy and possibly other parameters should depend on the interaction energy scale $E$ and this should be implemented to even the metric level.  As a consequence, the metric solution of a Schwarzschild-analog black hole should look like 
\begin{flalign}
ds^{2} = & - f^{-2}\left(\frac{E}{E_\text{pl}}\right)\left( 1 - \frac{2GM}{r} \right) dt^{2} + g^{-2}\left(\frac{E}{E_\text{pl}}\right) \left( 1 - \frac{2GM}{r} \right)^{-1} dr^{2} \\ \notag &+ r^{2}g^{-2}\left(\frac{E}{E_\text{pl}}\right) d\Omega^{2},
\end{flalign}
where we have constraints that as $E \rightarrow 0$, the other functions should approach their values in general relativity, i.e., $f(E/E_\text{pl}), g(E/E_\text{pl}) \rightarrow 1$. Hence, by choosing a ``proper'' form of $f(E/E_\text{pl})$ and $g(E/E_\text{pl})$, and by identifying $E$ with the black hole mass $M$, one may obtain a remnant solution that goes to zero Hawking temperature limit \cite{ali, 0608061}. One potential problem of this idea is that there is no constructive principle to choose the functions $f$ and $g$, and hence one can construct [almost] any solution as one wishes.

For more discussions about remnants for various types of black holes in gravity's rainbow scenario, see \cite{1410.5706}.

\item[-- Ho\v{r}ava-Lifshitz gravity:]  Another energy-dependent approach, which is radically different from the above two proposals, involves modifying the gravity sector at the action level to render it non-Lorentz invariant at a sufficiently high energy scale. 
If one has a theory of gravity that recovers the usual GR at long distances [low energy regime], but is radically different at short distance scale [high energy regime] such that the theory is renormalizable, then such a theory can be a candidate for a UV completion of Einstein gravity, which would pave the way for a full quantum theory of gravity. [One outstanding problem of quantum gravity is the well-known fact that Einstein gravity is non-renormalizable, and hence a perturbative quantum gravity based naively on GR is impossible.]

To realize renormalizability -- at least at the power-counting level -- one interesting proposal is the Ho\v{r}ava-Lifshitz gravity \cite{Horava:2009uw}. 
For a review of the subject, see \cite{1010.3218}.
Ho\v{r}ava introduced a \textit{dynamical} scaling behavior of space and time. To be more specific, space and time are treated on equal footing [as in GR] for long distances [the critical exponent is $z=1$, and hence, relativistic], however space and time are treated differently for short distances, where the critical exponent is $z=3$ and hence non-relativistic. This evades the usual problems, such as ghost degrees of freedom induced by the higher order time derivatives, which render the theory non-unitary. 
Due to the non-relativistic nature at short length scale, the theory may also allow a stable black hole solution, which would serve as a remnant. However we should remark here that, the horizon radius of a black hole in Ho\v{r}ava-Lifshitz gravity generically depends on the energies of test particles \cite{1105.4259}, therefore it is not clear ``when
and for whom they are black'' \cite{0910.5487}.

A good way to discuss Ho\v{r}ava-Lifshitz gravity is to first write the metric in the standard ADM-decomposed form,
\begin{eqnarray}
ds^{2} = -N^{2} dt^{2} + h_{ij} \left( dx^{i} - N^{i} dt \right) \left( dx^{j} - N^{j} dt \right),
\end{eqnarray}
where, as usual, $N$ is the lapse function and $N^i$, with $i \in \left\{1,2,3\right\}$, are the components of the shift vector.

The extrinsic curvature is defined by
\begin{eqnarray}
K_{ij} = \frac{1}{2N} \left( \frac{\partial}{\partial t} h_{ij} - \nabla_{i}N_{j} - \nabla_{j}N_{i} \right),
\end{eqnarray}
while the Cotton tensor is defined by
\begin{eqnarray}
C^{ij} = \epsilon^{ikl} \nabla_{k} \left( R^{j}_{\;l} - \frac{1}{4} R \delta^{j}_{\;l} \right).
\end{eqnarray}
In terms of these quantities, one suggested form of the Ho\v{r}ava-Lifshitz action is as follows \cite{Horava:2009uw}:
\begin{eqnarray}
S &=& \int dt d^{3}x~ \sqrt{h}N \left[ \frac{2}{\kappa^{2}} \left( K_{ij}K^{ij} - \lambda K^{2} \right) + \frac{\kappa^{2}\mu^{2} (\Lambda_{W} R^{(3)} - 3 \Lambda_{W}^{2})}{8 (1-3\lambda)} \right. \nonumber\\
&&+ \frac{\kappa^{2}\mu^{2}(1-4\lambda)}{32(1-3\lambda)} \left(R^{(3)}\right)^{2}\left. - \frac{\kappa^{2}}{2\omega^{4}} C^{ij}C_{ij} + \frac{\kappa^{2}\mu}{2\omega^{2}} \epsilon^{ijk}{R^{(3)}}_{il}\nabla_{j}{R^{(3)l}}_{\;k}  \right. \nonumber \\
&&- \left. \frac{\kappa^{2}\mu^{2}}{8} {R^{(3)}}_{ij} {R^{(3)ij}}+ \mu^{4} R^{(3)} \right],
\end{eqnarray}
where $R^{(3)}$ denotes the 3-dimensional [spatial] scalar curvature, while  $\kappa$, $\lambda$, $\omega$, $\mu$ and $\Lambda_W$ are all parameters of the theory. 
We do not need to know the details of these parameters for the purpose of this review.
Of particular importance is the running coupling $\lambda > 1/3$, which at low energy limit is expected to flow to the value $\lambda=1$, at which point GR is recovered. 
The speed of light depends on energy scale in this theory, and is given by
\begin{equation}
c=\frac{\kappa^2 \mu}{4}\sqrt{\frac{\Lambda_W}{1-3\lambda}}.
\end{equation}

For a static and spherically symmetric background, we can further write down the metric ansatz as
\begin{eqnarray}
ds^{2} = -N^{2}(r) dt^{2} + \frac{1}{f(r)} dr^{2} + r^{2} d\Omega^{2},
\end{eqnarray}
and for a choice of parameters for an asymptotically flat background, the solution becomes \cite{Kehagias:2009is, Park:2009zra}
\begin{eqnarray}
N^{2} = f = 1 + \omega r^{2} - \sqrt{r (\omega^{2} r^{3} + 4 \omega M)}.
\end{eqnarray}
If $r \gg (M/\omega)^{1/3}$, then it is approximately the Schwarzschild solution.
This model has two horizons
\begin{eqnarray}
r_{\pm} = M \left( 1 \pm \sqrt{1 - \frac{1}{2\omega M^{2}}} \right).
\end{eqnarray}
Note that there exists a lower bound $M^2 \geqslant (2\omega)^{-1}$.
The Hawking temperature is
\begin{eqnarray}
T = \frac{2 \omega r_{+}^{2} -1}{8\pi r_{+} (\omega r_{+}^{2} + 1)}.
\end{eqnarray}
As expected, there exists a parameter space that allows a stable solution with zero Hawking temperature.
The radius of the extremal black hole is $r_{\text{ext}}=M=1/\sqrt{2\omega}$. This is of the order of the length scale at which the short distance physics becomes dominant. Due to this behavior, there could be an interesting connection to the generalized uncertainty principle \cite{Myung:2009va}.

\end{description}

\subsection{Semi-Classical Framework: Modification of Gravity Sector}\label{grav}

Various remnants can be introduced from not only the quantum gravitational framework, but also the semi-classical framework. Especially, from string theory background, there will be various corrections and one can find many examples of remnants. To realize this within the semi-classical framework, one can modify either the gravity sector or the matter sector of the action. In this subsection we will focus on the first option, and leave the second option to the next subsection. We first clarify what we mean by modifying these sectors:
\begin{description}
\item[-- Modifying the gravity sector:] This means that we add higher curvature terms to the action or modify the Einstein sector [Ricci scalar term in the action]. These modifications will change the solutions that may induce remnants.
\item[-- Modifying the matter sector:] This means that we add some hypothetical matter fields [motivated from, e.g., string theory]. These new matter fields may provide new conserved charges that may in turn, induce long-lived remnants.
\end{description}
Note that in general, if regular black holes naturally tend toward their extremal state under Hawking evaporation, then this implies the existence of remnants, but the converse statement is not true -- remnants need not necessarily be regular black holes! So, in this section, we shall cover various other models which give rise to remnants.

By adding higher curvature terms or other terms to the gravity sector, one can modify the metric solution. For example, if the metric function $g_{tt}=f(r)$ with spherical symmetry admits a series expansion of the form
\begin{eqnarray}
f(r) = 1 - \frac{2M}{r} + a_{2} \left( \frac{M}{r} \right)^{2} + a_{3} \left( \frac{M}{r} \right)^{3} + ... + b_{1}\left( \frac{r}{M} \right)+ b_{2} \left( \frac{r}{M} \right)^{2} + ...;
\end{eqnarray}
then the Hawking temperature of the corresponding black hole can be expanded as
\begin{eqnarray}
T = \frac{c_{0}}{M} \left( 1 + c_{1} M + c_{2} M^{2} + ... + \frac{c_{-1}}{M} + \frac{c_{-2}}{M^{2}} + ... \right)
\end{eqnarray}
and the time derivative of the black hole mass yields \cite{Barrow:1992hq}
\begin{eqnarray}
\frac{dM}{dt} = -\sigma T^{4} A = -\frac{d_{0}}{M^{2}} \left( 1 + \frac{d_{-1}}{M} + \frac{d_{-2}}{M^{2}} + ... + d_{1} M + d_{2} M^{2} + ... \right),
\end{eqnarray}
where $a_{i}$, $b_{i}$, $c_{i}$, and $d_{i}$ are complicated functions of the model parameters of the correction terms to the Lagrangian. By adjusting these model parameters, if we can achieve $T \rightarrow 0$ or $dM/dt \rightarrow 0$ as $t \to \infty$, then this may indicate the formation of remnants.

We now briefly list down some possible realizations of such a remnant scenario.

\subsubsection{Dilaton Field with Higher Curvature Terms}

From string theory, the gravity sector and the dilaton sector can be expanded into the following form \cite{Gasperini:2007zz}:
\begin{eqnarray}
S &=& \frac{1}{2\lambda_{s}^{D-2}} \int d^Dx \sqrt{-g} \left[ e^{-2\phi} \left( R + 4(\nabla \phi)^{2}  + \alpha' \mathcal{O}(R^{2}) + \alpha'^{2} \mathcal{O}(R^{3}) + ... \right) \right. \nonumber \\
&& \;\;\;\;\; \left. + \left(R + ... \right) + e^{2\phi} \left(R + ...\right) + ... \right],
\end{eqnarray}
where, for example,
\begin{eqnarray}
\mathcal{O}(R^{2}) &=& a_{1} R_{\mu\nu\rho\sigma}^{2} + a_{2} R_{\mu\nu}^{2} + a_{3} R^{2} + a_{4} R^{\mu\nu}\nabla_{\mu}\phi\nabla_{\nu}\phi + a_{5} R(\nabla \phi)^{2} + a_{6} R\nabla^{2}\phi \nonumber \\
&& + a_{7} (\nabla^{2}\phi)^{2} + a_{8} \nabla^{2}\phi(\nabla\phi)^{2} + a_{9} (\nabla \phi)^{4},
\end{eqnarray}
with some constants $a_{i}$'s. Here $D$ is the spacetime dimension, and $\lambda_{s}$ is the string tension parameter that is related to the $D$-dimensional Newton's constant. The first line is the $\alpha'$ expansion that includes higher curvature terms, while the second line is the genus expansion of the dilaton field that includes higher coupling terms of the dilaton field.

Of course, in general it is not possible to solve the most general effective action of string theory. Therefore we need to either (1) approximate this expansion, for example by introducing the form-factors \cite{Gasperini:2007zz} and deal with the action effectively; (2) cut off the expansion up to a certain order; or (3) consider the solutions in a perturbative way. For example, in the paper of Callan, Myers and Perry \cite{Callan:1988hs}, the authors considered a model with a dilaton field $\phi$ and a higher curvature term $R_{\mu\nu\rho\sigma}R^{\mu\nu\rho\sigma}$, with the action
\begin{eqnarray}
S = \frac{1}{2\lambda_{s}^{D-2}} \int d^{D}x \sqrt{-g} e^{-2\phi} \left( R + 4 \left(\nabla \phi\right)^{2} + C \alpha' R_{\mu\nu\rho\sigma}R^{\mu\nu\rho\sigma} \right),
\end{eqnarray}
where $C$ is a constant. This model allows various solutions that modify the Hawking temperature, at least up to $\alpha'$ order.

Explicitly, the Hawking temperature is modified to be
\begin{eqnarray}
T = \kappa_{1} \frac{D-3}{M^{1/(D-3)}} \left[ 1 - \kappa_{2} \frac{(D-2)(D-4)\alpha'}{M^{2/(D-3)}} \right],
\end{eqnarray}
where $\kappa_{1}$ and $\kappa_{2}$ are dimension-dependent constants. Therefore, in $D > 4$ dimensions, there will be significant corrections to the Hawking temperature in the small mass limit. The final mass scale is given by
\begin{eqnarray}
M_{\mathrm{min}} = \left[ \kappa_{2} (D-2)(D-4)\alpha' \right]^{(D-3)/2} \propto \mathcal{O}\left( \alpha' \right)^{(D-3)/2}.
\end{eqnarray}

\subsubsection{Lovelock Gravity and General Second Order Gravity}

If we include more higher curvature terms such as in Lovelock gravity \cite{Lovelock1, Lovelock2}, black hole solutions will of course be modified \cite{Myers:1988ze}. In general, Lovelock gravity  is described by the complicated Lagrangian
\begin{eqnarray}
\mathcal{L} &=& \sum_{m=0}^{k}c_{m} \mathcal{L}_\mathrm{m}, ~~\text{where}\\
\mathcal{L}_{\mathrm{m}} &=& 2^{-m} \delta^{\alpha_{1} \beta_{1} ... \alpha_{m} \beta_{m}}_{\gamma_{1} \epsilon_{1} ... \gamma_{m} \epsilon_{m}} R^{\gamma_{1}\epsilon_{1}}_{\;\;\;\;\;\;\alpha_{1}\beta_{1}} ... R^{\gamma_{m} \epsilon_{m}}_{\;\;\;\;\;\;\alpha_{m}\beta_{m}},
\end{eqnarray}
in which $\delta$ is totally anti-symmetric in both sets of indices. Each term $\mathcal{L}_{\mathrm{m}}$ corresponds to the Euler density of $2m$-dimensional manifolds.

The simplest expansion is to include the second order terms up to the well-known Gauss-Bonnet term $(\lambda/2) R_{\mathrm{GB}}^{2} := (\lambda/2) (R_{\mu\nu\rho\sigma}R^{\mu\nu\rho\sigma}-4 R_{\mu\nu}R^{\mu\nu}+R^{2})$, where $\lambda$ is a coupling constant of dimension $(\text{length})^2$, which is presumed small so as to recover GR at low energy [small curvature] limit. In this case, the Hawking temperature is given by 
\begin{eqnarray}
T = \frac{D-3}{4\pi r_{h}} \left( \frac{r_{h}^{2} + \lambda (D-5)(D-4)/ 2}{r_{h}^{2} + \lambda (D-3) (D-4)} \right),
\end{eqnarray}
in which the apparent horizon $r_{h}$ should satisfy the polynomial equation
\begin{eqnarray}
r_{h}^{D-3} + \frac{\lambda (D-3) (D-4)}{2} r_{h}^{D-5} - \frac{8\pi \Gamma(N/2)}{(D-2)\pi^{N/2}} M = 0.
\end{eqnarray}

Note that this correction term does not imply the existence of a remnant in general. However, for example, if we choose $D=5$, then as the black hole shrinks $r_{h} \rightarrow 0$, the Hawking temperature tends toward zero.  In general, Lovelock gravity does not imply remnant scenario; however, Myers and Simon \cite{Myers:1988ze} had shown that, at least, the remnant solution is possible for $D = 2k + 1$ dimensions with the terms up to $\mathcal{L}_{2k}$ included.

Similar analysis was carried out by Whitt \cite{Whitt:1988ax} in the context of general second order gravity theories \cite{Wheeler:1985qd}, in which the following form of Lagrangian is considered:  
\begin{eqnarray}
\mathcal{L} = \sum_{m=0}^{N} \frac{L_{m}}{(D-2m)(D-2)\!} \epsilon^{\mu_{1} \cdots \mu_{D}} R_{\mu_{1}\mu_{2}} \wedge R_{\mu_{3}\mu_{4}} \wedge \cdots \wedge e_{\mu_{(2m+2)}} \wedge \cdots \wedge e_{\mu_{D}},
\end{eqnarray}
where the $e_{\mu}$'s are as usual, the co-frame dual to the vierbein. Again, there are some solutions that eventually lead to zero Hawking temperature. However, this is not a generic phenomenon.

We would also like to mention an interesting and important fact regarding Lovelock gravity --- it suffers from shock formation due to the nonlinear nature of its transport equation \cite{RTW1, RTW2}. This occurs because different polarizations of the graviton correspond to different characteristic [hyper-]surfaces, i.e., gravity can propagate either superluminally or subluminally, which corresponds to timelike characteristics and spacelike characteristics surfaces, respectively [c.f. the case in GR in which characteristic surfaces have to be null]. Due to the shock formation, cosmic censorship may be violated \cite{RTW1}. What is even more intriguing is the possibility that, in Lovelock gravity, due to the nontrivial characteristics, information may be able to escape the horizon defined by the naive null generators \cite{izumi}. However, it is more likely that such superluminal nonlinear shocks are also a sign that the theory may be problematic. This was indeed argued to be the case in \cite{camanho}, in which the authors showed that higher derivative corrections [with large coupling constants] to GR are strongly constrained by causality.  Gauss-Bonnet theory for example, is therefore ruled out, unless there exist also infinite tower of massive higher spin particles, as in the case of string theory.

\subsubsection{String-Inspired 4D Effective Model}

Up to now, we have not discussed $D=4$ dimensional remnants in this subsection. One reason is that, the Gauss-Bonnet term, being a total boundary term in 4-dimensions, does not affect the local geometry. However we could start from higher dimensions and proceed to compactify the string action to yield a 4D effective action. One may further restrict the model such that the leading $\alpha'$ correction should include the coupling to the dilaton field; in addition, if one further imposes the validity of the no-hair theorem, then one may drop some complicated dilaton terms of $\mathcal{O}(R^{2})$. One suggested form is [in the Einstein frame] \cite{Kanti:1995vq} 
\begin{eqnarray}
S = \frac{1}{16\pi } \int d^{4}x \sqrt{-g} \left( R + \frac{1}{2}\left(\nabla \phi\right)^{2} - C \alpha' e^{\phi} R_{\mathrm{GB}}^{2} \right),
\end{eqnarray}
where $C$ is a constant with the ratio $2:0:1$ for Bosonic, Type II, and Heterotic, string theories, respectively \cite{Ratioc}. 

In these models, there is no exact analytic solution; but the existence of a static solution was checked by numerical methods \cite{Kanti:1995vq} assuming metric ansatz of the usual form
\begin{eqnarray}
ds^{2} = - f(r) dt^{2} + \frac{1}{g(r)} dr^{2} + r^{2} d\Omega^{2}.
\end{eqnarray}
Using numerical technique, one may solve $f(r)$, $g(r)$, and $\phi(r)$ by varying $M$, and obtain $f(M,r)$, $g(M,r)$, as well as $\phi(M, r)$ as numerical data. From this, one can fit these functions, for example, as follows \cite{Alexeyev:2002tg}:
\begin{eqnarray}
f(r) = 1 - \frac{2M}{r} \left( 1 - \frac{\epsilon_{1}}{M} - \frac{\epsilon_{2}}{M^{2}} + \frac{\epsilon_{3}}{M^{3}}- \frac{\epsilon_{4}}{M^{4}} \right),
\end{eqnarray}
where the $\epsilon_{i}$'s should be tuned by numerical solution and this will be true only up to a certain numerical regime; one can proceed similarly for $g(M,r)$ and $\phi(M, r)$. From these analysis, Alexeyev, Barrau, Boudoul, Khovanskaya and Sazhin \cite{Alexeyev:2002tg} observed that these string-inspired 4D black holes may approach a remnant solution with zero Hawking temperature.

\subsubsection{Quadratic Palatini Gravity}

In Einstein's gravity, the metric is the only dynamical variable. In the Palatini approach \cite{Olmo:2011uz} of gravity, both the metric and the connection are independent variables\footnote{The Palatini formalism is actually due to Einstein in 1925. See the discussion in \cite{FFR}.}. This Palatini formalism can be extended to modified gravity, for example up to the quadratic term: 
\begin{eqnarray}
S = \frac{1}{16\pi }\int d^{4}x \sqrt{-g} \left( R + aR^{2} + R_{\mu\nu}R^{\mu\nu} \right) + S_{m}, 
\end{eqnarray}
where $a$ is a constant and $S_{m}$ is the matter sector action \cite{Olmo:2011np}. Due to the correction terms obtained from the variation of the connection, one can potentially model new physical phenomena using the new degrees of freedom not available in a purely metric theory of gravity like GR. In \cite{Olmo:2011np}, the authors found the following solution for a spherically symmetric charged black hole: 
\begin{equation}
 ds^{2} = g_{tt}dt^{2} + g_{rr}dr^{2}+r^{2}d\Omega^{2},
\end{equation}
where
\begin{eqnarray}
g_{tt} = - \frac{A(z)}{\sigma_{+}}, \;\; g_{rr} = \frac{\sigma_{+}}{\sigma_{-}A(z)}, \;\; A(z) = 1 - \frac{1+\delta_{1} G(z)}{\delta_{2} z \sigma_{-}^{1/2}},
\end{eqnarray}
and $z^{2} := r^{2}/(q\sqrt{2})$ with charge $q$, $\sigma_{\pm} := 1\pm 2q^{2}/r^{4}$, and also
\begin{eqnarray}
\delta_{1}^{2} := \frac{1}{16M_{0}^{2}} \left(\sqrt{2}q\right)^{3},\;\; \delta_{2}^{2} := \frac{\sqrt{2}q}{4M_{0}^{2}},
\end{eqnarray}
with the black hole mass $M_{0}$. In addition $G(z)$ is a function that satisfies the ODE
\begin{eqnarray}
\frac{dG}{dz} = \frac{z^{4} + 1}{z^{4} \sqrt{z^{4} - 1}}.
\end{eqnarray}
This model in general has a singularity at $z = 1$ but this can be avoided if we have sufficient charge such that $M_{0}^{2} \lesssim \alpha^{3/2} N_{q}^{3}$, where $\alpha$ is the fine-structure constant and $N_{q}$ is the number of charges. That is, a black hole that already possesses a sufficient amount of charge can be singularity-free. Moreover, if the charge is smaller than a certain critical limit, then it becomes a microscopic massive object that can be interpreted as a remnant \cite{1311.6487}.

\subsection{Semi-Classical Framework: Modification of Matter Sector}

In the previous discussion, we considered the case with only one conserved charge -- the mass of the black hole, $M$. However, if we add other matter fields, then we can in principle allow various conserved charges [hence, various black hole hairs], say the electric charge $Q$, the magnetic charge $P$, and so on. Therefore, similar to the case of a Reissner-Nordstr\"om black hole in GR, these various conserved charges can allow extremal black hole configurations.

In general, for any static spherically symmetric black hole with a metric tensor of the form
\begin{equation}
ds^2 = -f(r) dt^2 + f(r)^{-1}dr^2 + r^2(d\theta^2 + \sin^2\theta d\varphi^2),
\end{equation}
its Hawking temperature is given by the well-known formula
\begin{eqnarray}
T = \frac{1}{4\pi} \left|\frac{df}{dr}\right|_{r_{+}},
\end{eqnarray}
where $f(r_{+}) = 0$ is the outer horizon. However, if there are a number of charges, then it can have more than one horizon. Let us denote the outer horizon and the inner horizon by $r_{+}$ and $r_{-}$, respectively. For the extremal limit, we have $r_{+} \to r_{-}$. For the specific value of $r$, say $r_{0}$, that satisfies $df/dr|_{r_{0}} = 0$, the inequality $r_{-} \leqslant r_{0} \leqslant r_{+}$ is satisfied. Hence, in the extreme limit, $r_{+} = r_{-} = r_{0}$ should be satisfied. Therefore, in the extremal limit, the Hawking temperature is zero [in fact this is sometimes taken as a working definition for an extremal black hole]. These extremal black holes can be candidates for remnants, \emph{if the evolution of the black holes bring them toward the extremal limit}. An extremal remnant may also be \emph{not} absolutely stable. These are very important points worthy of detailed discussion, and so we will postpone the discussion to subsec.(\ref{concluding remarks}). 
Note also that it is possible for an extremal black hole to be quite large, instead of the very small sized extremal remnants in the regular black hole proposals.

\subsubsection{Extremal Black Holes and the Fuzzball Picture}

There are a lot of extremal black hole solutions in string theory. Most of these solutions are motivated to investigate black hole entropy by using a string duality \cite{Strominger:1996sh}. To employ such a duality, one typically needs to rely on supersymmetry and this restricts the corresponding solutions to be either extremal or near extremal. To keep the length of this paper reasonable, we will not discuss all of the string inspired extremal black holes.

One important observation is that, there are \emph{many} string-inspired hairy solutions. This was the motivation of the fuzzball picture \cite{Mathur:2005zp}. Although the detailed description of the fuzzball picture is beyond the scope of this paper, we shall provide a short summary. The fuzzball picture relies on the fact that there exist a large class of hairy fuzzball solutions, which are asymptotically similar to ordinary black hole solutions but have no event horizon. Hence, for a number of states $N$, if we can construct $N$ different kinds of fuzzballs, then this may solve the information loss paradox. A black hole is thus nothing more than a superposition of these different geometries in a short time scale [compared to the lifetime of the ``black hole'']. Each fuzzball has neither an event horizon nor a singularity, and hence the whole process should conserve information. This is argued to resolve the information loss paradox.

There are nevertheless  some criticisms on the fuzzball picture. Firstly, in order for this process to work, we need to find enough number of fuzzball solutions, which may not be feasible. Secondly, only the [near-]extremal fuzzballs are well understood \cite{1208.3468}, but for this proposal to be realistic, we also need to consider the splitting of states for generic non-extremal black holes. Thirdly, if  black holes are just superpositions of geometries that could in principle split off into different observed states, the status of the in-falling observer is not very obvious. According to the fuzzball complementarity \cite{Mathur:2012jk, Chowdhury:2012vd}, as long as the free-falling observer has sufficiently large energy $E > T$ where $T$ is the Hawking temperature, then the observer will experience just the usual free fall as in general relativity. However, once this kind of free fall is allowed, then this will cause inconsistency anyway with firewall \cite{apologia, Hwang:2012nn}. Lastly, even if these problems are resolved, it is still not entirely clear how to relate the original information with the radiation emitted by the fuzzballs. Perhaps, this relies on the ``splitting process'' that is yet to be clearly worked out.

An interesting discussion on fuzzball and string theory in the context of information loss can be found in \cite{bena}.

\subsubsection{More Ado about Charges and Hairs}

Let us focus on extremal black holes that are more realistic. Once again, these extremal black holes arise from the introduction of various charges or black hole hairs. 
Before proceeding, let us remind the readers the rationale for discussing black hole hairs. The main idea is that an additional hair could provide also an additional parameter for a black hole to become extremal. Naively, an extremal black hole could live indefinitely and serves as a remnant.  

From the no-hair theorem, we know that in GR there is no other [fundamental] hairs except the tensor field $g_{\mu\nu}$ and the vector field $A^{\mu}$ \cite{Teitelboim:1972qx}. The tensor field is related to gravity while the vector field is related to the electric charge. [There is also another conserved quantity, namely the angular momentum, but it is not related to some fields.]

To find more hairy black holes, we have at least two strategies.
\begin{description}
\item[-- More vector fields:] The first option is to introduce more vector fields. The easiest way to do this is by introducing a new and larger gauge group, whose gauge bosons are vector fields. Of course, the color charge of quarks can be a candidate but it is not that realistic as a black hole charge [see, however, \cite{0806.4605}]. 
Even with the color charge, there are still not enough degrees of freedom to resolve the information loss paradox.
Perhaps, a larger gauge group can come from the string inspired models. As an example, Coleman, Preskill and Wilczek \cite{Coleman:1991sj1, Coleman:1991sj2} investigated the case of $\Bbb{Z}_{N}$ gauge symmetry. The $\Bbb{Z}_{N}$ electric hairs lead to non-perturbative quantum effects and hence it is not a trivial matter to reach a conclusive result. However, for the case of $\Bbb{Z}_{N}$ magnetic hairs, quantum effects are well controlled and allows zero Hawking temperature solutions with the constraint
\begin{eqnarray}
M^{2} = \frac{2\pi}{e^{2}} \frac{n(N-n)}{N},
\end{eqnarray}
where $M$ is the corresponding black hole mass, $e$ is the coupling constant, and $n = 1, ..., N-1$ is the $\Bbb{Z}_{N}$ charge. Due to the diversity of gauge fields, even for the extremal black holes, the mass parameter may not be unique.

\item[-- Violation of no-hair theorem:] If we want to consider scalar field hair instead of a vector field one, we have to violate some of the assumptions that go into the no-hair theorem \cite{Sudarsky:1995zg}. There are various ways to implement this. The simplest way is to introduce a potential $V(\phi)$ of the scalar field $\phi$. For example, if $V(\phi)$ can have negative values, then this may allow a hairy black hole solution \cite{Anabalon:2012ih}. However, although this possibility remains open, the relationship between such a scalar field and the conserved charges remains obscure. In particular it is not guaranteed that with a scalar field one can obtain an extremal black hole [which serves as a remnant] in the limit. 
Such a difficulty is illustrated, for example, by the so-called ``stealth solution'' on BTZ background [a stealth field is a non-trivial self-interacting, non-minimally coupled scalar field that does not contribute to the gravity sector, due to the vanishing energy-momentum tensor] \cite{stealth}. For some reasonable parameters, the stealth field eventually decays into the black hole.
Perhaps, a more interesting approach is to couple a scalar field to the gauge field, an option which we will discuss next. Recently it was also pointed out that, notwithstanding the no-hair theorem, a rotating black hole in GR can allow ``bristles''  -- extremely short-range stationary scalar configurations near its horizon \cite{1411.2609}.
\end{description}

\subsubsection{String Inspired Model: Dilaton and Charge Coupling}

In a large class of string-inspired models, there could exist a non-trivial coupling between the dilaton field and the electric/magnetic field. For example, it is permitted to have a dilaton $\phi$ and a gauge field $F_{\mu\nu}$ coupled together via a term that goes like
\begin{eqnarray}
\sim e^{-\beta \phi} F_{\mu\nu}^{2}
\end{eqnarray}
where $\beta$ has the ratio $0:1/2:1$ for Type~II, Type~I, and Heterotic string theory, respectively [see sec.(8.1) of \cite{BBS}].  These can give rise to different dynamics \cite{Hansen:2014rua} and thus have various implications. In addition, string-inspired models allow various classes of extremal black holes. This issue was investigated by Gibbons and Maeda \cite{Gibbons:1987ps} [see also, \cite{Torii:1993vm}] for the specific model given by the following action:
\begin{flalign}
S = &\int \sqrt{-g} d^{D}x \left[\frac{R}{16\pi} - \frac{\left( \nabla \phi \right)^{2}}{2\pi (D-2)} - U(\phi) - \frac{e^{-\frac{4}{D-2}g_{2}\phi}}{4} F_{2}^{2}\right. \nonumber \\ &\,\,\,\left.- \frac{e^{-\frac{4}{D-2}g_{D-2}\phi}}{2(D-2)!} F_{D-2}^{2} \right],
\end{flalign}
where $\phi$ is the dilaton field, $F_{2}$ is the 2-form field with coupling $g_{2}$ [which corresponds to an electric field], $F_{D-2}$ is the $(D-2)$-form field with coupling $g_{D-2}$ [which corresponds to a magnetic field], and $U(\phi)$ is the potential of the dilaton field. Strictly speaking the dilaton is \emph{not} an additional hair of the black hole. The reason is that it is coupled to the electromagnetic field in such a way that if $F_{\mu\nu} \equiv 0$, then the dilaton has to be constant. 

In the literature, among the class of solutions [see, e.g., \cite{1001.3739}], a specific one receives much emphasis due to its various interesting properties. This solution is the famous Garfinkle-Horowitz-Strominger [GHS] solution \cite{Garfinkle:1990qj, dark, gregoryharvey} in the 4-dimensional Heterotic case. The charged black hole in this theory has a metric tensor of the form 
\begin{flalign}
ds^{2} = &- \left( 1 - \frac{2M}{r} \right) dt^{2} + \left( 1 - \frac{2M}{r} \right)^{-1} dr^{2}  \\ \nonumber &\,\,\,\,+ r \left( r - \frac{Q^{2} e^{-2\phi_{0}}}{M} \right) d\Omega^{2},
\end{flalign}
where
\begin{eqnarray}
e^{-2\phi} &=& e^{-2\phi_{0}} \left( 1 - \frac{Q^{2}}{Mr} e^{-2\phi_{0}} \right);\\
F &=& Q \sin \theta d\theta \wedge d\varphi.
\end{eqnarray}
Here $Q$ is the magnetic charge and $\phi_{0}$ is the asymptotic dilaton field value. This model includes the extremal ``black hole'' solution at $Q^{2} = 2 M^{2} e^{2\phi_{0}}$. The metric tends to a naked \emph{null} singularity in the extremal limit and thus will cease to be a black hole in the usual sense [note that the 2-sphere metric shrinks to zero size in this limit; also one should take caution in that $r$ does not play the usual role of an areal radius here.]

A very interesting aspect of the charged dilaton black holes is the fact that strings do not experience the above mentioned geometry [``Einstein metric''] but rather the \emph{string frame metric}, which is related to the Einstein metric by a conformal factor: $\hat{g}_{\text{string}}=\exp(2\phi)g_{\text{Einstein}}$. The string world sheet has minimal surface area with respect to the string frame metric.
In the string frame, things are in fact quite different. By absorbing a factor of $\exp(2\phi_0)$ into the coordinates: $te^{\phi_0} \longmapsto t$ and $r e^{\phi_0} \longmapsto r$ we could write the string frame metric as
\begin{equation}
\hat{g}_{\text{string}} = -\frac{1-\frac{2Me^{\phi_0}}{r}}{1-\frac{Q^2e^{-\phi_0}}{Mr}} dt^2 + \frac{dr^2}{\left(1-\frac{2Me^{\phi_0}}{r}\right)\left(1-\frac{Q^2e^{-\phi_0}}{Mr}\right)} + r^2d\Omega^2.
\end{equation}
In the extremal limit, $Q^2 \to 2M^2e^{2\phi_0}$ and the string metric reduces simply to
\begin{equation}
\hat{g}_{\text{string}} \to -dt^2 + \frac{dr^2}{\left(1-\frac{2M}{r}\right)^2} + r^2d\Omega^2.
\end{equation}
One notes that the naked null singularity in the Einstein frame is no longer a single point in the string frame but a spherical surface, which is also the extremal horizon. 
 
The most interesting property is that in this limit, the extremal black hole \emph{as seen by the strings} is geodesically complete without any event horizon or singularity [they have been pushed to an infinite proper distance]. The geometrical structure looks like an infinitely long throat; this has an infinite amount of volume and hence the internal structure can potentially store an infinite amount of information. Furthermore the pair-production rate of such charged black holes from the vacuum is bounded \cite{Banks:1992is}. However, such an infinite throat solution only arises when the black hole is magnetically charged, but \emph{not} when it is electrically charged; despite the fact that the metric looks the same in the Einstein frame, they are different in the string frame.

\begin{figure}
\begin{center}
\includegraphics[scale=0.6]{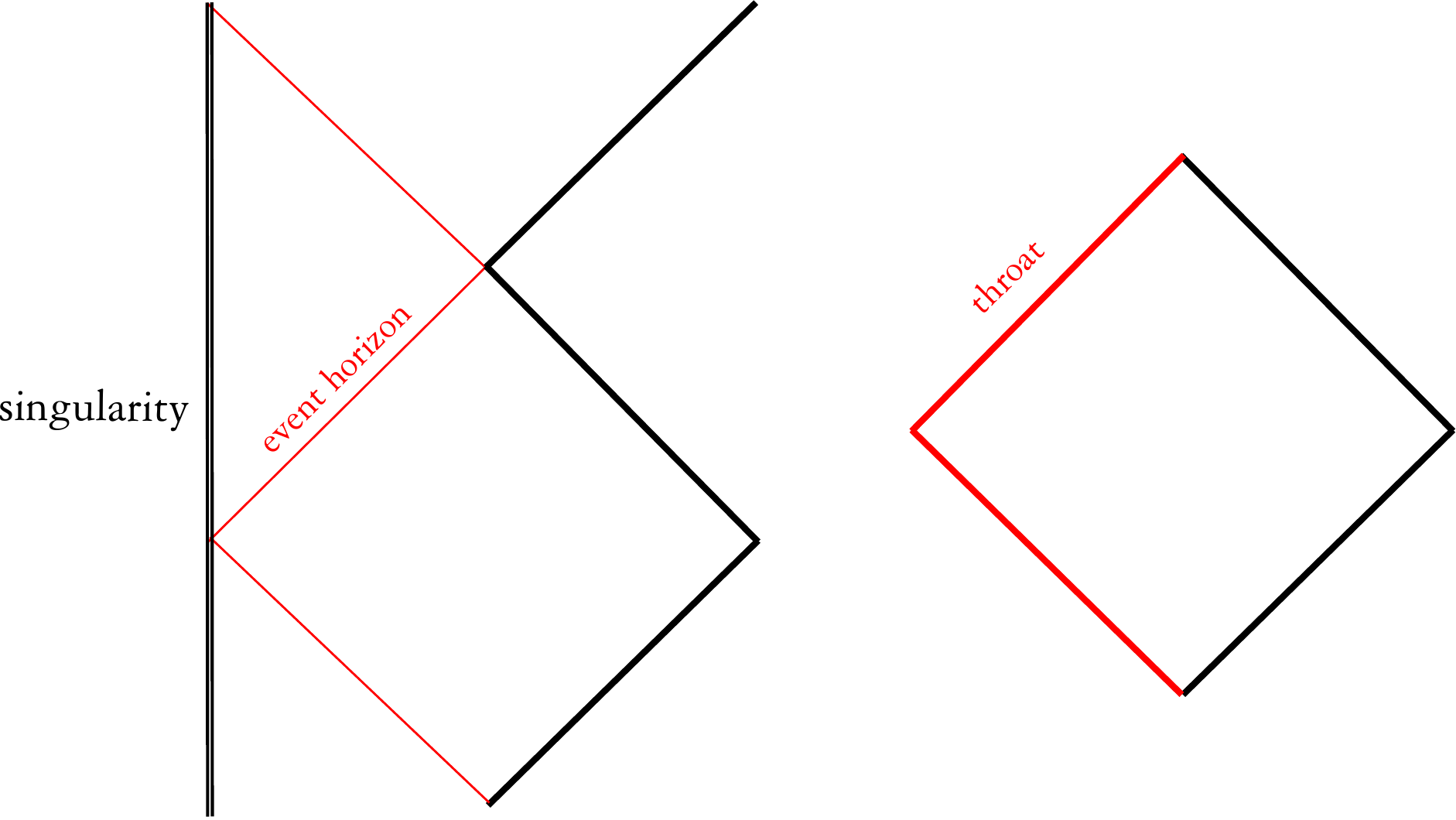}
\caption{\label{fig:extreme}\textbf{Left:} (Color online) The causal structure of an extremal Reissner-Nordstr\"om black hole, where thin lines denote the event horizon; \textbf{Right:} The causal structure of the extremal Garfinkle-Horowitz-Strominger [magnetically charged] black hole in the string frame. The left side of the diamond is the ``throat'' of the geometry.}
\end{center}
\end{figure}

One has to take note that this infinite tube geometry is very different from the more familiar infinite tube structure of the extremal Reissner-Nordstr\"om black hole in GR [see Fig.(\ref{fig:extreme})]. In the case of an extremal charged dilaton black hole, the horizon [which is singular] is at an infinite \emph{proper} distance to any causal observer, whereas for the extremal Reissner-Nordstr\"om black hole, the horizon is only at an infinite \emph{spatial} distance away. This difference is crucial -- it means that for the dilaton black hole, the extremal horizon behaves like another null asymptotic infinity, but no such interpretation can be given to the extremal Reissner-Nordstr\"om geometry\footnote{A question closely related to the issue of information storage capacity is whether [near-]extremal Reissner-Nordstr\"om black holes contain degenerate states. See \cite{0012020}.}. 

This model suggests an interesting possibility that black hole possesses a ``large interior'', a concept that we will return to in later sections in this review. At this point, however, one should point out that it is not clear how this scenario could help solve the information loss paradox -- should the string frame metric indeed be the fundamental one, even though when we describe gravitational collapse in GR we work in terms of the Einstein metric? Physical predictions are arguably the same for both metrics \cite{0111031} [see, however, \cite{9806032}] when they both make sense, but when one of the metrics break down due to singularities, it would be nice to have a more detailed picture of the equivalence. 
Also, the fact that the aforementioned large interior only arises in the presence of magnetic charge also casts some doubts to its utility in resolving the information loss paradox.

Finally, we comment on the singularity $r=0$ in the string metric of an \emph{electrically charged} GHS black hole. We note that in the limit $r \to 0$, the string coupling becomes \emph{weak} since $e^\phi \to 0$ in this case [the coupling becomes strong for the magnetically charged case, since $\phi \to -\phi$ under the electromagnetic duality that relates a magnetically charged solution to an electrically charged one]. Since the GHS solution is derived from a low-energy effective action of the full string theory, one should not trust the solution near high curvature regime, such as in the neighborhood of the curvature singularity. Nevertheless, as pointed out in \cite{dark},
\begin{quote}
``[...] \emph{it's difficult to resist speculating about what it might mean if the exact classical solution had
a similar behavior. It would suggest that, contrary to the usual picture of large quantum
fluctuations and spacetime foam near the singularity, quantum effects might actually be
suppressed. The singularity would behave classically!}''
\end{quote}
In other words, there remains the possibility that [at least some] curvature singularities may not be cured by quantum gravity, as we discussed in sec.(\ref{regular}).

\subsection{Remnant Scenarios in Lower Dimensions}

Black hole physics in various dimensions has always been an interesting topic. Since the information loss paradox occurs not only in 4 dimensions but also in other dimensions, the investigation in general dimensions may shed some lights on the essential properties of the information loss paradox and the remnant picture. In some of the previous subsections we have already discussed a few higher dimensional examples. In this subsection, we shall discuss the lower dimensional case.

In lower dimensions, as one might expect, black hole physics is simpler than that of four dimensions. However, because of the simplicity of gravity in lower dimensions, to obtain a black hole solution, some special conditions need to be imposed. [In pure GR without a cosmological constant, no black hole solution is possible in lower dimensions]. For example, in order to render black hole solutions in two dimensions dynamical, one must couple the dilaton field to the gravity sector \cite{Callan:1992rs}. For the three dimensional case, to find a non-trivial solution without a conical singularity, one has to turn on the negative cosmological constant \cite{Banados:1992wn}. We briefly summarize the status of the information loss paradox and discuss the connection to the remnant picture in these dimensions.
\begin{description}
\item[-- Two dimensional case:] A dilaton black hole model was constructed by Callan, Giddings, Harvey and Strominger \cite{Callan:1992rs}. This is the famous CGHS model. One can turn on the renormalized energy-momentum tensor that describes Hawking radiation \cite{Davies:1976ei}, but there is some freedom to choose this renormalized energy-momentum tensor. If one imposes a global symmetry, then one can even find exactly solvable models, such as the ones found by de Alwis \cite{deAlwis:1992hv}, Bilal and Callan \cite{Bilal:1992kv}; as well as Russo, Susskind and Thorlacius [RST] \cite{Russo:1992ax}.

The original CGHS model with Hawking radiation turned on can be analyzed using computer simulations \cite{Piran:1993tq}. According to the recent calculations \cite{Ashtekar:2010qz}, the authors reported that the end point of evaporation seems to approach a constant Bondi mass: $M \sim 0.86 (N/24) M_{\mathrm{P}}$, where $N$ is the number of matter fields and is assumed to be large. Motivated from their numerical calculations, the authors argued that this final endpoint should be interpreted as a remnant \cite{Almheiri:2013wka}; in addition, they argued that Hawking radiation is not pure, but instead entangled with the long-lived remnant.

This is an interesting proposal, but we need to be careful when interpreting the numerical results of CGHS. For example, it is theoretically possible to invoke the freedom of choosing the renormalized energy-momentum tensor for some analytic calculations. In the case of the RST model, the endpoint of evaporation can be calculated analytically where it has a \emph{negative} mass parameter \cite{Russo:1992ax}. It is not easy to interpret this as a remnant. The canonical interpretation is that, at the end stage of Hawking evaporation, the one-loop order approximation breaks down, and the black hole emits all of its remaining energy [which could be either positive or negative], and finally becomes a flat spacetime.

\item[-- Three dimensional case:] In three dimensions, we find a black hole solution in anti-de Sitter background, the famous Ba\~{n}ados-Teitelboim-Zanelli [BTZ] solution \cite{Banados:1992wn}. Although BTZ black holes have non-zero Hawking temperature, they do not completely evaporate due to their positive specific heat, which enables them to attain thermal equilibrium with their own Hawking radiation [unlike AdS black holes in higher dimensions, a BTZ black hole does not exhibit Hawking-Page transition -- they are always thermodynamically favored over pure AdS geometry]. Hence, these black holes are eternal. 

Since there is no complete evaporation, it seems that the black hole stores information forever and there is no information loss. Furthermore, these black holes are defined on an anti-de Sitter background and it is believed that the boundary field theory should be unitary \cite{Maldacena:1997re} [see, on the other hand, \cite{vafa}]. Nevertheless, correlation functions between the outside and the inside of the event horizon seems to decay to zero as time passes. This problem was discussed by Maldacena \cite{Maldacena:2001kr}. According to Maldacena, this demonstrates \emph{information loss even for an eternal black hole}. Perhaps, the loss correlation can be recovered by a non-perturbative ``trivial history''\footnote{By a ``trivial history'', Maldacena meant the trivial topology solution. } that contributes to the entire path integral for the boundary observer of anti-de Sitter space and this will resolve the tension between information loss of the eternal black hole and the AdS/CFT correspondence \cite{Maldacena:1997re}.

Although this is not quite the same as black hole remnants that we have been discussing, it nevertheless tells us that: \emph{the existence of a stable object may not ensure the conservation of information}.
\end{description}


\subsection{The Role of Electrical Charge in Black Hole Evolution}\label{concluding remarks}

Now let us remark on the various proposals that involve remnants as some kind of stable extremal black holes that arise as the end state of Hawking evaporation. This actually has several problems. The first problem is that a large class of extremal black holes are actually \emph{not} stable. The idea that they are stable comes from the fact that many extremal configurations are what known as BPS objects in string theory. However, it has been mathematically proved that certain perturbations on the horizon of extremal black holes would blow up in a finite time, signaling the instability of the geometry, even at the \emph{classical} level. This is the Aretakis instability \cite{1110.2006, 1206.6598, 1208.1437, 1307.6800}.  [An illuminating reference on the Aretakis instability is that of Bizo\'n and Friedrich \cite{1212.0729}.] Of course, this may not be an issue since black holes most probably never become \emph{exactly} extremal although they might come arbitrarily close to extremality. The Aretakis instability however only happens to an \emph{exactly} extremal black hole. Therefore if a black hole never becomes exactly extremal, it will not be subject to Aretakis Instability. 
The relevant and essential question is, instead, the following: \emph{Does a given black hole evolve naturally toward its extremal configuration?} 

Let us investigate the case in which the extremal condition is provided by the electrical charge [as in an electrically charged Reissner-Nordstr\"om solution in GR]. 
Asking such a question of course pre-supposes that the black hole is in isolation. In a realistic universe in which the black hole is surrounded by matter fields, one does not expect  geometrically interesting electrical charges to be present. However, this is a crucial question to ask because one can certainly formulate the information loss paradox even for such idealistic black holes. Investigating the evolution of charged black holes would then allow us to either rule out or support some proposed solutions to the information loss paradox. [For example, Harlow and Hayden \cite{HH} proposed that firewalls can be evaded by considering the enormously long time that one needs to ``decode'' Hawking radiation, so long in fact, that the black holes would have already evaporated by then. They were correct to worry about the charged case, since it \emph{appears} that a charged black hole may have infinite life time, and so their proposal would seemingly not work. This was shown \emph{not} to be the case in \cite{OMC}, via a careful analysis of a certain type of charged AdS black hole.]

In the literature on the information loss paradox, most works only consider either the Schwarzschild black hole, or its neutral counterpart in modified gravity or the regular black hole scenario. The usual justification given is that, as we have just mentioned: astrophysical black holes are not expected to possess any significant amount of electrical charge. Even at the level of purely theoretical considerations, it is commonly thought that electrical charges should not play any critical role in the issue of information loss paradox -- after all, it is permissible to consider a chargeless black hole to begin with. It is argued in \cite{OMC, OC}, however, that this is an over-simplication. The reason is, even an isolated neutral black hole can pick up a significant amount of electrical charge \emph{over its long lifetime}, as long as a theory allows a Maxwell-field. This build-up of electrical charge is from the fact that Hawking radiation does produce charged particles, albeit exponentially suppressed if the initial data consist of a neutral black hole. However, a typical black hole lifetime is extremely long, and therefore the effect accumulates and should not be ignored entirely. Of course, since the probability of producing positively and negatively charged particles are equal, by symmetry alone, one would be tempted to conclude that \emph{a priori} there is no reason that the evaporation process would favor one type of charge accumulation over the other. However, the charge loss rate is coupled to the mass loss rate [which also involves neutral particles], and therefore the evolution is actually non-trivial, and was only settled by a careful analysis of Hiscock and Weems in 1990 \cite{kn:HW}.

\begin{figure}[!h]
\centering
\mbox{\subfigure{\includegraphics[width=2.2in]{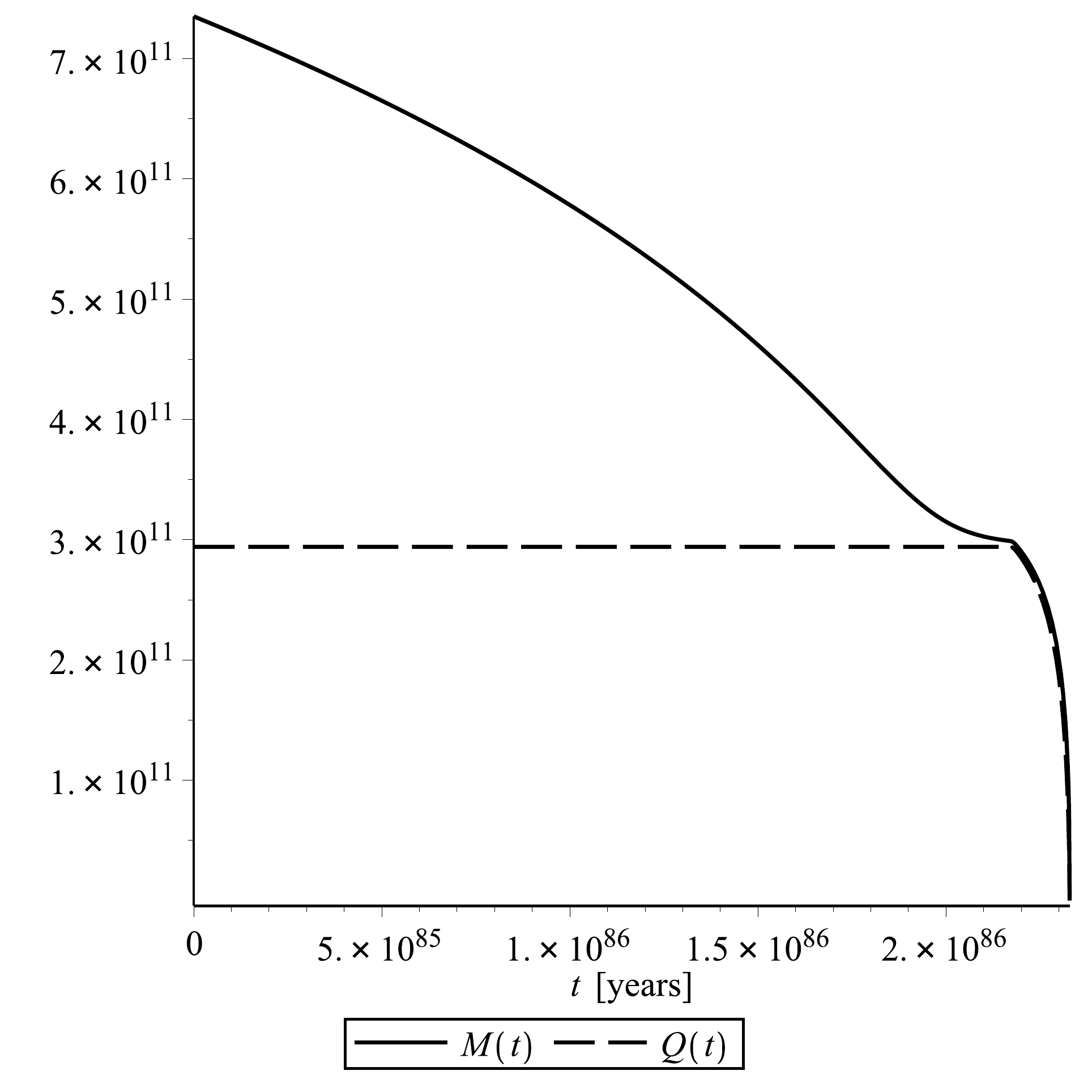}}\quad
\subfigure{\includegraphics[width=2in]{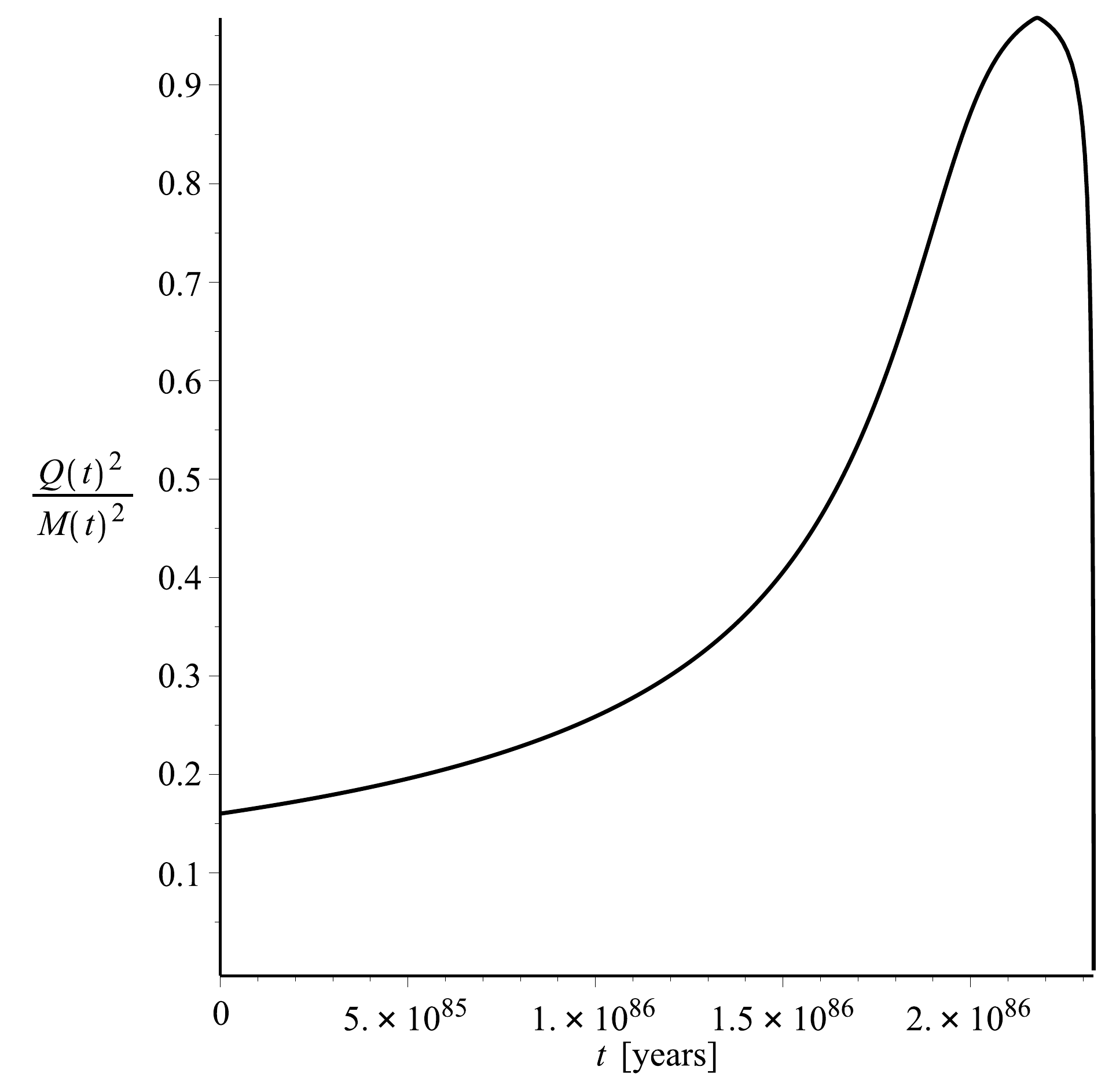} }}
\caption{\textbf{Left:} The evolution of mass and charge of an asymptotically flat Reissner-Nordstr\"om black hole with a very low value of the initial charge-to-mass ratio. The initial mass and charge are set to be $M=7.35 \times 10^{11}$ cm and $Q=2.94 \times 10^{11}$ cm, respectively. \textbf{Right:} The evolution of the charge-to-mass ratio of the same black hole. Note that \emph{initially} the charge-to-mass ratio increases, and in fact gets very close to the extremal limit, but \emph{eventually} turns around and decreases toward the Schwarzschild limit. See \cite{OC} -- from which these diagrams are taken -- for more detailed discussions. \label{RNevol}}
\end{figure}

As shown in Fig.(\ref{RNevol}) the evolution of a sufficiently large\footnote{This assumption was required in \cite{kn:HW} since the authors modeled charge loss via Schwinger effect of particle creation, and thermal Hawking radiation only emits neutral particle. This requires a \emph{cold} black hole, which, in asymptotically flat geometry, means a massive black hole.} asymptotically flat Reissner-Nordstr\"om black hole in GR is such that, even if the initial charge-to-mass ratio may be very small, it can approach the extremal limit during the long course of its lifetime \cite{kn:HW, kn:SP}. Nevertheless, the charge-to-mass ratio eventually turns around and approaches the Schwarzschild limit [note that the model eventually breaks down when the black hole gets sufficiently hot as it shrinks toward the Schwarzschild limit]. Therefore [near-]extremal black holes do \emph{not} play the role of black hole remnant as far as GR is concerned, if we assume the usual thermal picture of the Hawking radiation. [This statement is of course also true for a realistic universe in which black holes are not isolated.]

We recall that in many discussions about modified black hole solutions [regular black holes, non-commutative black holes etc.], one often argues that since the \emph{neutral} black hole solution possesses two horizons, they would naturally tend toward the extremal limit as they Hawking evaporate away. 
\emph{As far as information loss is concerned}, this argument needs a further, explicit calculation to justify. This is because neutral black holes are too special -- black holes do not stay neutral even if they are neutral to start with, so we really need to consider charged black holes\footnote{We remark that
electrical charge sometimes plays a crucial role in some remnant solutions. For example, 
electrically charged solutions within the Eddington-inspired Born-Infeld theory of gravity are not only singularity-free [the singularity is replaced with a wormhole supported by the electric field], but could also allow remnant \cite{1311.0815}. There is also the possibility of ``solitonic remnants'' -- sourceless geons containing a wormhole generated by the electromagnetic field \cite{1306.2504}. In addition, Markov and Frolov considered an ``almost-closed'' Friedmann-Lema\^itre-Robertson-Walker [FLRW] universe linked by a Reissner-Nordstr\"om wormhole back in 1971 \cite{MM0}. Such ``friedmon'' has nontrivial interior geometry but can look like a tiny particle from the outside.}, albeit with generic [i.e., not too high] charge-to-mass ratio. It would be interesting to study the evolution of charged modified black holes and see if they do tend toward extremal limit.  

We emphasize again that this discussion only holds true for idealistic \emph{isolated} black holes, not realistic black holes in our universe. The point is, the information loss paradox can be considered under the same assumptions, and thus a proposed resolution of the paradox should be examined to see if it holds also for the charged case. 

One might of course argue that the statement that Reissner-Nordstr\"om black holes tend to the Schwarzschild limit depends on the specific model of Hawking evaporation. This is certainly true. For example, in \cite{kn:kimwen}, it is argued that the extremal limit is an attractor, and so it would be natural for [near-]extremal Planck-size black holes to play the role of black hole remnant, which is an exact opposite of the conclusion obtained above. However, as was argued in \cite{OMC, OC}, due to the large uncertainty concerning Planck scale physics, we should try to approach the issue regarding the end state of black hole evaporation using a better understood quantum gravitational system --- such as the AdS/CFT correspondence. Note that in order to stay within the regime in which AdS/CFT is relatively well understood, one should work with AdS black holes with the property that the curvature at the horizon never becomes too large throughout its evolution. This should be contrasted with an asymptotically flat Schwarzschild black hole, whose curvature [specifically, the Kretschmann scalar $R_{\mu\nu\rho\delta}R^{\mu\nu\rho\delta}$] becomes unbounded as the horizon shrinks. 

In the context of charged black holes, this correspondence is best understood when the boundary is a flat spacetime, and thus the corresponding black hole in the bulk has a flat horizon [i.e. the horizon is either planar or a flat torus]. It was then shown that unlike the asymptotically flat case, these asymptotically locally AdS black holes with flat horizons do appear to approach the extremal limit without turning back toward the neutral limit [at least in the regime of validity of the model, which requires a sufficiently large AdS curvature length scale $L$]. However, these black holes are always eventually destroyed by a kind of stringy instability [Seiberg-Witten instability; see \ref{B}.], or by a phase transition from the black hole state to a type of solitonic state \cite{kn:surya, kn:hormy}. Furthermore, the curvature at the horizon of a flat black hole remains small --- it is $144/L^2$ in the extremal limit. 
This suggests that once we take the electrical charge into consideration, it is unlikely that the black hole can survive long enough. That is, from this point of view, \emph{remnants are unlikely to exist}. This result, even if correct, definitely does \emph{not} rule out all remnants. In particular, black holes in our real universe may still end up as remnants. The result does however suggest that we should not expect \emph{all} black holes to become remnants. We will discuss this futher in sec.(\ref{discussion}). 

This, however, raises another question: What happens to the information when a [charged] black hole is destroyed by the Seiberg-Witten instability? One may be tempted to think that with the destruction of the horizon, the information will just be released and be accessible to anyone who initially stayed outside the black hole. However, we would like to emphasize again that we need to know exactly what happens to the black hole singularity in a full quantum theory of gravity, without which we cannot conclude with much confidence about the fate of the information that has fallen into the black hole [although resolving the singularity does not by itself guarantee that we can solve the information loss paradox].

\subsection{A Short Summary}

In this section wee see that the remnant picture appears in many models and theories of either the phenomenological approach [including GUP], the semi-classical approach or ``full fledged'' quantum gravity candidates, as well as modified theories of gravity.  Let us now summarize these various approaches together with their advantages as well as disadvantages. 

\begin{description}
\item[1. Quantum gravitational principles:] Remnants are allowed by an effective approximation of  some quantum gravitational principles.
\begin{description}
\item[-- Strong points:] These remnants are consistent with the generalized uncertainty principle; remnants are of the Planck scale since quantum gravitational effects are turned on around that energy scale. This agrees with our intuitions on quantum gravity [if indeed our intuitions work well in this subject].
\item[-- Weak points:] For these examples, there are no rigorous field theoretical derivations, but rather only a toy model from the effective descriptions of quantum gravitational principles. Therefore, the dynamical generalization would be difficult. Can this approach be extended to include dynamical descriptions so that these models are at least perturbatively stable?
\end{description}
\item[2. Modification of gravity sector:] Remnants can arise from the introduction of higher curvature terms in the gravity action.
\begin{description}
\item[-- Strong points:] The introduction of higher curvature terms is very natural and is motivated by various theories, e.g., string theory.
\item[-- Weak points:] In these models, one needs to select the higher curvature terms very carefully to allow for remnant. This is a fine-tuning problem. The existence of remnant is not entirely trivial. However, in the Planck scale regime, all higher curvature terms should be equally important. Is there a guiding principle that may help to select the terms that will guarantee the existence of a remnant solution? In addition, the introduction of higher curvature terms can also be problematic \cite{camanho}.
\end{description}
\item[3. Modification of matter sector:] The inclusion of new charges allow the existence of new black hole hairs, which can accommodate extremal black holes as remnants.
\begin{description}
\item[-- Strong points:] This can be motivated from various particle physics models or string theory, even within the perturbative regime.
\item[-- Weak points:] In these models, one needs to choose the initial condition carefully to have a certain amount of charge. In addition, it is unclear whether the subsequent quantum effects would induce the decay of charge [similar to the Schwinger effect], and therefore push the solutions away from extremality.
\end{description}
\end{description}

\section{Challenges for Remnants}\label{challenges}

The idea that black holes eventually stop evaporating and remain as Planck-sized remnants, although well-motivated as we have seen from the previous section, still face some serious challenges. There are at least two types of challenges. The first is about the \emph{existence of remnants}, as we briefly touched upon in the concluding remarks in subsec.(\ref{concluding remarks}). The second challenge is: assuming that remnants exist, do they play any significant role in resolving the information loss paradox? We will review these issues in this section. 

\subsection{Over-Production of Remnants?}\label{overproduce}

Let us start with the first challenge. In addition to the point already raised in the preceding section about Seiberg-Witten instability which is based on a string theoretic argument, we would now explore whether Planck-sized remnants are consistent with known \emph{established} physics, especially with what we know about quantum field theory.

Consider the formation of a black hole via gravitational collapse and its eventual evaporation via Hawking radiation. Suppose also that the process is completely unitary \emph{but} Hawking radiation is completely thermal, in the sense that it does not contain any information about the matter that collapsed to form the black hole. If the end state of the evaporation is a remnant, then there are two possibilities: either the remnant is completely stable and stays in such form indefinitely, or that it is unstable and will eventually decay [``quasi-stable remnant'' or ``long-lived remnant'']. It has been argued that purity of the final state cannot be restored in the ``final stage'' of a remnant since there are too few quanta [only $E \sim M_\text{P}$ left] to encode the necessary correlations. Furthermore,
these quanta cannot be all emitted simultaneously in a burst; instead the information has to come out slowly, one particle at a time.
Therefore the lifetime of a remnant, even if it is not infinite, must be very long, exceeding the current age of the universe \cite{ACN}. That is, they are ``quasi-stable'' with life time proportional to $M^4$ \cite{preskill}. [See also, \cite{Giddings1, CW}.] For all practical purposes then, this suggests that we need not distinguish between a stable remnant and a quasi-stable one. 

It has been argued that both types of remnants lead to the infinite production problem [see, e.g., \cite{Giddings3} and the references therein]: since information is stored in the remnant, and since we can form a black hole from an arbitrary initial state, it seems that there will be essentially an infinite number of different remnant species. It is then argued that, at the level of effective field theory with cutoff scale $\lambda_c \gg L_{\text{P}}$, these remnants can be treated as ordinary point particles. Now, since effective field theory cannot resolve any structure with a scale smaller than the cutoff scale $\lambda_c$, these remnants should therefore have the same production cross-section. Since there are essentially an infinite number of species, even if the production rate for each of these species is exponentially suppressed, overall we still have an infinite pair-production problem --- the vacuum should be unstable and produce copious amounts of black hole remnants. This is of course not observed in Nature. 

To be more detailed, for a black hole with mass $M$, its Bekenstein-Hawking entropy is of the order $M^{2}$, and hence the naive probability for the production of such a remnant is $P \sim N e^{-M^{2}/M_{\text{P}}^2}$, where $N$ is the number of different species of the remnant. The exponential factor is the famous term in the Gross-Perry-Yaffe-Kapusta [GPYK] theory of black hole formation \cite{grossperryyaffe, kapusta}. We would like to comment that the GPYK black hole nucleation rate assumes a background temperature $T$ of the universe. This rate is therefore the rate of black hole production due to the vacuum fluctuations under this temperature. Black holes so produced will have their masses $M \sim 1/T$. 

Hence, $N \sim \exp S_{\mathrm{st}}$, where $S_{\mathrm{st}}$ is the statistical entropy of the remnant [which is \emph{not} the same as the (coarse-grained) Bekenstein-Hawking entropy]. The problem is that, to contain an arbitrarily large amount of information in such a remnant, $S_{\mathrm{st}}$, and therefore $N$, should be arbitrarily large. One might be tempted to argue that, for a Planck sized object, $M \sim M_{\text{P}}$, so the exponential suppression is not very effective. Therefore, it seems that the quantum vacuum should generate essentially an infinite [or finite but extremely large] numbers of Planck-sized remnants. This statement is however misleading, since $M \sim 1/T$ and so the only time when $M \sim M_{\text{P}}$ can naturally happen is during the Planck era, and the copiously produced black holes would be much diluted by inflation, similar to what happened to the magnetic monopoles. In late time universe the temperature is much lower than the Planck temperature and therefore such black hole production is exponentially smaller. The only problem here is that $N$ can be so large that it could in principle overcome the exponential suppression. 

However, perhaps one could also argue that if the statistical entropy [which could be different from the thermal entropy] of a black hole is sufficiently large, the production probability of a remnant should be proportional to $e^{-S_{\mathrm{th}}}$ and not $e^{-M^{2}/M_{\text{P}}^2}$. That is to say, perhaps one has to differentiate between two types of black holes: primordial black holes and black hole remnants. Although they are of the same size, their entropy contents may not be the same.
Furthermore, following standard QFT, it seems that the probability of finding a final state should be the total cross section divided by the ``incoming flux'', i.e., the number of the initial states. So we may not gain anything even if $N$ is large and $P \propto N e^{-M^{2}/M_{\text{P}}^2}$.

There are indeed numerous counter-arguments that attempted to explain why the above naive argument against the existence of remnants is not correct. These include\footnote{Another argument is that, at least in de Sitter spacetime, the number of remnant species could be finite \cite{0803.2467}. Of course, this argument would not solve the information loss paradox in a hypothetical asymptotically flat universe. As pointed out in the paper itself, even for de Sitter spacetime, it is not clear if a large but finite number of states is phenomenologically allowed.}:
\begin{itemize}
\item[(1)] Black hole remnants continue to have a large interior volume [``massive remnants''] although they appear as point particles to the exterior observers. Therefore one should not treat them as point particles that can be created from the quantum fields via local operators. See e.g., the ``cornucopion'' scenario [expanding ``horn'' of a black hole
remnant] of Tom Banks \cite{Banks}. See also, \cite{piran}.
\item[(2)] Pair production of remnants is suppressed due to strong effective couplings \cite{Giddings3}.
\item[(3)] Pair production of magnetic black holes in a weak magnetic field is estimated in a weakly-coupled semiclassical expansion about an instanton and found to be finite, despite the infinite degeneracy of states \cite{BLS}.
\item[(4)] Fluctuations from the infinite number of states lead to a divergent stress tensor. This spoils the instanton calculation that led to the conclusion that production rate is infinite \cite{Giddings01}. 
\item[(5)] Effective field theory may not be well-defined for remnants \cite{sabine}. 
\end{itemize}

Susskind argued that even if remnants possess large interiors, the existence of remnants in the thermal atmosphere of Rindler spacetime would still lead to the pair-production problem \cite{suss0}. Indeed, it is claimed that ``the entire Rindler space fills up with remnants out to arbitrarily large distances, even to where the acceleration is negligible''. However, in view of point (5) above, this still depends on whether effective field theory can be well-defined for remnants. In addition, Susskind's argument depends on treating the enormous entropy inside remnant on equal footing as the Bekenstein-Hawking entropy, i.e., the additional remnant entropy also contributes to thermodynamics. It is not obvious that this is necessarily the case. 

Let us further remark on the infinite-production problem. Central to this argument is the assumption that, if effective field theory holds, then one could have a cutoff scale $\lambda_c \gg L_{\text{P}}$. For Planck sized remnants, whose size is much smaller than the cutoff scale, they can thus be treated as point particles. Like all quantum particles one could compute the production rate of these remnants from the vacuum. It is argued that, even though the production rate of a specific remnant is exponentially suppressed, due to the huge number of ``species", the overall production rate of remnants should still diverge. Recall that in the context of cosmology there is the so-called ``Boltzmann Brain Problem'' [see, e.g., \cite{0610199}]: If the universe is in thermal equilibrium, a small random fluctuation that produces a sentient brain with certain thoughts like ``I think, therefore I exist'' before dissolving back into the environment would be far more likely than a large fluctuation that produces galaxy clusters which lead to sentient beings like us. So why aren't we ``typical'' observers like the Boltzmann Brains? The answer is that the universe is far from being in thermal equilibrium. Regardless, if one thinks about a human brain, it has extremely many variations -- the number of neurons, and the exact shape of the dendrites [branching of a neuron that propagates electrochemical signal along the neural network] can all vary from one brain to the next. Indeed no one has the same memory, and so each brain is completely unique. Of course the probability of fluctuating out a brain is exponentially suppressed, but since there are potentially infinite number of different brains, why don't we see Boltzmann Brains being copiously produced even in a universe which is \emph{not} in thermal equilibrium? This suggests that we should not treat objects which contains large amount of information, be they black hole remnants or disembodied Boltzmann Brains, in the same way as we treat elementary particles -- namely, if there are $N$ different types of an object $X$, the total probability of production is \emph{not} the naive one given by $N$ multiplied by the [exponentially suppressed] production rate of a single object $X$. Of course one could reason that brains are macroscopic and the cut-off scale of effective field theory for remnants is far smaller, so we should not compare Boltzmann Brains with remnants. {\color{black}In addition, perhaps such comparison is not correct since the number of brains of a given size should not be greater than the number of states for a black hole of the same size, i.e. of the order $\exp(A/4)$ for boundary area $A$, if one trusts that the entropy bound holds. That is to say, the statement that the number of distinct brains could potentially be infinite may be incorrect.}

All these boil down again, to the central issue -- whether one could apply effective field theory to an object that looks like tiny particle from the outside but has non-trivial geometry on the inside. We suggest that this question should be explored in greater details; again we emphasize that GR is a geometric theory, it is only fair that we pay attention to geometries. In the words of Ellis in \cite{ellis}, 
\begin{quote}
``\emph{This argument requires you to believe standard QFT holds for super-Planck scale black hole remnants. This seems highly questionable.
Indeed if it were true, it is not clear why one needs a full quantum gravity theory.} ''
\end{quote}
Ellis also argued that the enormous number of initial states \emph{eventually left spacetime} as the information hits the singularity. Those states should therefore \emph{not} be included in any propagator calculated from \emph{within} spacetime.
 
Furthermore, even if one assumes the applicability of an effective field theory, Itzhaki counter-argued that, 
since with the cutoff $\lambda_c$ one cannot distinguish between different states of the remnants, one actually should sum over all the possible states of the remnants and describe the Hawking radiation as a \emph{mixed state} [see footnote 1 of \cite{Itzhaki}]. That is to say, the exterior observers will never be able to recover information that has fallen into a black hole. 
We will come back to this point when we discuss the role of remnants in the context of information loss paradox. Indeed, one obvious way out of the infinite pair-production problem is to allow Hawking radiation to carry information, so that the information content in a Planck-sized remnant is small. That is, the number of internal states of a black hole is finite \cite{Giddings2}. \emph{Whether Hawking radiation carries information or not is therefore both consistent with the existence of black hole remnants}.

Finally, we would like to mention a recent work \cite{xavier} in which the author argued that as long as quantum
gravity preserves the symmetries of the low energy effective field theory, low energy experiments do not rule out remnants. To be more specific, the measurement of the anomalous magnetic moment of the muon gives a bound on the number of black hole remnant states as $N\sim 10^{32}$. This would of course still be a problem if $N$ is actually potentially infinite, as is usually claimed. The author however assumed that the production rate is not proportional to the number of species, by arguing that for external observers only the traditional hairs [mass, electrical charge, and angular momentum] are relevant, and the internal states behind the horizons should not be important for the quantum production of these remnants. [The author also pointed out an interesting possibility that black holes at the Planck scale could be \emph{non-thermal} objects.]

\subsection{Does Remnant Violate CPT Invariance?}

Hawking has argued that remnant is not feasible due to CPT invariance. The CPT argument is that: since one could form a black hole from, say, a concentration of just gravitons and photons, it should be able for a black hole to completely dissolve back into a collection of gravitons and photons, leaving nothing behind. In fact the most recent argument of Hawking applies CPT invariance to not only remnants, but also firewalls and event horizons \cite{1401.5761}. This argument is incorrect. 

First of all, the CPT theorem \cite{gerhart}, while rigorously proven for Minkowski spacetime, has not yet been conclusively established for generic curved spacetimes [although this is widely believed to be true]. However, most importantly, CPT symmetry concerns the interactions between particles at the very small scale; this does not contradict the fact that we observe an arrow of time [entropy almost always increases] in everyday life. For the same reason, while one could perhaps argue that mini black holes formed from quantum fluctuation should decay due to CPT, the same argument does not apply to macroscopic black holes which are formed from gravitational collapse. There, entropy does increase in accordance with the second law of thermodynamics. When a black hole is formed [the Bekenstein-Hawking entropy is much greater than the entropy content of the collapsing star -- more than 30 orders of magnitude \cite{clarify}], and as it evaporates away, the smaller black hole [and eventually the remnant] \emph{and} the Hawking radiation together form a system, which has even higher entropy than the initial black hole. Even for the case in which mini black holes are produced from the vacuum, it is possible that they too end up as remnants; the decay argument using CPT may be correct, but the time scale to decay may still be extremely long and so the remnant would be for all practical purposes, stable\footnote{In the early universe just after the Big Bang, the temperature was high enough to produce \emph{primordial [mini] black holes} \cite{carr} by a copious amount. One might worry that if they remain as remnants this would be a problem for cosmology \cite{tegmark}. However just like the issue of  the magnetic monopole, the over-production of primordial black holes is not a concern -- they are washed away by inflation. It is however not true that there cannot be physical significance of remnants due to primordial black holes. In fact, the number density of black hole remnants produced at the end of inflation can potentially play the role of dark matter \cite{pisinDM}. {\color{black} Scenarios involving black hole remnants production in the early universe can  of course arise from many models, see e.g., \cite{1102.5096}.}}. 

This again suggests that one should not treat black hole remnants, which are end states of gravitational collapse, as elementary particles. If indeed remnants can store enormous amount of information, they are much harder to be nucleated from the vacuum than the mini black holes, which carry only some minute bits of information. 

{\color{black}It is worth mentioning that although, as we have just mentioned, the second law of thermodynamics and the observed arrow
of time does not contradict CPT as a symmetry of the evolution, it may 
suggest that at least our component of the quantum state does not
itself satisfy CPT invariance.  Of course, this is just a broken
symmetry, usually interpreted to be a case in which the quantum state
does not share the symmetry of the dynamics. It is possible that even though our observed component of the quantum state is not CPT invariant, the superposition of the complete set of states would contain CPT-reversal of said component, and therefore as a whole obeys CPT invariance\footnote{\color{black}We thank Don Page for informing us about this point, as well as the point raised in the next paragraph.}. }

{\color{black}Another argument against the claim that remnants are unstable due to CPT invariance is the following. Consider a gravitational collapse, which results in a black hole. Along its evolution toward a remnant state, the black hole has emitted a lot of Hawking radiation. The CPT-reversal of this process is \emph{not} that the remnant can turn back into a star, but that the remnant, \emph{together} with the CPT-reversal of all the Hawking particles, can in principle turn back into a star. Therefore, unless one can feed those CPT-reversed Hawking particles into the remnant, it could not simply turn back into a star. CPT symmetry therefore does not imply that a black hole remnant is unstable.

Finally, we remark that it remains an open question whether the theory of quantum gravity truly respects CPT symmetry.}

\subsection{The Implications of Remnants to Information Loss and Firewall}\label{3.3}

If Hawking radiation does not carry any information, then as a black hole gets smaller and smaller, it becomes clear that a Planck-sized remnant must store an arbitrarily large amount of information. This can be achieved with a massive remnant as we mentioned above. However, in such a scenario we lose the nice interpretation of the Bekenstein-Hawking entropy as a measure of the information content [or of the underlying quantum gravitational degrees of freedom] of a black hole. However, in addition to this so-called ``strong form'' of Bekenstein-Hawking entropy interpretation, there exists a weaker form which asserts that the Bekenstein-Hawking entropy is only related to the degrees of freedom on the horizon, in particular it is not dependent on the interior of a black hole \cite{sabine, Smolin-1}. If one were to take the position that all information is contained inside the remnant, then one has to give up the strong form of the Bekenstein-Hawking interpretation and subscribes to the weak form instead. See also the discussion in \cite{inout} and \cite{marolf}.  

We shall now discuss in more details what implication does black hole remnant have toward the information loss paradox. This of course closely depends on whether the strong form or the weak form of the Bekenstein-Hawking entropy interpretation is correct. If the strong form is true, i.e., if the area is proportional to the statistical entropy of the underlying degrees of freedom of the black hole, then Hawking radiation should contain information. Indeed, if one collapses a pure state to form a black hole, the end state would seem to be a
remnant and the Hawking radiation, which though together form a pure state, are separately in a mixed state. Purity is not totally recovered since some information remains locked up in the [quasi-]stable remnant. Nevertheless, in such a picture, by allowing information to be carried out by the transfer of quantum entanglement means that we still have the firewall paradox. After all, the firewall appears at least at around the Page time\footnote{The Page time is roughly the time scale at which point the black hole has lost half of its entropy. We will review this in \ref{A}.}, at which point a black hole can be still very large and far away from reaching its final remnant phase. As long as we adhere to the strong form of Bekenstein-Hawking entropy interpretation, the firewall paradox would raise its head. 

However, if we adopt the weak form instead, then the area is not proportional to the actual statistical entropy --- the horizon area is just the degrees of freedom available to the outside observer, and is not the real internal degrees of freedom. In this case there is no need for the Hawking radiation to contain any information. 
The information could be entirely contained in the black hole. As we have emphasized many times, it would be crucial to know what exactly happens at the singularity of a black hole, as classically the information would crash into the singularity and without having any control over the boundary conditions there we would lose the information. 

One may argue that the weak form of the Bekenstein-Hawking entropy interpretation would violate the covariant entropy bound \cite{Bousso:1999xy} [see also, Bekenstein's entropy bound \cite{287, 9307035}], loosely put\footnote{More precisely, a suitably defined ``entropy flux'' $S$ through any null hypersurface $\Sigma$ generated by geodesics with non-positive expansion starting from some spacelike 2-surface of area $A$ must satisfy $S\leqslant A/(4G)$.} as the statement that the entropy of a system with area $A$ cannot consume more information than $A/(4G)$. However the entropy bound requires the validity of the Null Energy Condition [NEC] \cite{FMW}, but we know that the NEC is definitely violated during black hole evaporation. Therefore, the entropy bound argument is not a conclusive refutation. In addition, it is possible that at the Planck scale, the concept of entropy no longer makes sense. 
Furthermore, even at the classical level, one could contemplate matter configurations with arbitrarily large entropy bounded by a fixed small area as viewed from the exterior. These so-called ``monsters'' \cite{hsu1,hsu2}, which we will discuss later in subsec.(\ref{goldbh}), though violate the \emph{spacelike} entropy bound, still obey the covariant entropy bound, 
since the light-sheets off of their boundary surface are truncated by black hole singularities [monsters inevitably collapse to form black holes] \cite{hsu2, bousso}. Again, we see that the role of singularities should not be neglected in discussing the information loss paradox. In fact, according to Bekenstein \cite{9307035}, 
\begin{quote}
``[...]\emph{ if a remnant had an information capacity well above that indicated by [the] bound 
\begin{equation} \left[S \leqslant \frac{ 2\pi R E}{\hbar c}\right] \nonumber  \end{equation}
 in terms of its external
dimension, it would cause a violation of the GSL [Generalized Second Law], were it dropped into a large black hole. This point might be
countered if the infall of the remnant produces a singularity, thus making the GSL moot. However, in that case
the whole role of remnants as stable information repositories is put in doubt.''}
\end{quote}

\begin{figure}
\begin{center}
\includegraphics[scale=0.45]{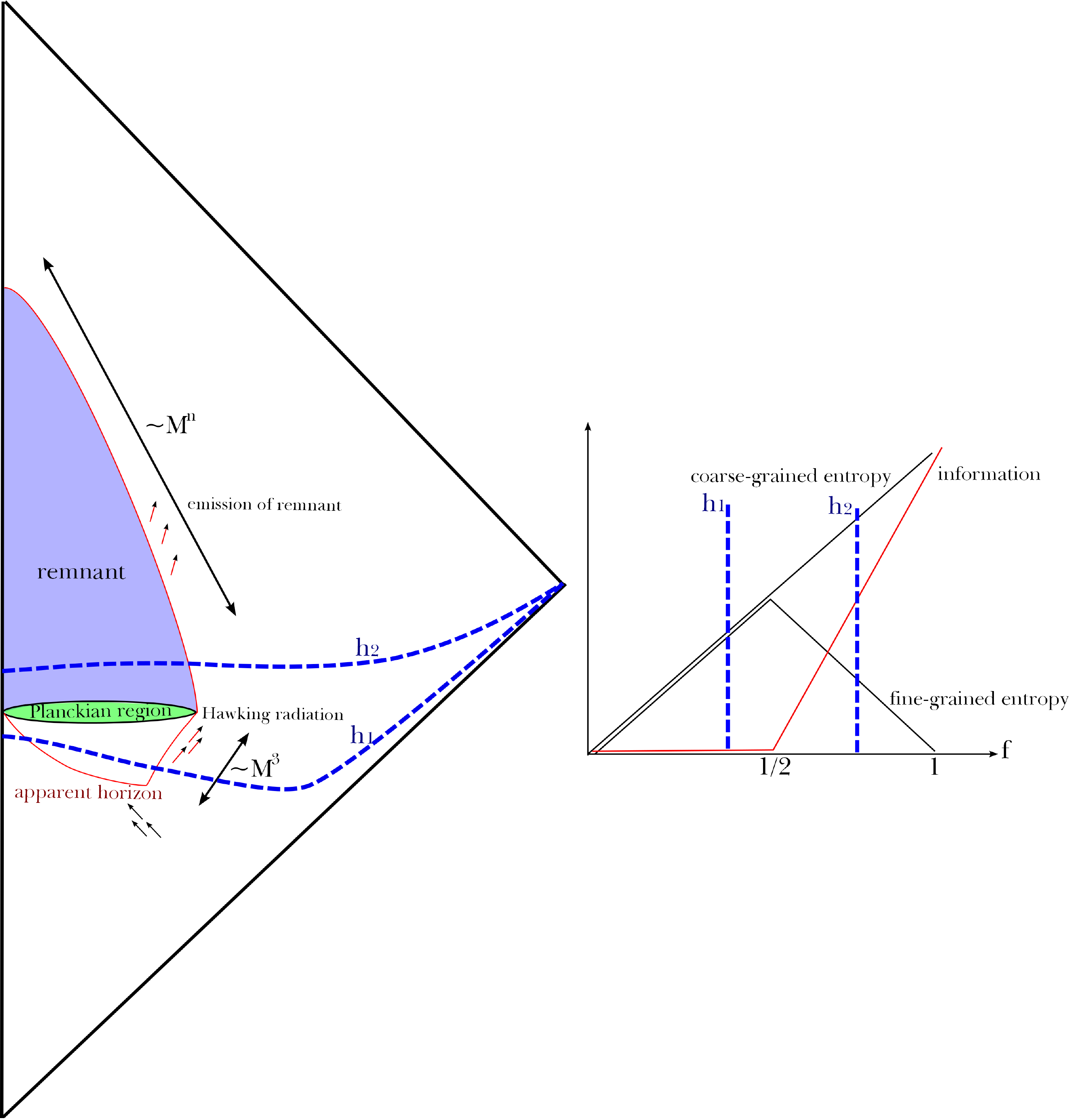}
\caption{\label{fig:regular_remnant}(Color online) The typical causal structure of a regular asymptotically flat neutral black hole that turns into a meta-stable remnant that very slowly evaporates away. Almost all information is slowly emitted [with lifetime proportional to $M^n$, with $n$ at least equals to 4] via the radiation of the remnant. Therefore, at the surface $h_{1}$, there is only a negligible amount of 
information content; at the surface $h_{2}$, emitted particles contain information. The small figure shows the Page curve. Note that information starts to come out after the Page time. The vertical axis of the Page curve diagram is  $\log S_{\text{BH}}$; and the horizontal axis is the fraction $f$ of the [logarithm of] evaporation time. Note that due to the gray-body factor, the turning point is not exactly at $f=1/2$ \cite{page2}.}
\end{center}
\end{figure}

Of course, one can also contemplate the possibility that Hawking radiation still carries [perhaps some] information out from a black hole even though we assume only the weak form of the Bekenstein-Hawking entropy interpretation. In such a scenario, as long as the black hole contains a finite statistical entropy [though it can be much larger than the apparent entropy given by the Bekenstein-Hawking value], once the statistical entropy decreases to about half of the original value, Page's argument should follow through and information will start to be released [see Fig.(\ref{fig:regular_remnant})]. This Page time will be enormously longer than the one in the no-remnant picture, proportional to $M^4$ instead of $M^3$. One may be tempted to argue that, by that time the black hole remnant is so near to the Planck scale that new physics should be taken into account and this would prevent a firewall from setting in. 

\begin{figure}
\begin{center}
\includegraphics[scale=0.45]{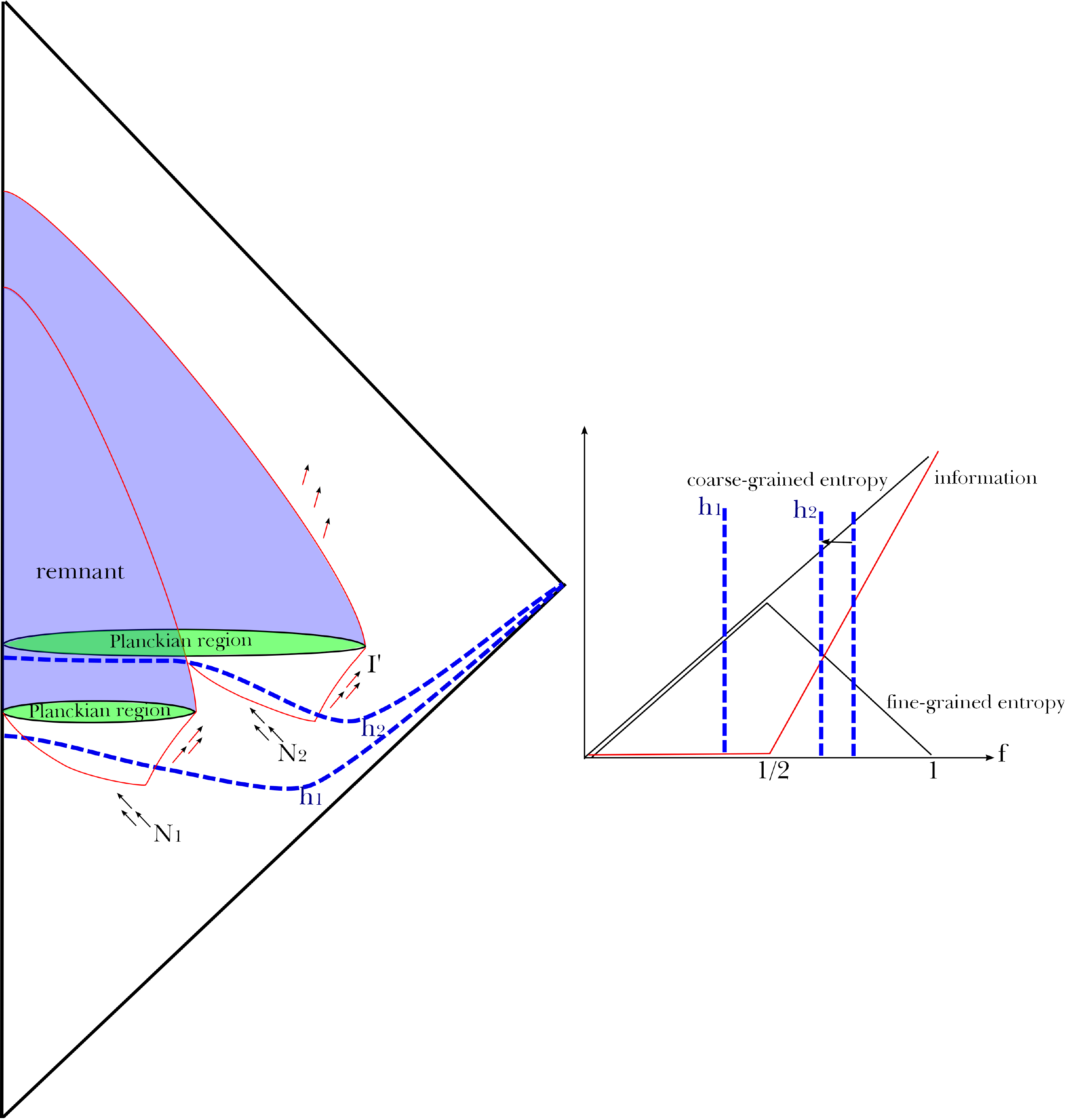}
\caption{\label{fig:regular_remnant_duplication}(Color online) If we add mass $m \gg M_{\text{P}}$ to the remnant around $h_{2}$ with states $N_{2} \ll N_{1}$, where $N_{1}$ is the number of states of the original black hole, then the state of $h_{2}$ in the graph of Fig.(\ref{fig:regular_remnant}) will be shifted. However, if $N_{2} \ll N_{1}$, then the new configuration is still on the decreasing part of the Page curve, and the new information will be quickly scrambled and then carried out via Hawking radiation.}
\end{center}
\end{figure}

Nevertheless this does not entirely evade firewalls. Consider the following question: what will happen if we throw some matter or concentrate a beam of energy [with mass/energy much greater than the Planck mass] into the remnant [Fig.~\ref{fig:regular_remnant_duplication}]? One expects that a semi-classical black hole will form. One important point is that, although the mass is greater than the Planck mass, the statistical entropy of the matter/radiation can be much smaller than the initial statistical entropy of the remnant [which contains an arbitrarily large, though presumably finite, entropy]. Then, since we have a semi-classical black hole after the information retention time; and the original configuration already emitted half of the original [as well as the ``new'' -- since the newly added matter/radiation is assumed to have a very small entropy] statistical entropy, the newly formed semi-classical black hole should begin to emit the added information soon after the scrambling time $\sim T^{-1} \log (S/M_{\text{P}}^2)$ \cite{Hayden:2007cs} via Hawking radiation [Fig.(\ref{fig:regular_remnant_duplication})]. Then it seems that the firewall problem would once again, reappear. 

The conclusion is this: even if one subscribes to the point of view that a remnant can contain an arbitrary large [but finite] amount of information, as long as we allow Hawking radiation to carry information [in the manner that satisfies all the assumptions in \cite{amps}, with the subtlety that the statistical entropy need not be equal to the Bekenstein-Hawking entropy], we would face the menace of the firewall. That is, in order for remnants to play \emph{the} role in resolving the firewall paradox, one has to subscribe to the belief that the weak form of the Bekenstein-Hawking entropy is the correct interpretation, \emph{and} that Hawking radiation should not carry information. There is of course one obvious way out of this argument --- a remnant with \emph{infinitely} large degrees of freedom\footnote{This statement should not be confused with the case in which the \emph{Bekenstein-Hawking} entropy is infinitely large, e.g. for planar black holes [``black branes''], in which case the relevant quantity is actually the entropy \emph{density}, not entropy \emph{per se}. See e.g. \cite{kn:son}.} would presumably never reaches its Page time. However, it is hard to justify how this could be the case if gravitational collapse starts with a typical \emph{ordinary} star with \emph{finite} degrees of freedom. 

Another possible way out of our conclusion is the information locking scheme proposed by Smolin and Oppenheim \cite{SmoOpp}. In that scenario, the authors pointed out that even if information does not escape during the semi-classical evaporation process, it still does not mean that all the
information should remain in the black hole until the final stages of evaporation. This is due to information not necessarily being additive; a small number
of quanta can ``lock up'' a huge amount of information. In other words, one can restore the information only after the evaporation has finished. This is an interesting idea. If it is true, then we do not need to require large entropy for a remnant. One problem is that such ``locking'' mechanism requires a type of ``capacity'': the remaining small black hole [or remnant] that ``locks'' the information until quantum effects dominate needs to contain quantum information which grows like $3 \log(\log N)$, where $N$ is the number of states, but its lifetime goes only like $(\log M)^2$ instead of $M^4$ [the lifetime of  a ``long-lived remnant'' \cite{CW}; see Fig.(\ref{fig:regular_remnant})]. Therefore for sufficiently large $M$, the ``lock'' fails: it is not large enough to contain all the information. Smolin and Oppenheim argued that this possible failure of the locking mechanism is unlikely to be a problem in practice, but theoretically it certainly remains one.

\subsection{Bag-of-Gold Geometries and the Interior of Black Holes}\label{goldbh}

While much effort has been focused on understanding the near-horizon region of a black hole, its singularity has received relatively less attention. This is understandable --- in the neighborhood of the horizon of a semi-classical black hole, we believe that general relativity and ordinary quantum field theory should be sufficient to understand the physics [of course the issue of firewall claims that there are subtleties even in this low curvature regime], whereas in the near singularity region, spacetime curvature is so large that we should expect quantum gravitational effects to become important, and is thus beyond our current knowledge of physics. However the singularity, as well as the \emph{interior} of a black hole as a whole, \emph{is} important if we want to understand what happens to the information that falls into the black hole. After all, a typical black hole is \emph{not} like a black box; its interior is a dynamical spacetime region\footnote{One should not therefore talk about the ``mass'' being concentrated at the ``center'' of a Schwarzschild black hole, for example. That ``center'' $r=0$ is spacelike, and not necessarily a ``point''. Any ``derivation'' involving the [Compton] wavelength of a particle confined inside a black hole [including \cite{pisin}], or tunneling from the interior of a black hole to the exterior spacetime, is therefore strictly \emph{heuristic}, and should be taken with a healthy dose of skeptism.}, and curved spacetime can behave in non-intuitive ways. In fact, since general relativity is a \emph{geometrical} theory, we should pay more attention to understand the implications of a non-trivial geometry with a large interior volume to black hole physics. It is worth commenting here that the spatial volume of a spherical black hole cannot be defined naively as $(4/3)\pi r_{{h}}^3$, where $r_{{h}}$ denotes the horizon of the hole, since the coordinate $r$ does not have geometrical meaning as a ``distance'', and furthermore as we remarked in Footnote 2, $r$ plays the role of time in the interior of a black hole. 

The fact that one could even smoothly glue a nontrivial geometry to a Schwarzschild throat such that the manifold still satisfies the Einstein field equations is actually nontrivial. 
One may suspect that if the geometries are different, due to the long range nature of gravity, asymptotic observers should in principle be able to detect the deviation away from a Schwarzschild manifold with ``trivial'' interior. This is not true. 
A theorem \cite{corvino}, albeit only for metrics that satisfy certain technical assumptions, guarantees that asymptotic observers could not tell the difference:
\begin{quote}
\textbf{Theorem [Corvino (2000)]:} Let $g$ be any smooth, asymptotically flat and scalar-flat metric on $\Bbb{R}^n$, such that the geometry is conformally flat at infinity, with positive ADM mass $M$. Given any compact set $\mathcal{K}$, there exists a smooth scalar-flat metric on $\Bbb{R}^n$ which is asymptotically Schwarzschild and agrees with the original metric $g$ inside $\mathcal{K}$. 
\end{quote}
In other words, take any metric tensor satisfying the technical assumptions of the theorem, one could cut a ball out of the manifold and replaces the exterior region with that of the Schwarzschild geometry;
the details of the gravitational configurations inside the ball is essentially inaccessible to the asymptotic observers\footnote{One recalls from freshmen physics that in Maxwell electromagnetism, one could hide much details about the charge configurations by placing said charges inside a conductor. Such shielding effect does not arise in gravity since there is no ``negative mass'' analogous to negative charge, however Einstein equations are far richer than Maxwell's, and so it turns out that ``shielding'' in the form discussed here is allowed. }. This result has since been extended to the Kerr geometry in \cite{corvino2}. 

\begin{figure}
\centering
    \includegraphics[width=0.75\textwidth]{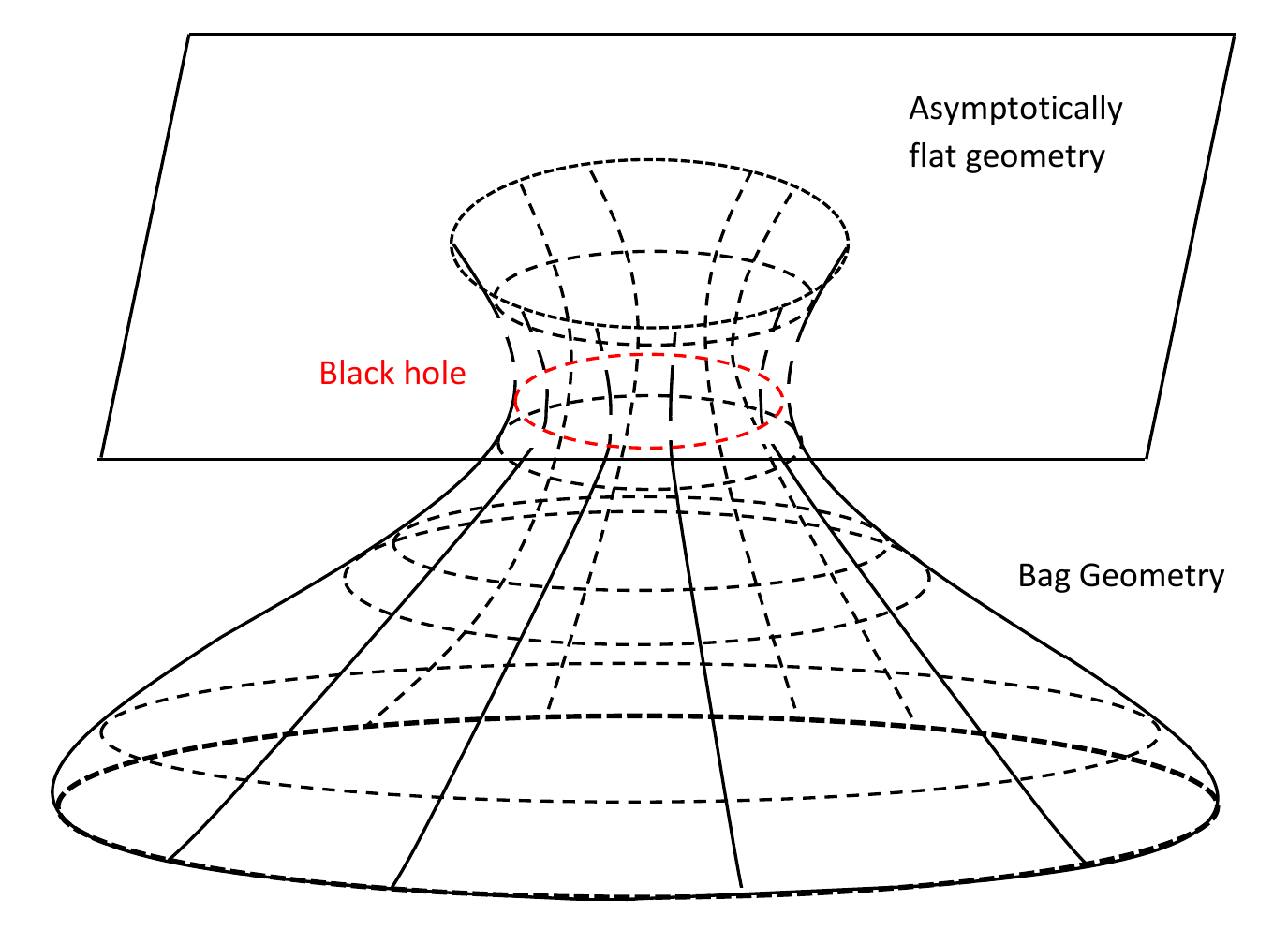}
    \caption[A ``bag-of-gold''-type spacetime.]{(Color online) A ``bag-of-gold''-type spacetime consists of an asymptotically flat geometry glued to a bag across the black hole horizon. That is, what appears to be a black hole to an exterior observer actually contains a potentially unbounded amount of spacetime region inside. Generalizations of the bag-of-gold to spacetimes with other asymptotic geometries can also be made. \label{bag}} 
\end{figure}

For physicists, gluing constructions [not necessarily subject to the technical assumptions of the aforementioned theorem] to build various spacetimes are used frequently.
An example of non-trivial geometry is Wheeler's ``bag-of-gold'' \cite{bagofgold}: what appears as an ordinary black hole to an exterior observer, actually has an \emph{entire universe} inside\footnote{Technically, it is a closed [and thus finite] FLRW universe, but it can be arbitrarily large.}. Such an idea has since been revitalized many times, e.g., Lee Smolin proposed that black holes \emph{create} new universes \cite{smolin1, smolin2} [see also, \cite{Frolov:1988vj, FMM1, 9511136}]. We will thus refer to such geometries, regardless of whether the interior region is finite or infinitely large, as ``bag-of-gold geometries'', or simply as ``bag geometries'' [Fig.(\ref{bag})].

Therefore, one possible way out of the information loss paradox is that the information is simply stored in the large interior within, despite the fact that evaporating black holes eventually become very small with respect to the exterior observers. At the final stages of the Hawking evaporation, as we already mentioned, there is no reason to trust effective field theory to continue to be valid, and therefore it is not out of the question that the evaporation will stop with some kind of ``massive remnant''; or if the evaporation is complete, the two spacetime regions across the horizon may get completely separated, such as in some scenarios involving bubble universes [which we will review in the next section]. Such an idea that the information simply ``ends up somewhere else'' behind the horizon goes as far back as Markov and Frolov \cite{MM0}, and Dyson \cite{dyson}.

More recently, Stephen Hsu and David Reeb \cite{hsu1, hsu2} coined the term \emph{monsters} to describe configurations that possess entropy much greater than their area would suggest [i.e., they possess more entropy than a black hole of the same ADM mass, due to their potentially unbounded interior volume.] Unlike the ``bag-of-gold'', monsters are \emph{not} black holes, i.e., they do not [yet] have a horizon. Perhaps one could argue that they are the kind of configurations that would potentially collapse into a ``bag-of-gold''-type geometry with a horizon. Do ordinary stars somehow eventually develop such monstrous geometry before they collapse into a black hole, which in turn has large interior\footnote{Hsu and Reeb had argued that monsters cannot be formed from any realistic initial data; but they cannot rule out the possibility that a monster can be formed via quantum mechanical fluctuation.}? If so we may be able to store information inside a black hole.

Unfortunately the existence of monsters and ``bag-of-gold'' seems to threaten the AdS/CFT correspondence [for more discussions see \cite{marolf}]. This is because if there can be such objects in AdS spacetime, then the corresponding field theory on the boundary will not have enough degrees of freedom to describe them. [On similar note, Don Page proposed gravitational fireball or ``grireball'', which is a configuration with more entropy than a black hole of the same mass \cite{grireball}. {\color{black}Grireballs have no infinite production problem since they have finite numbers of states up to some mass and size, although a lot more than black holes would have.}] It is thus an important question to investigate -- especially since such configurations are potential solutions to the information loss paradox  -- if they are indeed allowed by physics. 

It turns out that monsters are allowed in general relativity, although this usually means that we have to violate some energy conditions, which is not very surprising since energy conditions are specifically meant to prevent ``unrealistic configurations''. However at the quantum level, energy conditions can be violated [remember that this is how black holes can evaporate in the first place --- Hawking radiation (more precisely the quantum average of the energy-momentum tensor $\left\langle T_{\mu\nu}\right\rangle$) needs to violate the Weak Energy Condition, due to the negative energy of the ingoing Hawking flux; as well as the Null Energy Condition, due to the shrinking of the horizon area.] Therefore, if one wants to claim that monsters are somehow not allowed by the laws of physics, one must look for a prevention mechanism in the quantum theory.


Just such a mechanism -- the Seiberg-Witten instability [\ref{B}] -- was found to render asymptotically locally AdS monsters with spatially flat geometry [e.g. with toral topology] unstable \cite{ycmonster}.

Of course, since monsters are already unstable at the classical level, one has to somehow argue that the Seiberg-Witten instability would actually preclude the \emph{existence} of monsters in the full quantum theory of gravity. The argument is as follows: if the brane action of a configuration is unbounded below and there is no parameter that can be tuned by a finite amount to render the action positive, then physically this suggests that such configuration cannot exist, since no backreaction can bring it to a more stable, new configuration.

In the case of monsters, there is no such parameter, but they can still escape the menace of Seiberg-Witten instability if they collapse to form black holes fast enough, since the instability argument depends on probe brane geometry in the Euclidean version of the spacetime, which does not care about the \emph{interior} of the black hole \emph{horizon}. [{\color{black} It would be interesting to ask what happens in the Lorentzian picture. However in that case one has to be very careful when dealing with the interior of a black hole, due to the non-static nature of the spacetime there. Also, even for ordinary black hole -- i.e., not one formed from the gravitational collapse of a monster -- its interior can still be infinite (more on this point in later discussion), and so the analysis of brane action is delicate, if not impossible to carry out.}] However, it was argued in \cite{ycmonster} that the time scale for the Seiberg-Witten instability to set in is of the same order as that of gravitational collapse to form a black hole. Consequently we cannot be confident that gravitational collapse can save the day [and allow the existence -- however brief -- of a monster]. It was therefore concluded that [sufficiently large] monsters with flat spatial geometry probably do not exist in string theory\footnote{Monsters with spatially hyperbolic geometry are also ruled out. Positive curvature seems to protect against the Seiberg-Witten instability, however we do not know how conclusive this result is. From the view of the AdS/CFT correspondence, quantum field theory is best understood in flat spacetime, in which case its gravitational dual has flat spatial geometry.}. 

Nevertheless, even if monsters do not exist in a full theory of quantum gravity, this does not logically rule out the existence of a non-trivial geometry with large volume inside of a black hole, that somehow only formed \emph{after} a star has completely collapsed into a black hole. 

Lastly we shall discuss another possibility that is perhaps more natural: the interior of a black hole could be ``naturally'' large even without sewing a bag-of-gold through the Einstein-Rosen Bridge. We could very naively see how this might be possible -- the Schwarzschild coordinate $t$ and $r$ exchange roles inside the horizon, i.e. $\partial_t$ is now spacelike. It would therefore be a mistake to think of black hole as a sphere with radius $r_h$. [See also the argument in \cite{mathurped}]. To make any quantitative argument at all, one of course has to really compute the spatial \emph{volume} inside a black hole horizon. Note that the 3-volume is \emph{not} a well-defined concept in GR since it depends on the choice of spacetime slicing. We should therefore \emph{not} assign the naive geometric volume $(4/3)\pi r_h^3$ to a spherical black hole. 
After all, if we look at the classical maximally extended Kruskal-Szekeres spacetime, there is an entire universe [a second asymptotically flat region] inside of the black hole; there is no need to artificially attach a closed FLRW universe as a bag-of-gold to achieve a large interior. The relevant question is of course: what about a realistic black hole which is formed via stellar collapse?

In a recent work \cite{MCCR}, M. Christodoulou and C. Rovelli showed that the statement  ``there exists a 3-dimensional spacelike hypersurface $\Sigma$ bounded by a 2-dimensional surface $\mathcal{S}$ with 3-volume $V$'' has two equivalent statements in Minkowski spacetime $\Bbb{R}^{3,1}$, namely:
\begin{itemize}
\item[(1)] The spacelike hypersurface $\Sigma$ is on the same simultaneity surface as $\mathcal{S}$.
\item[(2)] The spacelike hypersurface $\Sigma$ is the largest spherically symmetric hypersurface bounded by $\mathcal{S}$.
\end{itemize}
Even though the first definition does not extend to curved spacetimes in a unique way, the second does. For Schwarzschild black hole, the authors showed that in terms of the null coordinate $v$,
which is related to polar coordinates in $\Bbb{R}^{3,1}$ by $v = r + t$  at past infinity, the volume inside the sphere $S(v)$, denoted by $V(v)$, is an increasing function of $v$. In fact, 
\begin{equation}
\lim_{v \to \infty} V(v) = 3\sqrt{3} \pi M^2 v.
\end{equation}
Their analysis suggests that for a stellar black hole, the interior volume can in fact be \emph{larger than the observable universe}.

{\color{black}We would also like to remark that, for a baby universe nucleated inside a black hole [see the next subsection], it is connected to an asymptotic infinity [assuming open topology]. Since the 3-volume [with the usual FLRW-type foliation] is infinite, it is not so unreasonable to use our usual intuitions of entropy in flat spacetime, in the sense that infinite volume allows infinite entropy. However, the volume in this current context is calculated on a hypersurface which does not have an asymptotic infinity end. Therefore, even though the volume is large, it is unclear whether we can still invoke the same intuition of entropy, and thus it is also not known if this can provide sufficient entropy to explain the information loss paradox. In short, does curved spacetime mess with our notion of entropy?}

Once again, the lesson is that:  although it is tempting to use flat spacetime intuition and treat mini black holes and remnants as particles, \emph{we should instead pay attention to curved spacetime and take geometry very seriously}.

\subsection{The Bubble Universe Scenario: Does Information Leak to Elsewhere?}

As mentioned above, it is possible that information is stored by a universe that is created inside the black hole. The point is that the interior of a black hole can be ``larger'' than the outside. This can be realized if the interior spacetime undergoes inflation, or the speed of expansion there is faster than that of  the exterior region. This can be achieved in numerous ways, such as via nucleation and tunneling of a small\footnote{By this we mean the false vacuum bubble is smaller than the size of the background horizon.} false vacuum bubble \cite{Blau:1986cw,Farhi:1989yr}. However, unless we consider scalar-tensor gravity or modified gravity \cite{Kim:2010yr}, Einstein gravity does not allow such a small false vacuum bubble to be a regular instanton. Regardless, it is not so difficult to imagine -- although rather vaguely -- that some sort of bubbles can be created from a regular and unitary process via certain mechanisms. [A recent proposal suggests that in the presence of torsion, black holes could harbor a regular, homogeneous, and isotropic universe on the other side of its horizon \cite{Poplawski}.] Of relevance to our discussion here, a recent work by Novikov, Shatskiy, and Novikov investigated the possibility of information exchange between different universes via wormholes \cite{NSN}.

Once we adopt the bubble geometry that is approximately a de Sitter geometry inside but  remains Schwarzschild on the outside [with asymptotically flat \cite{Blau:1986cw}, de Sitter \cite{Aguirre:2005nt}, or anti-de Sitter \cite{Freivogel:2005qh} geometries], there are two kinds of bubble dynamics: either the bubble collapses to form a black hole or the bubble bounces back, expands and eventually inflates. For such a bouncing and inflating bubble, it has two properties: (1) the expanding bubble should be located inside the Schwarzschild wormhole \cite{Blau:1986cw} unless it violates the null energy condition [e.g., \cite{Lee:2010yd}] and (2) the expanding bubble should either begin to form a singularity or it is past incomplete, due to the Farhi-Guth singularity theorem \cite{Farhi:1986ty}. Therefore, in principle, we can contemplate collapsing bubble which does not violate any [known] singularity theorem and hence this can be generated by a regular and unitary process. Such a bubble would be called ``buildable'' [i.e., in principle can be built by an arbitrary advanced civilization; note that monsters are \emph{not} buildable \cite{hsu1, hsu2}]. On the contrary, for bouncing and inflating bubble, it is subject to the Farhi-Guth singularity theorem and hence we can only rely on quantum gravitational process to produce such a bubble, and hence it is dubbed ``unbuildable'' \cite{Freivogel:2005qh}.

\begin{figure}
\begin{center}
\includegraphics[scale=0.5]{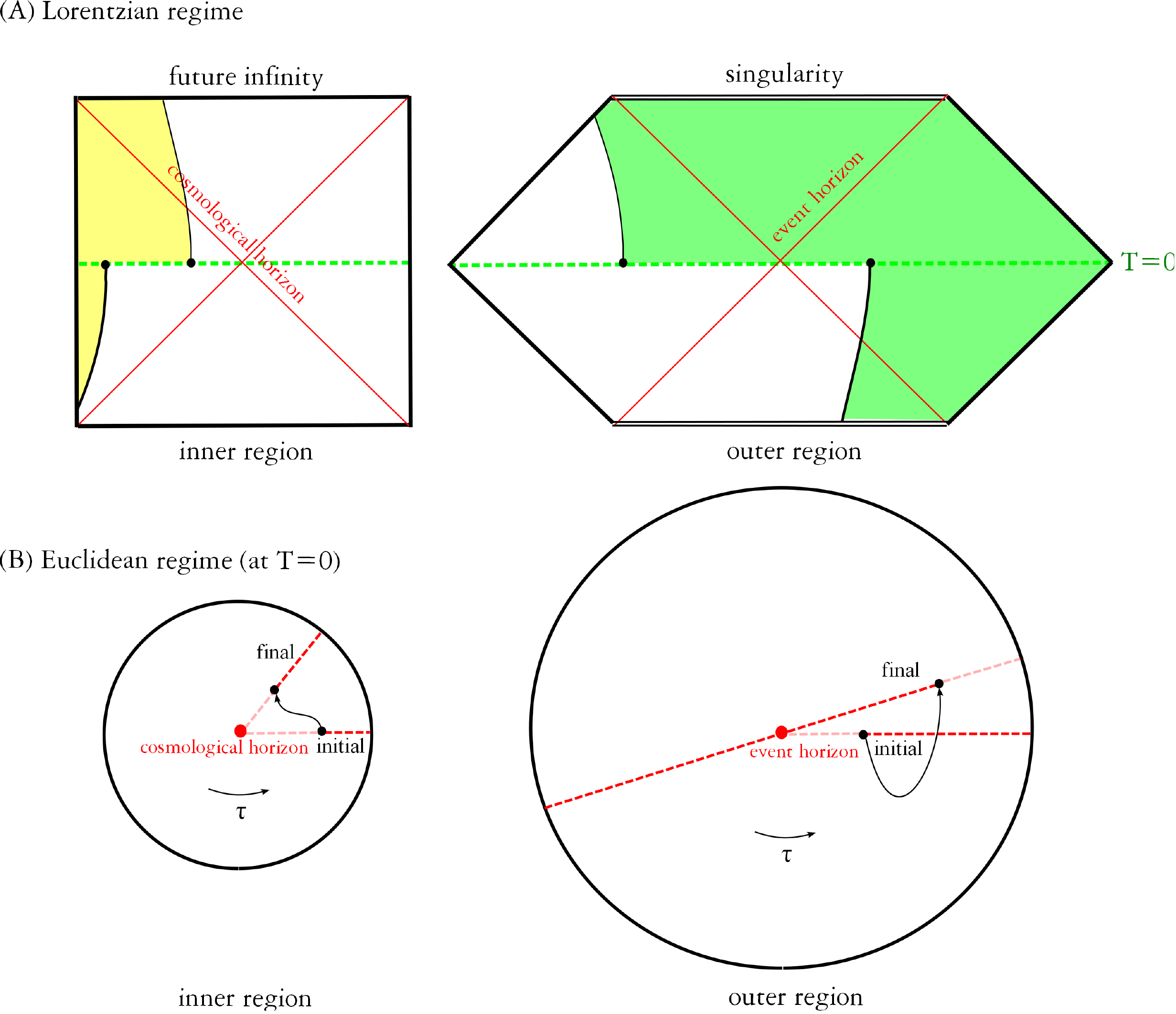}
\caption{\label{fig:FGG}(Color online) The Farhi-Guth-Guven/Fischler-Morgan-Polchinski tunneling in the Lorentzian and Euclidean pictures:  (A) the causal structure of a de Sitter [interior]-Schwarzschild [exterior] system; the two geometries are connected by a thin-shell. The curves denote the trajectory of the shell. For the interior de Sitter space, only the left shaded region is physical. For the exterior Schwarzschild geometry, only the right shaded part is physical. Farhi-Guth-Guven and Fischler-Morgan-Polchinski considered a tunneling from a time-symmetric collapsing shell [that begins from zero size, approaches the maximum radius $r_{1}$, and shrinks back to zero in the end] to a time-symmetric bouncing shell [that begins from infinity, approaches the minimum radius $r_{2} > r_{1}$, and moves to infinity in the end]. The tunneling is defined on the time-symmetric $T = 0$ hypersurface [dashed slices]. (B) On this hypersurface, quantum tunneling is possible. The tunneling can be approximated by an Euclidean instanton that connects $r_{1}$ to $r_{2}$ through the Euclidean manifold: On the left is the Euclidean de Sitter interior geometry [the center is the cosmological horizon and the boundary of the circle is $r = 0$] and on the right is the Euclidean Schwarzschild exterior geometry [the center is the event horizon and the boundary of the circle is $r =\infty$]. It can be shown that the outer Schwarzschild part eventually covers not only the right but also the left side of the maximally extended Schwarzschild spacetime. This implies that after the tunneling, the shell moves across the Einstein-Rosen bridge.
}
\end{center}
\end{figure}

\begin{figure}
\begin{center}
\includegraphics[scale=0.4]{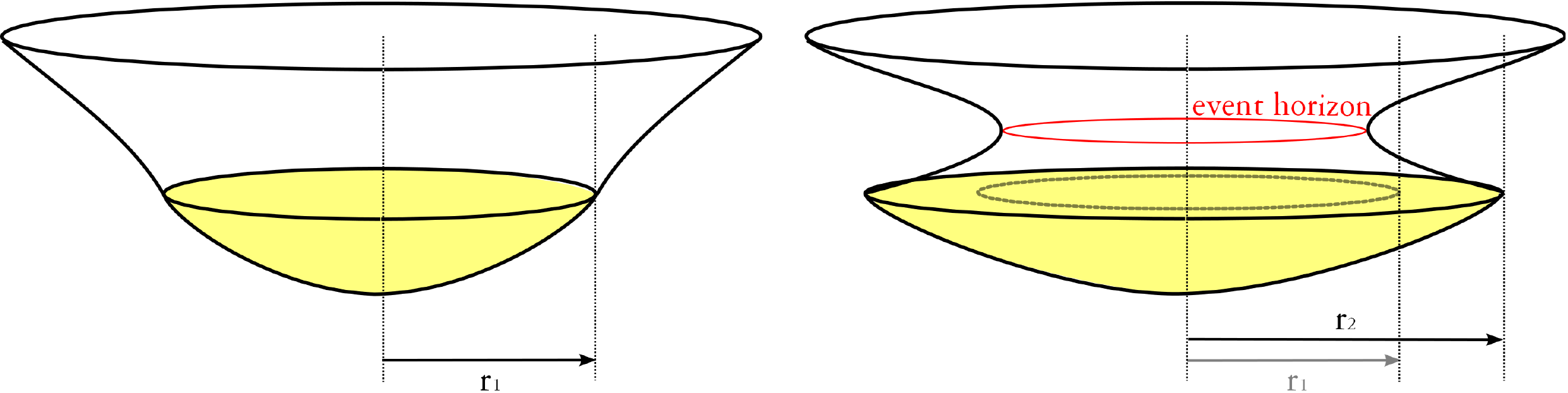}
\caption{\label{fig:FGG_2} (Color online) A conceptual figure for the Farhi-Guth-Guven/Fischler-Morgan-Polchinski tunneling: 
\textbf{Left:} Before the tunneling, inside the shell one finds a de Sitter space [shaded region] with radius $r_1$ and outside of the shell is a Schwarzschild geometry. \textbf{Right:} After the tunneling, the shell increases to a size $r_2$. Eventually, the location of the shell moves beyond the Einstein-Rosen bridge. There exists an event horizon at the throat of the Einstein-Rosen bridge, indicated by the thin curve.}
\end{center}
\end{figure}

The tunneling process that connects a buildable bubble to an unbuildable one was investigated by Farhi, Guth, and Guven [FGG] \cite{Farhi:1989yr} [see also Fischler, Morgan, and Polchinski \cite{FMP1, FMP2}]. See Fig.(\ref{fig:FGG}) and Fig.(\ref{fig:FGG_2}). Although it is suppressed, the probability is non-zero and there is no strong reason to rule out the existence of such a process. On the other hand, once there exists such a process, it seems to violate unitarity for the asymptotic observer \cite{Freivogel:2005qh}; in the sense that a free-falling particle that goes into the second asymptotic region inside the black hole [assuming the spatial section has non-compact topology] will lose its correlation to the original universe outside of the black hole. Hence, if we trust AdS/CFT and unitarity of the AdS asymptotic observer, this process should not be allowed. Therefore, some authors \cite{Freivogel:2005qh} think that FGG tunneling should be prohibited for AdS backgrounds, although there is no constructive argument in the bulk picture of why this should be the case.

Of course, even concerning ``buildable'' universes, the relevant question in the context of information loss is whether, given generic initial conditions, a black hole would, during the final phase of gravitational collapse, ``re-explode'' into some other region of spacetime. This is problematic, as Misner, Thorne and Wheeler \cite{MTW} explained:
\begin{quote}
``\emph{Such a process requires that the `exploding' end of the wormhole be built into the initial conditions of the universe, with mass and angular momentum (as measured by Keplerian orbits and frame dragging) precisely equal to those that go down the black-hole end. This seems physically implausible. So does the `explosion'}.''
\end{quote}

Regarding the information loss paradox, we are faced with the following critical questions:
\begin{enumerate}
\item \textit{Does such tunneling process happen in principle?} There is no strong argument \emph{against} the possibility of FGG tunneling. Theoretically, with less assumptions imposed, such an inflating universe can indeed be allowed. For example, this happens in scalar-tensor gravity based models \cite{Kim:2010yr,Lee:2010yd} or for the case in which space is filled with negative energy photons  generated by false vacuum bubbles \cite{Yeom:2009mn1, Yeom:2009mn2, Yeom:2009mn3}. 
\item \textit{Does the process happen for every black hole?} If such process is \emph{the}  resolution of the information loss paradox, then this should happen for every black hole, but this does not seem to be the case since the probability of such process happening is suppressed and remains very small. 
\end{enumerate}

In conclusion, a bubble universe inside the black hole may contain enough information and such process may indeed be possible in principle [at least, for an asymptotically flat or de Sitter geometries], though this is likely not going to happen inside a \emph{generic} black hole and hence does not seem to be a viable resolution of the information loss paradox.

Finally, we remark on the possibility that instead of giving rise to bubble universes inside, \emph{some} black holes could \emph{themselves serve as bubble nucleation sites} \cite{1401.0017}. This means that these black holes would end up as flat spacetime, losing their horizons in the process, and presumably allow information to escape.  Nevertheless, this process also does not seem to happen to all black holes.

\section{Discussion}\label{discussion}

Black hole evaporation, including its end-state, remains mysterious after 40 years since its discovery. Despite the more mainstream view that black holes should completely evaporate away, the existence of a remnant is well-motivated from various quantum gravity theories and models. Despite the usual reasons put forward against remnants, such as the infinite production problem, none of these objections are free of loopholes. In other words, black hole remnants remain a 
reasonable candidate for the end point of Hawking evaporation. Although naively [in asymptotically flat case] the curvature diverges at the endpoint of evaporation, this may be modified from quantum gravitational corrections; these lead to three if not more possibilities: (1) a naked singularity\footnote{{\color{black}Whether naked singularities exist, and how one might distinguish them from black hole candidates observationally, is of course an interesting question by itself. See, e.g., \cite{0710.2333, 1206.3077}.}}, (2) complete evaporation which ended with a flat spacetime or, (3) the Hawking temperature remains finite, so does the curvature, and the black hole evolution approaches a final well-defined end point. The remnant scenario is related to the last possibility. As we have seen, there are many models that point toward this direction. However, most of them do not consider a more realistic scenario in which a black hole carries electrical charge and/or angular momentum. As we argued in subsec.(\ref{concluding remarks}), the presence of even a minute amount of electrical charge could affect the evolution of an isolated black hole significantly over its long lifetime. Although these are unlikely to be important in the real universe, the point is that the information loss paradox [as well as the firewall paradox] can be formulated in such an idealized world, and thus any proposal of resolution to the paradoxes needs to be examined for the charged case to see if it still holds.  
In conclusion, the investigation of the remnants as well as the study of their phenomenological implications will be an important direction for future research.

We also still do not know the true nature of Hawking radiation, that is, whether the radiation contains any information of the in-falling matter. The information loss paradox remains unsolved today, and is in fact made worse by the firewall debate. Amidst all these discussions, the remnant scenario has received little attention. There are probably two reasons for this omission: firstly, it has to do with the fact that remnants had been argued to be problematic, and secondly, the way a remnant solves the information loss paradox is not in line with what most workers in the field would interpret as ``solving'' the problem. We now summarize our analysis for each of these reasons.

For problems regarding the existence of remnants, there are basically two major arguments -- the over-production of remnants and the violation of CPT invariance. The latter argues that since it is possible to form a black hole by concentrating nothing more than photons or gravitons, by CPT symmetry it should also be possible for a black hole to completely dissolve into photons and gravitons and leaves nothing behind. This argument is wrong for the same reason that CPT invariance at the fundamental level does \emph{not} preclude \emph{practically} irreversible processes in daily life in accordance with the second law of thermodynamics; after all, our universe has an arrow of time. Furthermore, CPT invariance actually requires one to feed the black hole remnant with the CPT-reversal of the Hawking particles before it can potentially turn back into a star [or other initial configurations that formed the black hole].

The over-production problem is non-trivial, but it is not without loopholes. The usual argument is that, since there are potentially infinite number [or at least an extremely large number $N$] of possible remnants [due to the various different initial conditions that could collapse into a black hole], even though the production rate of any specific remnant from the quantum vacuum is exponentially suppressed, this rate is multiplied by an arbitrary large number $N$ and therefore the total production would be sizable. The fact that we do not see remnants being copiously produced thus rules out remnant scenario. The validity of this objection boils down to whether we could treat remnants as particles in the quantum field theoretical sense, in light of the possibility that they may actually store a large amount of information, i.e. whether effective field theory is well-defined for remnants, especially if they have non-trivial interior geometries \cite{sabine}. Central to this argument is also the assumption that the suppression for production rate goes like $\exp(-M^2)$ even for remnants, but it could be that it is the statistical entropy, $S_{\text{st}}$,instead of the Bekenstein-Hawking entropy [$S_{\text{bh}} \propto M^2$, and $S_{\text{st}} \gg S_{\text{bh}}$] that should enter the exponent, i.e., $\exp(-S_{\text{st}})$. In addition, following the standard QFT formulation, the production rate should be divided with the ``incoming flux'', i.e., normalized properly with the number of initial states.
In any case, we feel that the over-production of remnants requires a more detailed investigation.

{\color{black} If black hole remnants exist, a relevant question is whether they are absolutely stable. Perhaps only a very few species of particles or
other entities within the universe are absolutely stable, such as massless gravitons, massless photons, the lightest neutrino, the lightest supersymmetric
particles, and perhaps the lightest magnetic monopole\footnote{\color{black}The authors thank Don Page for raising this important issue.}. There might be a small additional number of stable entities if there are other
absolutely conserved quantities besides 4-momentum, angular momentum,
electric charge, magnetic charge, and R-parity. The latter may not even be absolutely conserved once black hole formation and evaporation is taken into account. Also, given enough time, it is possible for a black hole remnant to tunnel into a black hole or other configurations, releasing massless particles in the process.}

Regarding the information loss paradox and firewall, we find that even if black hole remnants exist, it does not seem to ameliorate the problem of information loss. If remnants satisfy the weak form of the Bekenstein-Hawking entropy interpretation and has potentially infinite capacity of information, then it can potentially overcome the usual problems posed against remnants, e.g. the infinite pair-production problem. However, such an object violates the strong form interpretation of Bekenstein-Hawking entropy and thus does not fit well with the AdS/CFT correspondence, which is widely believed to be a salient feature of all consistent theories of quantum gravity. Black hole remnants also cannot evade the firewall unless Hawking radiation does not carry away any information.  
However this conclusion rests on taking ``solving the information loss paradox'' to mean that asymptotic observers should be able to recover all the information. In principle, however, there does not appear to have any problem if information is stored in a remnant which could well have a large interior geometry. This may not agree with the AdS/CFT correspondence, but we certainly should ask: what is the regime of validity for AdS/CFT, which after all, is still a conjecture? We know that AdS/CFT works very well for large black holes where gravity is weak, however very little is known in the case where the black hole is small in the bulk. If various approaches to quantum gravitational models [e.g., the generalized uncertainty principle] are indeed correct, then even the Bekenstein-Hawking area law receives correction terms [typically a logarithmic correction] and ceases to be strictly proportional to the horizon area. In short, we feel that it is premature to rule out remnants based on holographic argument. 


Having discussed monsters, bag-of-gold geometries, and even the argument that black hole interior is ``naturally large'', the main lesson is that we should not use flat spacetime intuition in dealing with black holes. \emph{Geometry is the central pillar of the edifice of general relativity}. It is only right that one pays more attention to what geometry has to say. 
We should also try to understand black hole singularities with a similar spirit, instead of sweeping it under the carpet with the hope that quantum gravity will eventually take care of that\footnote{\color{black}There  are of course some efforts already invested in achieving precisely this, see, for example, the recent review by Uggla \cite{1304.6905}.}.
Let this be a reminder as we celebrate the centenary of GR in 2015.

As the final point in this paper, it is also worthwhile to consider the following question: is there really such a thing as \emph{the} solution to the information loss paradox? After all, one sometimes speculates that the dark sector may actually be quite rich and there could be more than one type of dark matter. Similarly, there are various types of black holes with different dimensions, of varying topologies and distinct asymptotic geometries. It also seems that some black holes can be destroyed by quantum gravitational effect or tunneling, instead of ending as remnants. 
So should there indeed be a universal solution to the information loss paradox, or perhaps different black holes preserve unitarity in different ways? 

We shall end by stating the obvious: this review -- despite offering some new thoughts -- is clearly not a final say on the role of remnants in black hole physics. We hope that this work would stimulate the interest of the community in this fascinating topic.

\section*{Acknowledgement}
The authors thank Taiwan's National Center for Theoretical Sciences [NCTS], Taiwan's Ministry of Science and Technology [MOST], and the Leung Center for Cosmology and Particle Astrophysics [LeCosPA] of National Taiwan University for support. Part of this work was carried out in Paris while P.C was visiting Coll\`ege de France, Paris Diderot University's Astroparticle Physics and Cosmology Center [APC], and \'Ecole Polytechnique during the fall of 2014. Y.C.O thanks Nordita for support to finalize this paper in Paris with P.C. D-h.Y also thanks the JSPS Grant-in-Aid for Scientific Research (A) No. 21244033. The authors wish to thank Don Page, Bill Unruh, Gabriele Veneziano, Ingemar Bengtsson, Brett McInnes, Sabine Hossenfelder, Thomas Roman, and Piero Nicolini for various suggestions, comments, and discussions. The authors also thank Yi-Hsien Du for assisting in some literature search at the beginning phase of this work. 

\appendix

\section{A Plethora of Entropies}\label{A}

It is said that von Neumann once told Shannon to call his measure of missing information ``entropy'' because \cite{avery}
\begin{quote}``\emph{In the first place, a mathematical development very much like yours already exists in Boltzmann's statistical mechanics, and in the second place, no one understands entropy very well, so in any discussion you will be in a position of advantage.}''
\end{quote}

In the context of black hole physics, the most well-known concept of entropy is the Bekenstein-Hawking entropy, which is [at least for the leading term] proportional to the area of the black hole. Restoring all the constants, the expression in its full glory is
\begin{equation}
S_{\text{bh}}=\frac{k_B c^3A}{4G\hbar}.
\end{equation}
Although temperature and entropy are both thermodynamical concepts, the entropy of a black hole is perhaps far more mysterious than its temperature. The Hawking temperature can be understood as a consequence of QFT on curved spacetime, however the Bekenstein-Hawking entropy depends on the field equations of gravity and cannot be determined simply using QFT on a background metric [see Lesson 6 of \cite{Pad}].

It is a common belief in the literature that $S_{\text{bh}}$ measures the capacity of black holes to store quantum information. The fact that $S_{\text{bh}}$ is proportional to the area instead of volume is heralded as an important hint that quantum gravity is holographic. However, as pointed out in a recent review on black hole thermodynamics \cite{1402.5127}, such a behavior is also present in non-black hole systems:
\begin{quote}
``[...] \emph{without invoking anything as extreme as a black hole, that this is
something we should expect from General Relativity. Even a somewhat mundane
system like a sufficiently massive ball of radiation has an entropy that is not proportional
to its volume.''}
\end{quote}
One interesting feature in the expression of $S_{\text{bh}}$ is that, formally the role of $\hbar$ is only to give a finite value [and the correct dimension] to the entropy. Note that $\hbar$ cancels off in the first law of black hole mechanics, since it occurs in the numerator in the expression of the Hawking temperature. Indeed, it has recently been argued that black hole thermodynamics makes sense even without an underlying statistical mechanical interpretation \cite{curiel2}.

Despite the importance of the Bekenstein-Hawking entropy, we have to be open-minded regarding the possibility that it is \emph{not} the only entropy measure for a black hole. It was already pointed out by, e.g. Frolov \cite{facade}, that black holes may have \emph{statistical entropy} $S_{\text{st}}$ that is distinct from its thermal, Bekenstein-Hawking entropy $S_{\text{bh}}$. The interior of black holes may therefore contain information which is \emph{not} accounted for by $S_{\text{bh}}$, and therefore the information content can be larger than the bound set by $S_{\text{bh}}$. If this is true, then the Bekenstein-Hawking entropy is only a property of the horizon. This is known as the ``weak form'' interpretation. On the other hand, the ``strong form'' interpretation holds that Bekenstein-Hawking entropy is a property of both the horizon and the interior spacetime. Marolf \cite{marolf} provided a simple thought experiment that lends at least some support to the weak form interpretation:
\begin{quote}
``[...] \emph{one starts with a black hole of given mass $M$, considers some large
number of ways to turn this into a much larger black hole (say of mass $M'$), and then lets
that large black hole Hawking radiates back down to the original mass $M$. Unless information
about the method of formation is somehow erased from the black hole interior by the process
of Hawking evaporation, the resulting black hole will have a number of possible internal states
which clearly diverges as $M' \to \infty$. One can also arrive at an arbitrarily large number of
internal states simply by repeating this thought experiment many times, each time taking
the black hole up to the same fixed mass $M' > M$ and letting it radiate back down to $M$.
We might therefore call this the ‘Hawking radiation cycle’ example. Again we seem to find
that the Bekenstein-Hawking entropy does not count the number of internal states''}.
\end{quote}

From the perspective of quantum information theory, one natural interpretation for the Bekenstein-Hawking entropy is that it arises from the quantum entanglement between the interior spacetime and the exterior one \cite{9404039}, that is, it is an \emph{entanglement} entropy. In the usual picture of unitary evolution, one hopes that if a pure state collapses into a black hole, the end state should be pure as well. 
Suppose the black hole completely evaporates and we are left with a flat spacetime populated with Hawking particles at the end. Then purity can be restored if the radiation contains [highly entangled form of] information, by having the latter radiation to eventually purify the exterior -- earlier emitted -- radiation. Black hole evaporation will then be as unitary as the burning of a book; the information is not lost in principle [but lost in practice]. It can be shown that the entanglement entropy increases at first as Hawking radiation started, but eventually turns over about the mid point of the evolution, and reduces to zero as purity is recovered at the end. The plot of the entanglement entropy against time is known as the ``Page curve'' \cite{page1, page2}. The time it takes for the entanglement entropy to start decreasing is called the ``Page time''. If the black hole does not completely evaporate, then the interior and exterior states remain as mixed state when considered separately. Nevertheless if Hawking radiation carries information, the degree of entanglement would still evolve and one can still make sense of the Page curve and the Page time, as we have discussed in subsec.(\ref{3.3}). It was recently suggested that there is a form of quantum entanglement that has weight, since it affects the gravitational field \cite{DEB}. The implication of this result to the interpretation of black hole entropy as entanglement entropy has yet to be investigated. Of course, the correction due to this effect is extremely small and thus unlikely to change the aforementioned interpretation in any significant way. 

Lastly, we comment on the possibility that gravity itself may carry some kind of ``gravitational entropy''. Penrose asserted that the singularity in a black hole is fundamentally different from the Big Bang singularity: in a gravitational collapse one would expect that the Weyl curvature will generically be quite large, while Weyl curvature should vanish or at least stay small during the Big Bang. In Penrose's proposal, now known as the Weyl Curvature Hypothesis \cite{WCH}, the magnitude of Weyl curvature directly measures the gravitational entropy. This agrees with the second law of thermodynamics --- one should expect the entropy to grow during gravitational collapse, but remains small in the very beginning. There has since been various proposals to understand gravitational entropy, see e.g. \cite{1411.5708,0711.1656}. However, it is fair to say that not much is known about the subject at this point, not to mention the relationship between gravitational entropy and $S_{\text{bh}}$, or for that matter, the relationship between gravitational entropy and the information content of a black hole. The point is that, it is too early to conclude that only $S_{\text{bh}}$ measures the capacity of information storage.

\section{The Brane Action and Pair-Production Instability}\label{B}

Seiberg and Witten showed that if the brane action for a probe brane defined on the Wick-rotated spacetime becomes negative, the spacetime becomes unstable \cite{kn:seiberg, KPR, Barbon, 1411.1887}. 
Specifically, given a Bogomol'ny-Prasad-Sommerfield [BPS] brane $\Sigma$ in a Wick-rotated $D$-dimensional spacetime [$D=n+1$], the brane action in the appropriate unit is a function of the radial coordinate $r$ defined by
$
\mathcal{S}[\Sigma(r)] = \mathcal{A}(\Sigma) - (D-1)\mathcal{V}(\Sigma),
$
where $\mathcal{A}$ is proportional to the area of $\Sigma$ and $\mathcal{V}$ to its volume. [Of course one could also discuss the physics in the Lorentzian picture.]
That is,
\begin{equation}\label{SWaction}
\mathcal{S}[\Sigma(r)]=\Theta (\text{Brane Area}) - \mu (\text{Volume Enclosed by Brane}),
\end{equation}
where $\Theta$ is related to the tension of the brane and $\mu$ is related to the charge enclosed by the brane due to the background antisymmetric tensor field. The type IIB backgrounds
assumed here are of Freund-Rubin type \cite{FR}. That is to say, the $\text{AdS}_{n+1} \times S^{9-n}$ spacetime metric is supported by the background antisymmetric field strength. In string theory, the existence of such a flux field naturally induces compactification so that the full 10-dimensional spacetime reduces to a product manifold $\text{AdS}_{n+1} \times S^{9-n}$, where the factor $S^{9-n}$ is compactified\footnote{This can be generalized to any geometry of the form $X^{n+1} \times Y^{9-n}$, where $X^{n+1}$ is an $(n+1)$-dimensional non-compact asymptotically hyperbolic manifold [generalizing $\text{AdS}_{n+1}$] and $Y^{9-n}$ is a compact manifold [generalizing $S^{n+1}$].}.   

We see that the action will become negative if the volume term grows larger than the area term. The most dangerous case is when the charge $\mu$ reaches its maximal value: the BPS case with $\mu=n\Theta/L$. Explicitly, the Seiberg-Witten brane action is given by
\begin{equation}
\mathcal{S}[r]= \Theta ~r^{n-1} \int d\tau \sqrt{g_{\tau\tau}} \int d\Omega  - 
\frac{n\Theta}{L} \int d\tau \int^r dr' \int d\Omega ~r'^{n-1}\sqrt{g_{\tau \tau}}\sqrt{g_{r'r'}},
\end{equation} 
where $g_{\tau\tau}=-g_{tt}$, with $t$ Wick-rotated; and $L$ is the AdS length scale.
[The correspoding \emph{Lorentzian} action is defined in a straight-forward manner.]

The Seiberg-Witten instability is a non-perturbative instability that occurs due to an uncontrolled brane productions \cite{kn:seiberg, KPR}. Indeed, brane-anti-brane pairs are always spontaneously produced from the AdS vacuum. This phenomenon is analogous to the well-known Schwinger effect in quantum electrodynamics, with the rate of brane-anti-brane pair production being proportional to $\exp(-\mathcal{S})$, where $\mathcal{S}$ is the Seiberg-Witten brane action. If the brane action becomes negative, the AdS vacuum will nucleate brane-anti-brane pairs at an exponentially large rate [instead of being exponentially suppressed].

\end{document}